\documentclass[preprintnumbers,prd,twocolumn,showpacs,floatfix,
letterpaper,nofootinbib,superscriptaddress]{revtex4}

\usepackage[latin1]{inputenc}
\usepackage{graphicx}
\usepackage{amssymb}
\usepackage{color}
\usepackage{float}
\usepackage{amsmath}
\usepackage{amsfonts}
\usepackage{dcolumn}
\usepackage{hyperref}
\usepackage{bm}
\usepackage{verbatim}
\usepackage{setspace}
\usepackage{epsfig}

\begin{document}

\title{Higher dimensional non-Kerr black hole and energy extraction }

\author{Sushant G. Ghosh\footnote{Corresponding author.}}\email{sghosh2@jmi.ac.in, sgghosh@gmail.com}
\affiliation{Centre for Theoretical Physics,\\ Jamia Millia Islamia, New Delhi 110025, India}
\affiliation{Astrophysics and Cosmology Research Unit, School of Mathematical Sciences,\\University of KwaZulu-Natal, Private Bag 54001, Durban 4000, South
Africa}

\author{Pankaj Sheoran$^{1}$}\email{hukmipankaj@gmail.com}

\begin{abstract}

We investigate the properties of the horizons and ergosphere in a rotating higher dimensional (HD)
deformed Kerr-like black hole. We also explicitly bring out the effect of deformation parameter 
$\epsilon$ and the extra dimension on the efficiency of the Penrose process of energy extraction from a black hole.
It is interesting to see that the ergosphere size is sensitive to the deformation parameter $\epsilon$ as
well as spacetime dimensions $D$. This gives rise to a much richer structure of the ergosphere in a HD non-Kerr black hole, thereby
making the Penrose process more efficient compared with that of the four-dimensional Kerr black hole.
\end{abstract}


\maketitle

\section{Introduction}
The rotating black hole are formed when a star can no longer
support itself against its own gravitational collapse, thereby
compressing to a point. Energy extraction from a rotating black hole
interests us not only as engines of relativistic jets from active
galactic nuclei and quasars \cite{mei} but also as fundamentals of
black hole physics \cite{cs}. An interesting process called the Penrose
process allows one, in principle, to extract energy from a
rotating black hole \cite{pen} and relies on conservation of momentum
and energy. In the Penrose process, one shoots a massive particle
inside the ergosphere, and then it splits into two particles, one of
which has negative energy and one of which has positive energy.
Penrose showed that the negative energy particle would go down the
black hole, but the positive energy particle could escape, carrying with it
more energy than it came in with.  It turns out the negative energy
particle will slow down the spinning of the hole and reduce its
energy, and thus it provides an important method to extract energy from
a black hole \cite{pnfl,chr,bpt,bdd,pwdd}. Recently the Penrose process was
also extended  to the five-dimensional (5D) supergravity rotating black hole
\cite{pd}, to higher dimensional black holes and black rings \cite{nm},
to the Ho\v{r}ava-Lifshitz gravity black hole \cite{aa}, to the Kerr-NUT black hole \cite{aasra},
 and  also to a rotating black hole with a global monopole \cite{cja}.

Motivated by examining the no-hair theorem, Johannsen and Psaltis
\cite{jpc} recently applied the Newman-Janis \cite{nj} complex
transformation to the deformed Schwarzschild solution \cite{ys} and
constructed a Kerr-like black hole solution. In addition to $M$ and $a$,
this spacetime has at least one more parameter: it can be seen as a
deformation parameter $\epsilon$ that measures potential deviations
from the Kerr geometry. This rotating black hole possesses some striking
properties; e.g., as the deformation parameter $\epsilon >0$, the black hole
possesses two disconnected spherical horizons for a high rotation
parameter and has no horizon. When $\epsilon <0$, the horizon always
exists for the arbitrary $a$. Soon several researchers used a non-Kerr
black hole in various astrophysical applications
\cite{aasr,alc,bmp,bma,bmb,bmc,bmd,cjb,jt,jta,jtb,jtc,kz,lcj}. The
properties of the ergosphere and energy extraction by the Penrose
process in a rotating non-Kerr black hole were investigated \cite{lcj}. It
turns out that for $\epsilon>0$ it has been observed that a black hole
becomes more prolate than the standard Kerr black hole, whereas it is more
oblate for $\epsilon<0$ \cite{jpd}, thereby affecting the size of
the ergosphere and in turn the efficiency of the Penrose process \cite{lcj}.

It is rather well established that higher dimensions (HD) provide a
natural playground for the string theory and they are also required
for its consistency. It is interesting to study the HD extension of
Einstein's theory and, in particular, its black hole solutions \cite{resh}.
The HD generalization of Schwarzschild and Reissner-Nordstrom black holes
were obtained by Tangherlini \cite{ta} and the rotating black hole by Myers and
Perry \cite{mp}. There is a growing realization that the physics of
HD black holes can be markedly different and much richer than in four
dimensions \cite{mp,resh}. In this paper we focus our attention on
the energy extraction via the Penrose process in a HD non-Kerr black hole to study
the role of deformation parameter $\epsilon$ and extra dimensions in
the efficiency of the Penrose process.
We start with a review  of HD non-Kerr solutions in Sec. II; the subsection studies the behavior
of horizons and the ergosphere with respect to dimensions and the
deformation parameter; Sec. III analyzes the equations of motion of
particles and their motion at the equatorial plane in the vicinity
of the HD non-Kerr black hole.  We also obtain the negative energy states
for a test particle with a specific angular momentum, orbiting
around the black hole, as a function of the deformation parameter and
finally we succinctly summarize our main results and evoke some
perspectives to end the paper in Sec. IV.

\section{HD non-Kerr black holes} \label{spt}
The rotating black hole, in four dimension (4D) general relativity,  is
described by the Kerr solution \cite{ker}, which is completely specified
by the mass $M$ and angular momentum $a$. In 4D there is only one
possible rotation axis for a rotating black hole and only one angular
momentum. However, in its HD counterpart, the Myers-Perry rotating
black hole, there is a multitude of angular momentum parameters, each referring to
a particular rotation plane. Here, we focus on the simplest case for which
there is only one angular momentum parameter, namely $a$. The
rotating non-Kerr black hole metric \cite{jpa} was extended to HD by Ghosh
and Papnoi \cite{gp}. Beginning with a deformed HD Schwarzschild
solution and applying the Newman-Janis transformation, they
constructed a deformed HD Kerr-like metric with three parameters:
the mass $M$, one rotation parameter $a$, and the deformation
parameter $\epsilon$.  The metric in $(N+3)$ dimensions for the spinning non-Kerr black hole in
the standard Boyer-Lindquist-like coordinates \cite{gp} reads
\begin{eqnarray}\label{metric0}
ds^2 &=& g_{tt} dt^{2}+g_{rr} dr^{2}+g_{\theta\theta} d{\theta^{2}}\nonumber\\
   &+& g_{\phi\phi} d{\phi^2}+2 g_{t\phi}dt d{\phi}+r^2 \cos^{2}\theta d\Omega^2_{N-1},
\end{eqnarray}
with
\begin{eqnarray}
g_{tt} &=& -\left[1+h(r,\theta)\right]\left[1-\frac{\mu}{r^{N-2}\Sigma}\right], \nonumber\\
g_{rr} &=&  \frac{\left[1+h(r,\theta)\right]\Sigma}{\Delta+a^2 \sin^2 \theta h(r,\theta)}, \nonumber\\
g_{\theta\theta} &=& \Sigma,\nonumber\\
g_{\phi\phi} &=& \left[\left(r^2+a^2+\frac{\mu}{r^{N-2}\Sigma}a^2\sin^2\theta\right)\sin^2\theta\right. \nonumber \\
&  & + \left. h(r,\theta)a^2\sin^4\theta\left(1+\frac{\mu}{r^{N-2}\Sigma}\right)\right],\nonumber\\
g_{t\phi} &=& -\left[1+h(r,\theta)\right]\frac{\mu}{r^{N-2}\Sigma}a
\sin^2\theta,
\end{eqnarray}
with
\begin{equation}
\Sigma  =  r^2+a^2\cos^2\theta
\hspace{0.4cm} \Delta  =  r^2+a^2-\frac{\mu}{ r^{N-2}}.
\end{equation}
We define the function $h(r,\theta)$ in HD as
\begin{equation}
h(r,\theta) =  \frac{\epsilon\mu^{3}}{8 \Sigma^2 r^{3N-4}},
\end{equation}
and
\begin{equation}\label{ome}
d\Omega^2_{N-1}  =  d\chi_{1}^2 + \sin^2\chi_{1}\left(d\chi_{2}^2 +
\sin^2\chi_{2}\left(\cdots d\chi_{N-1}^2 \right)\right)
\end{equation}
is the metric of the unit $\left(N-1\right)$ sphere \cite{al}. 
The metric (1) is the HD generalization of the Johannsen and Psaltis metric
\cite{jpa}.  It becomes the well-known Myers-Perry black hole in the limit when  $h(r,\theta) $ vanishes.
In general relativity, the Einstein tensor of the HD non-Kerr metric is nonzero
unless $h(r,\theta)$ vanishes. Therefore, we regard the HD non-Kerr metric as a vacuum spacetime of
an appropriately chosen set of modified gravity field equations that are unknown but definitely different from
the Einstein equations for nonzero $h(r,\theta)$.  While this does not mean
that the metric does not make sense, it does, as  we justify the
nature of our metric, where we show that its properties are very similar to the ones of the Myers-Perry black
hole and the Kerr black hole (in 4D). In particular, we compute the location of the horizons and discuss their properties.

 The definition of the function $h(r,\theta)$ in HD is just an extension of its 4D
definition \citep{ys}. The square root of the determinant of the
metric (\ref{metric0}) reads as
\begin{equation}\label{detg}
\sqrt{-g} = (1+h) \sqrt{\Phi} \Sigma r^{N-3} \sin{\theta}
\cos{\theta}^{N-3},
\end{equation}
 where $\Phi$ is the determinant of the metric (\ref{ome}). Here $\mu$ is an integrating constant
that can be related to mass $M$ of the black hole via
\begin{equation}\label{M}
M = \frac{\left(N+1\right)A_{N+1}\mu}{16 \pi},
\end{equation}
$a$ is the angular momentum defined as
\begin{equation}\label{J}
J = \frac{A_{N+1}\mu\hspace{0.1cm} a}{8 \pi},
\end{equation}
and $A_{N+1}$ is the area of a unit $(N+1)$ sphere given by
\begin{equation}
A_{N+1} = \frac{2 \pi^{\frac{N+2}{2}}}{\Gamma \left(\frac{N+2}{2}\right)},
\end{equation}
From Eqs. (\ref{M}) and (\ref{J}), we get
\begin{equation}\label{jm}
\frac{J}{M} = \frac{2 a}{N+1}.
\end{equation}
In the 4D limit $(\mu = 2M$ and $N=1)$, the metric (\ref{metric0})
reduced to the non-Kerr black hole discovered in \cite{jpa}, and then the function
$h(r,\theta)$ becomes
\begin{equation}
h = \frac{\epsilon M^{3} r}{\Sigma^{2}},
\end{equation}
which is exactly the same as derived in \cite{ys,kz,jpa}. Further, we
discover the standard Kerr black hole in the general relativity limit $(\epsilon\rightarrow0$
and $N\rightarrow1)$. The standard Myers-Perry black hole \citep{mp} with a
single rotation parameter is recovered for vanishing deformation
parameter $\epsilon$. When the rotation parameter $a$ vanishes, one
may get a deformed Schwarzschild solution \citep{ys}. It may be noted that the 4D non-Kerr metric \cite{jpc} is not a 
solution of $R_{ab} = 0$. It is kind of a perturbative way in order to include various possible deviations from the Kerr solutions
in modified theories of gravity.

\subsection{Horizons and ergosphere}
 As $\epsilon = 0$, the black hole is
reduced to the typical Kerr black hole known in general relativity \cite{jpa}. Our aim here
is to discuss the effect of the extra dimension on the structure of
horizons and the ergosphere.  As $\epsilon = 0$, the black hole is reduced to
the usual Myers and Perry black hole .  Similar to the Myers and Perry black hole,
the above metric has two types of hypersurfaces or horizons: a
stationary limit surface or infinite redshift surface  and an event horizon. The static
limit  gets its name from the prediction that for radii smaller
than the Schwarzschild radius but greater than that of the horizon, an
observer cannot remain at rest and cannot stay static. It requires the
prefactor of $d{t}^2$ to vanish:
\begin{equation}\label{tls}
1 - \frac{\mu}{r^{N-2} \Sigma} =0\;\; \mbox{or}\;\;
r^{N} + a^{2}\cos^{2}\theta\; r^{N-2} - \mu = 0,
\end{equation}
where we have assumed that $1+h \neq 0$ as the surface defined by $1
+ h = 0$ is an intrinsic singularity  and cannot be the infinite
redshift surface  \cite{kz,lcj}. The surface of no return is
known as the event horizon. The event horizon must satisfy $g_{t\phi}^{2}-g_{tt}g_{\phi\phi}=0$, 
and $\epsilon > 0\; (<0)$ leads to more prolate (oblate) object than
 the 4D Kerr black hole \cite{jpc, kz, lcj}. The event horizon of the black hole is located at the outer root of
the
\begin{eqnarray}\label{eh}
&& \Delta +  a^2\; h\; \sin^2\theta=0,\; \nonumber \\ &&\mbox{i.e.},\; r^{N} + a^{2} r^{N-2} - \mu + h\; a^2 \sin^2\theta r^{N-2} = 0.
\end{eqnarray}
Clearly, the radii of the event horizon depend on $\theta$, which are
different from that in the  usual Kerr case, in which it is
independent of $\theta$. In 4D, for the small negative values of $\epsilon$, the spacetime has closed event horizon \cite{kz}.
On the contrary, $\epsilon>0$ may lead to disconnected event horizon \cite{kz}. Thus, the stationary limit surface and event
horizon depends on the spacetime dimension. However, it is seen that Eqs. (\ref{tls})
and (\ref{eh}) have at least one positive root for HD ($D\geq6$),
i.e., just one event horizon and stationary limit surface in HD independent of the magnitude of $a$.
This is a typical characteristic of the HD black hole and holds for the HD non-Kerr black hole as well.  In the limit $a \rightarrow 0$,
Eqs.  (\ref{tls}) and (\ref{eh}) coincide with the event horizons of the nonrotating black holes \cite{mp},
and they admit trivial solution $r_{+} = (\mu)^{1/N}$.   

Considering only the outer event horizon and stationary limit surface,  it can be verified that the
stationary limit surface always lies outside the event horizon in all dimensions. Hence, as in 4D, we
call the region between the stationary limit surface and the event horizon as the ergosphere.  The ergosphere is
the region that lies outside of a black hole. In the ergosphere it is possible
to enter and leave again, and the object moves in the direction of the
spin of the black hole. It has been shown that it is possible, at least
theoretically, to extract energy from the black hole in this region. The
ergosphere for the 4D Kerr black hole has an oblate spherical shape. Interestingly,
the HD non-Kerr black hole may have two horizons for small values of
deformation parameter $\epsilon$ even for $D\geq6$. Thus in contrast
to the Myers and Perry black hole has only one horizons for $D\geq6$.

\begin{figure*}
\begin{tabular}{ | c | c | c | c | }
\hline
\includegraphics[scale=0.3]{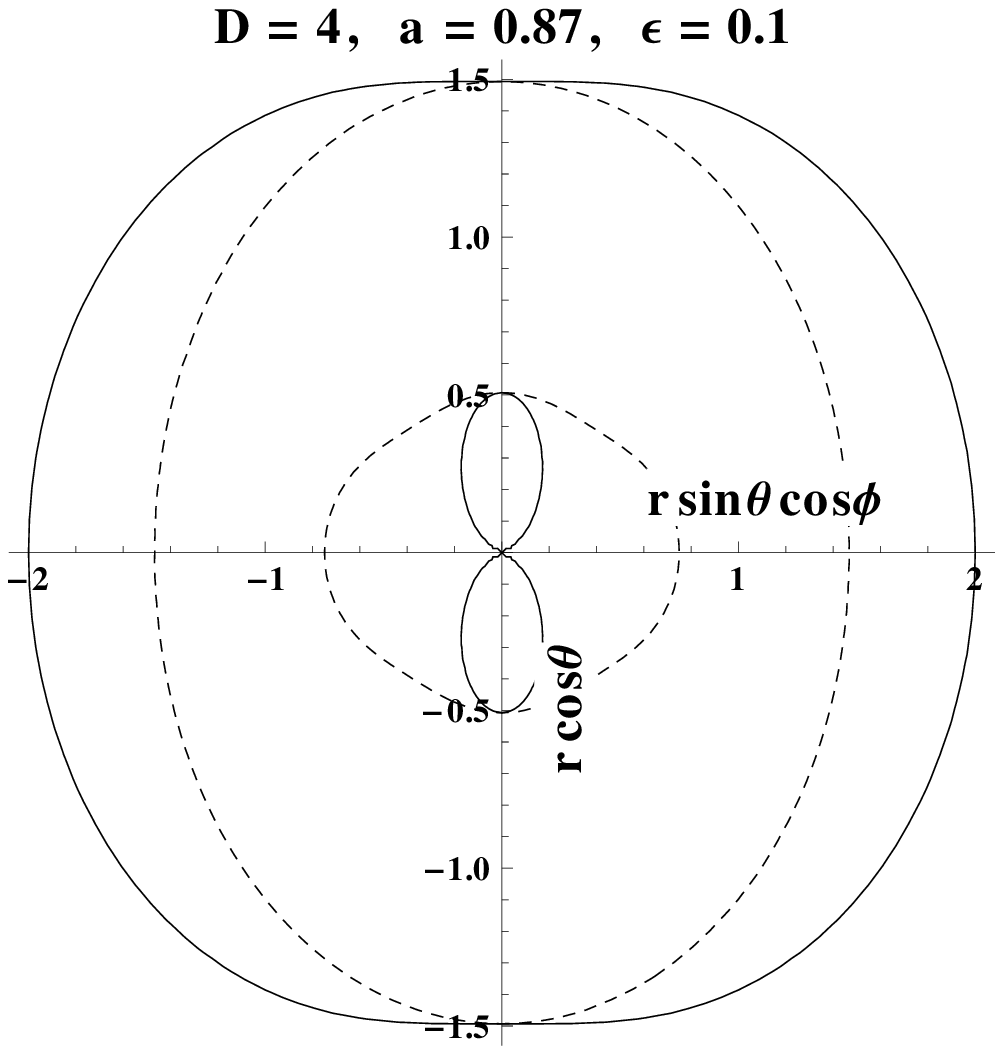}
& \includegraphics[scale=0.3]{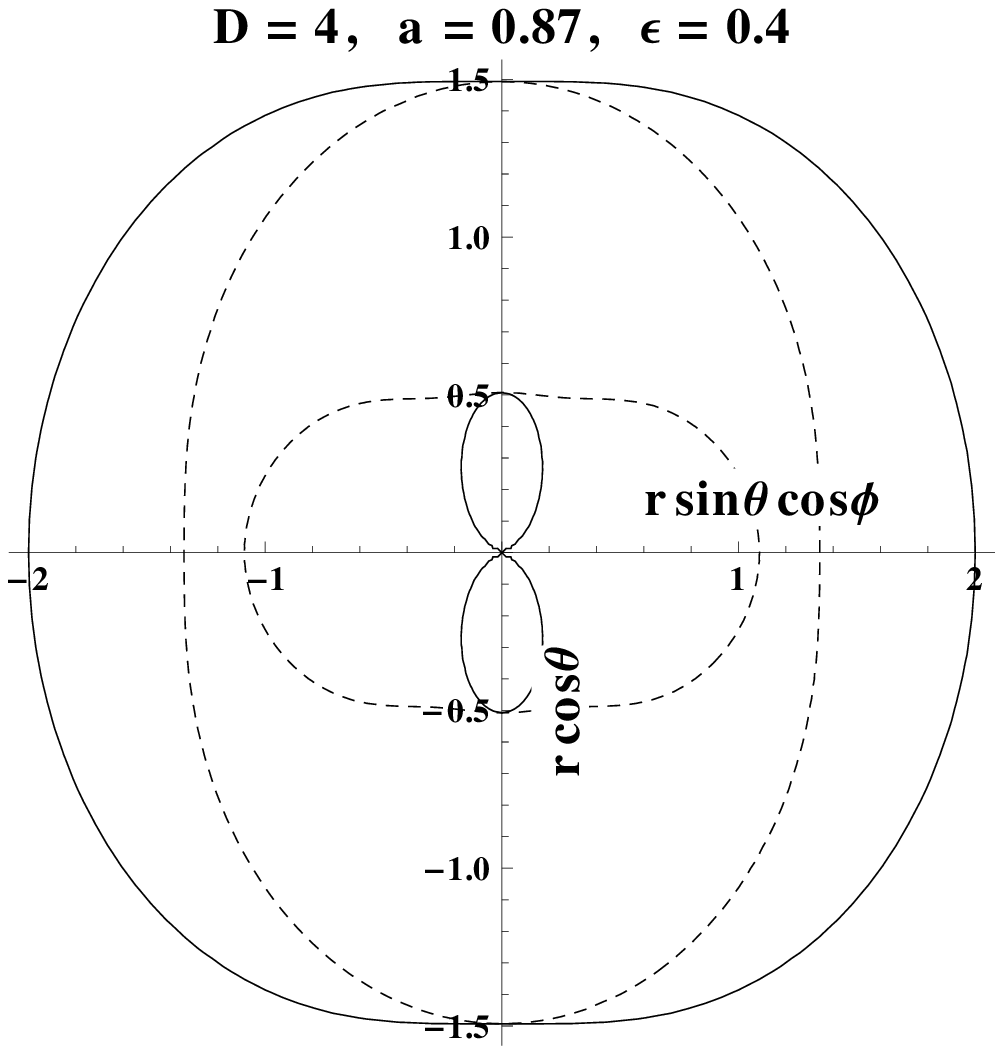}
& \includegraphics[scale=0.3]{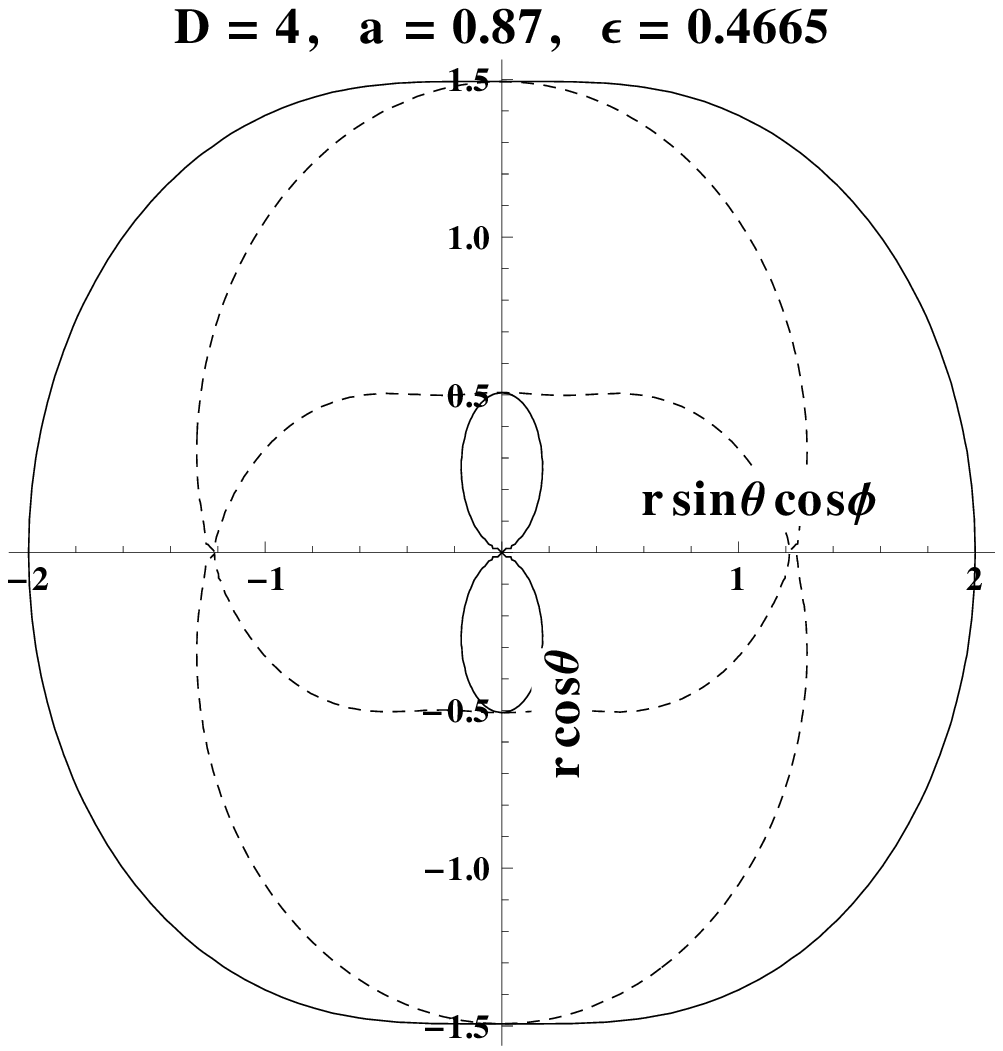}
& \includegraphics[scale=0.3]{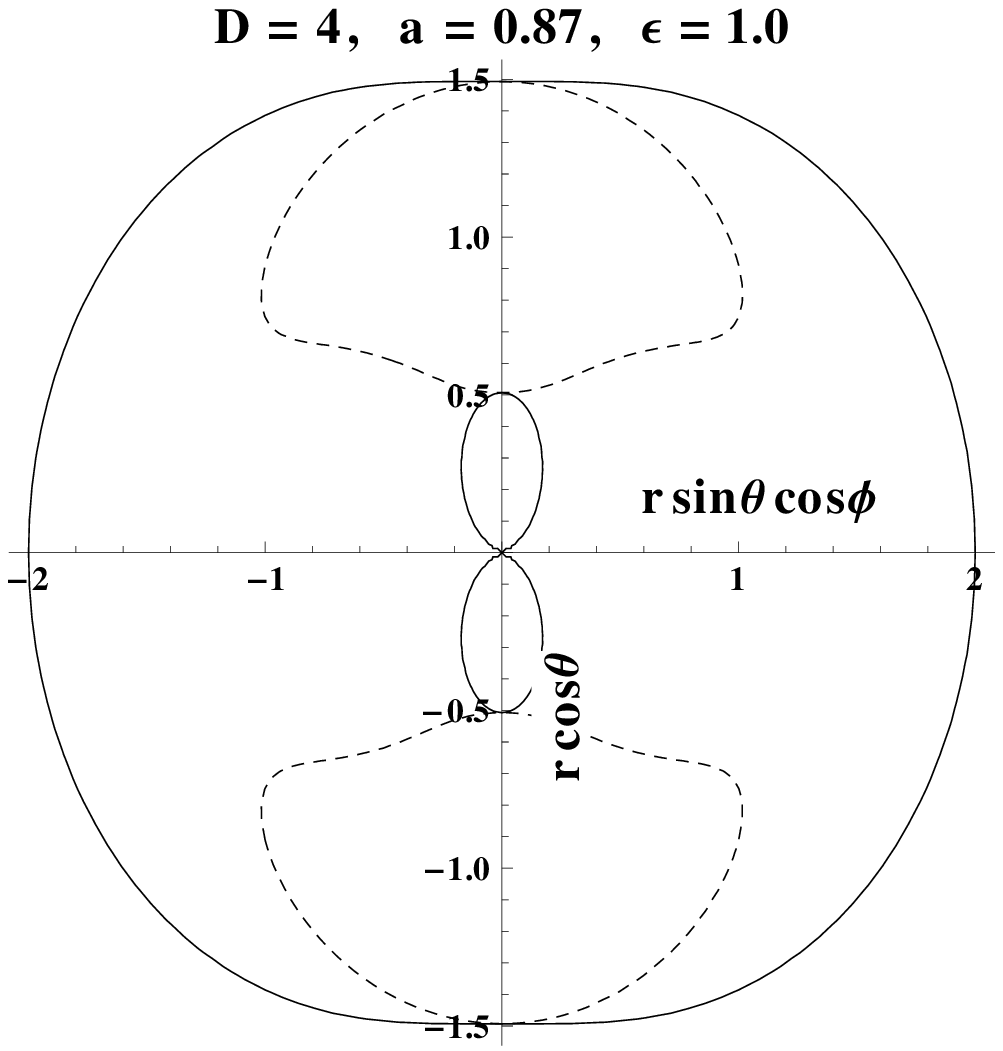}\\
\hline
\includegraphics[scale=0.3]{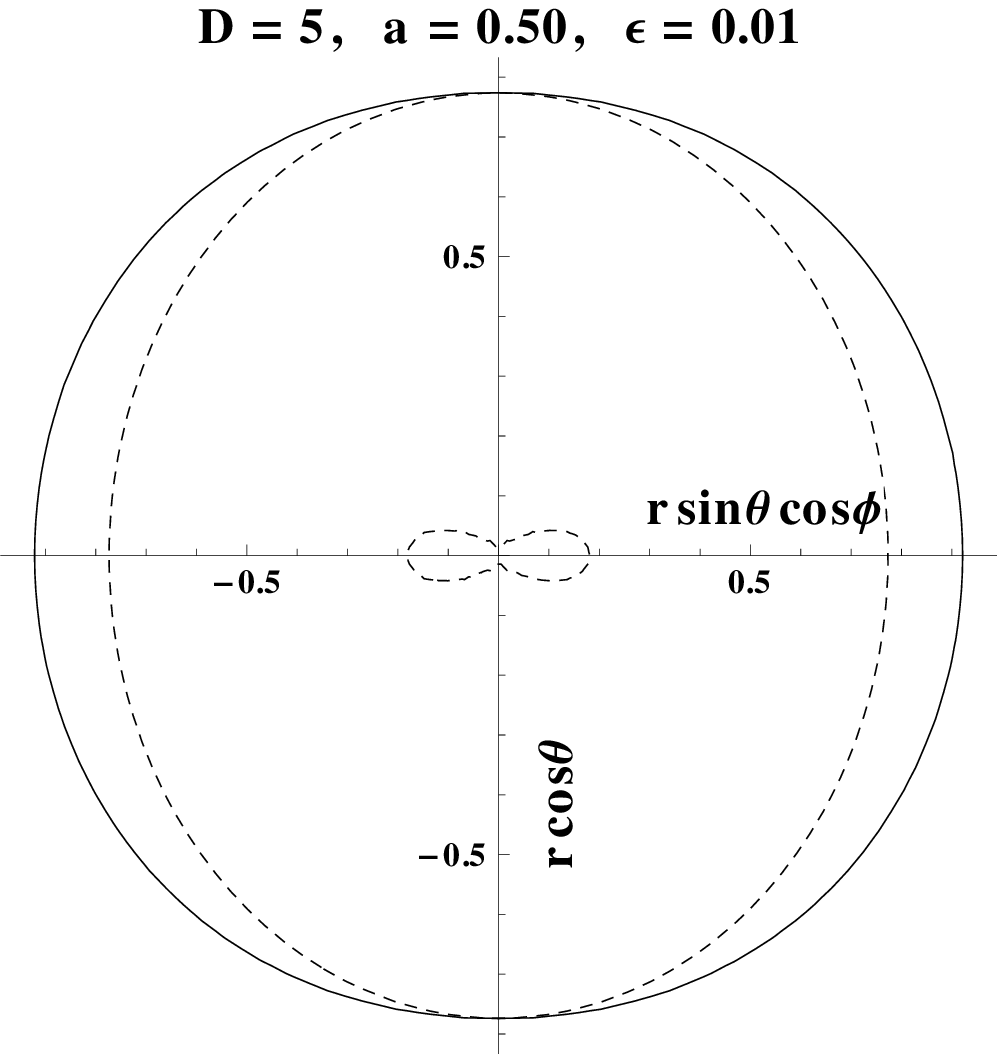}
& \includegraphics[scale=0.3]{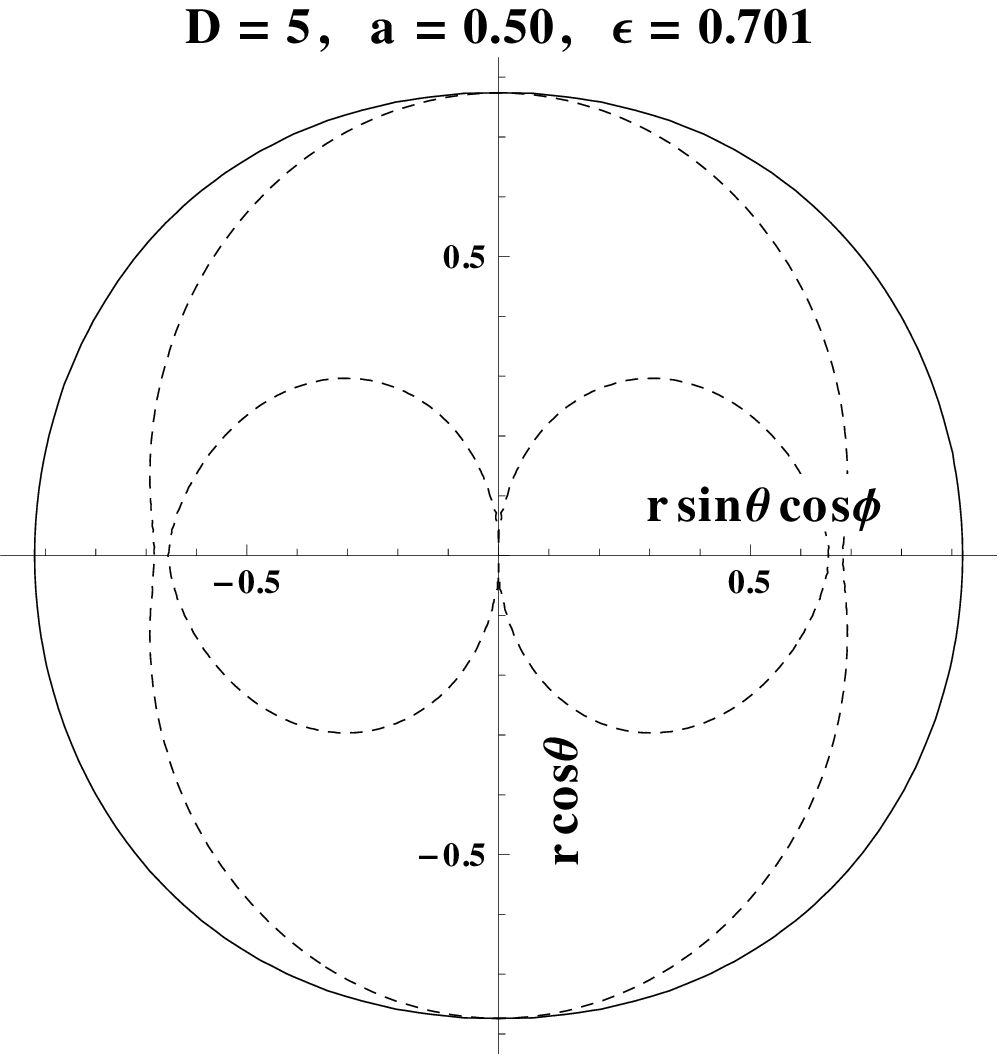}
& \includegraphics[scale=0.3]{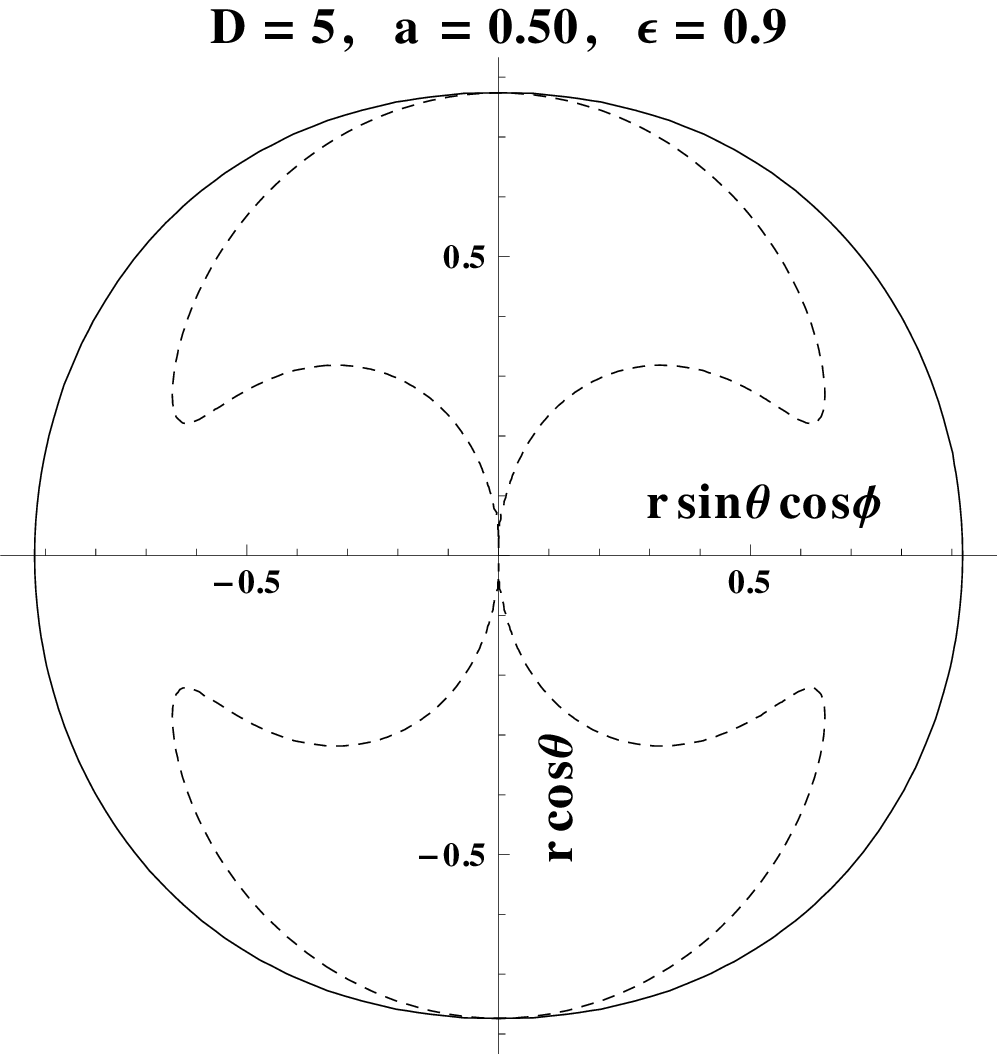}
& \includegraphics[scale=0.3]{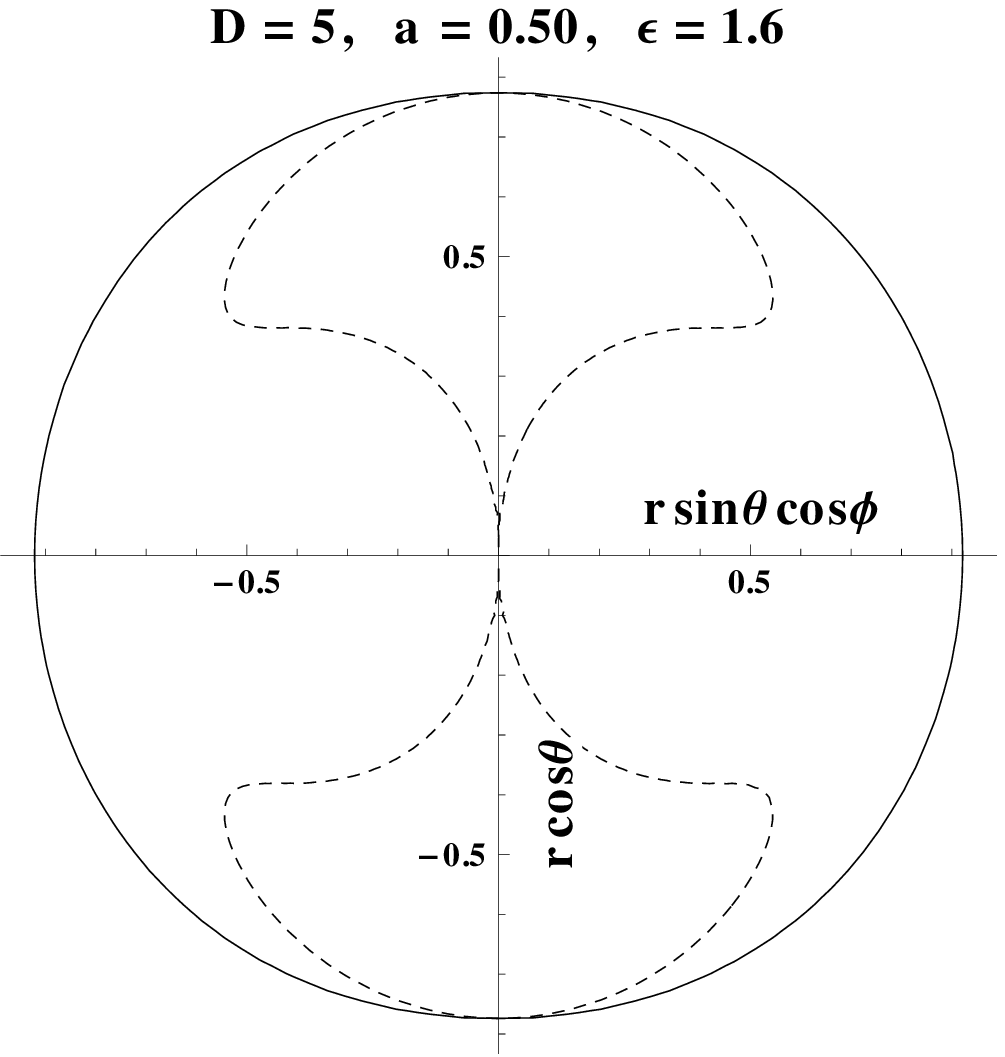}\\
\hline
\includegraphics[scale=0.3]{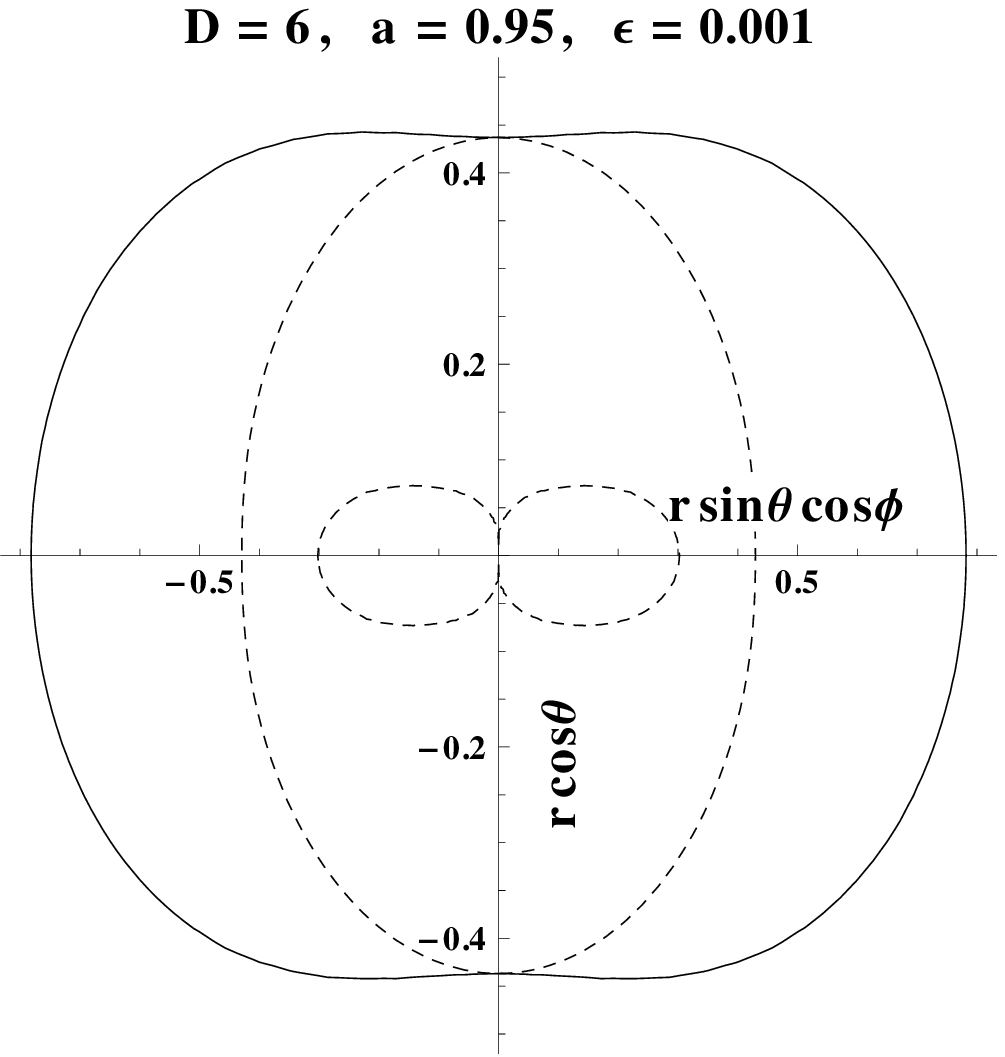}
& \includegraphics[scale=0.3]{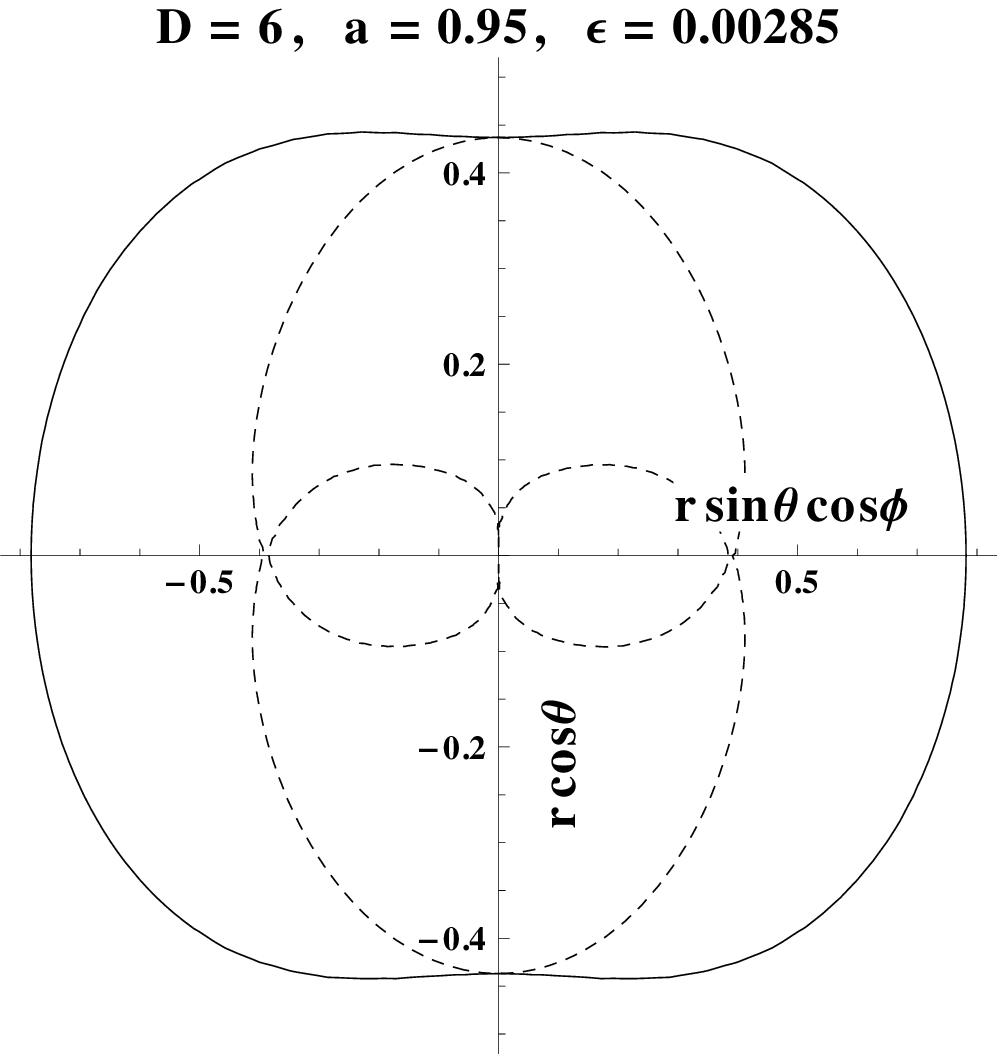}
& \includegraphics[scale=0.3]{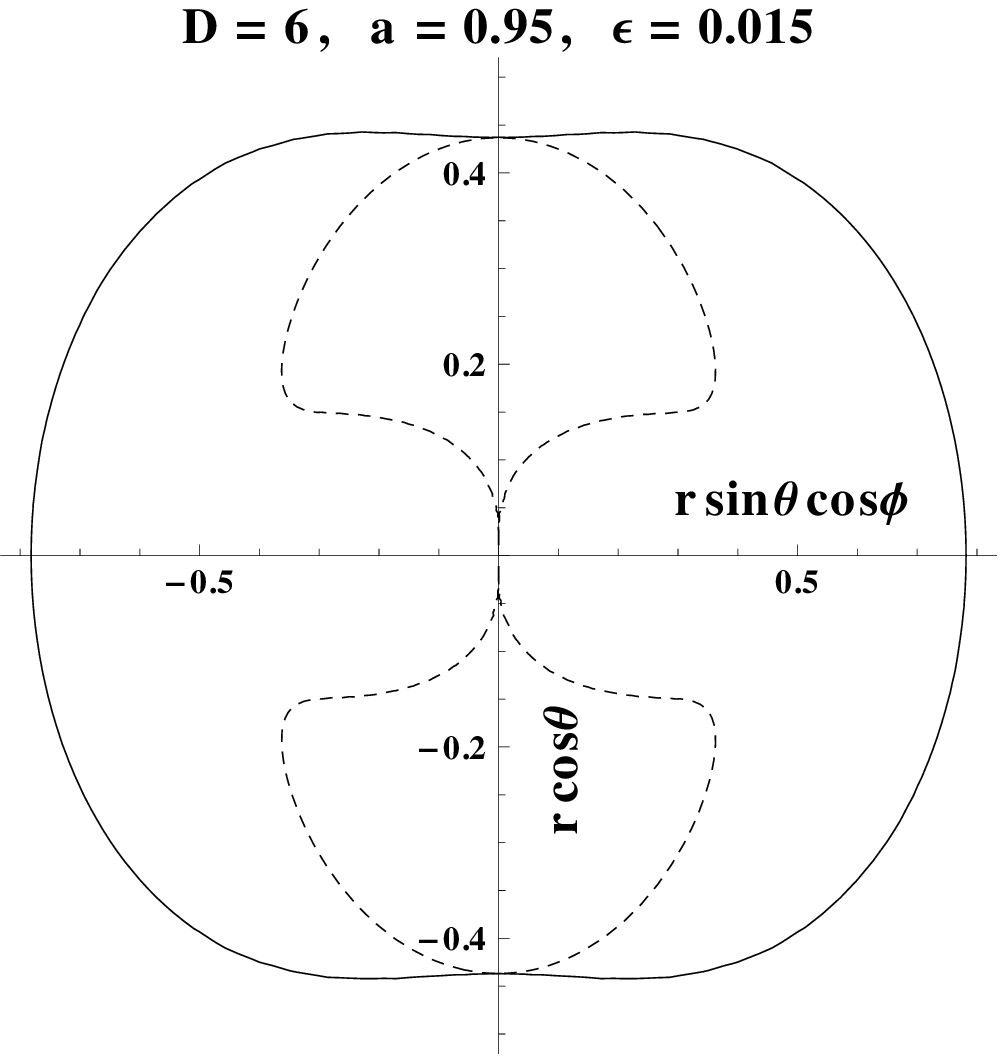}
& \includegraphics[scale=0.3]{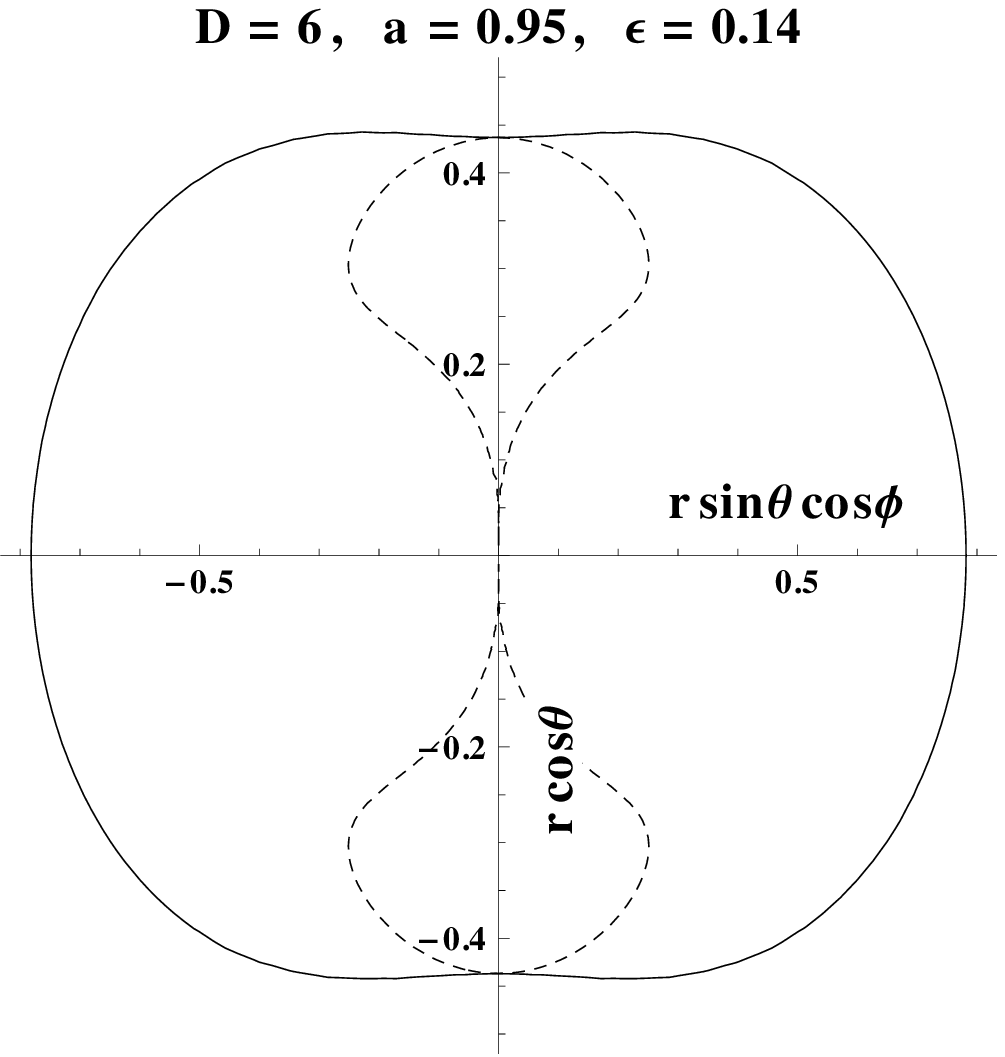}\\
\hline
\includegraphics[scale=0.3]{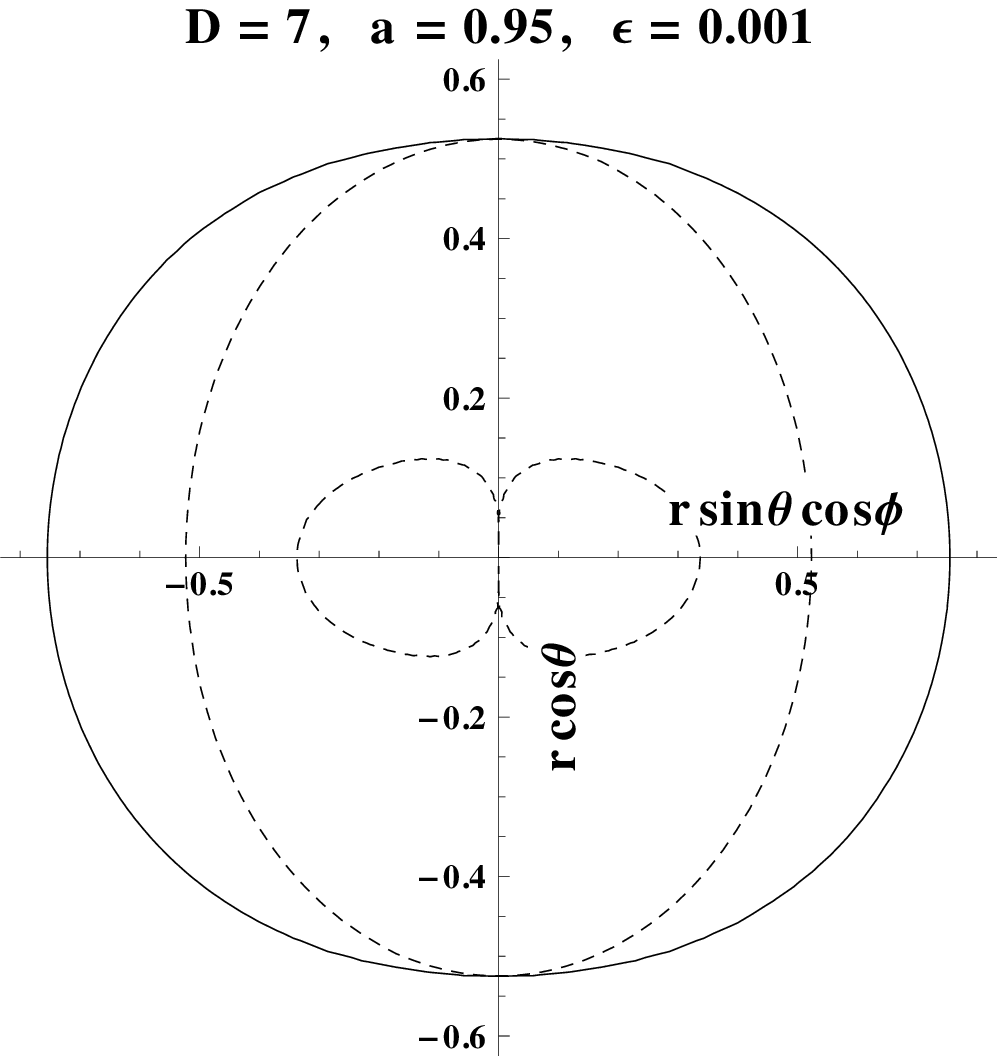}
& \includegraphics[scale=0.3]{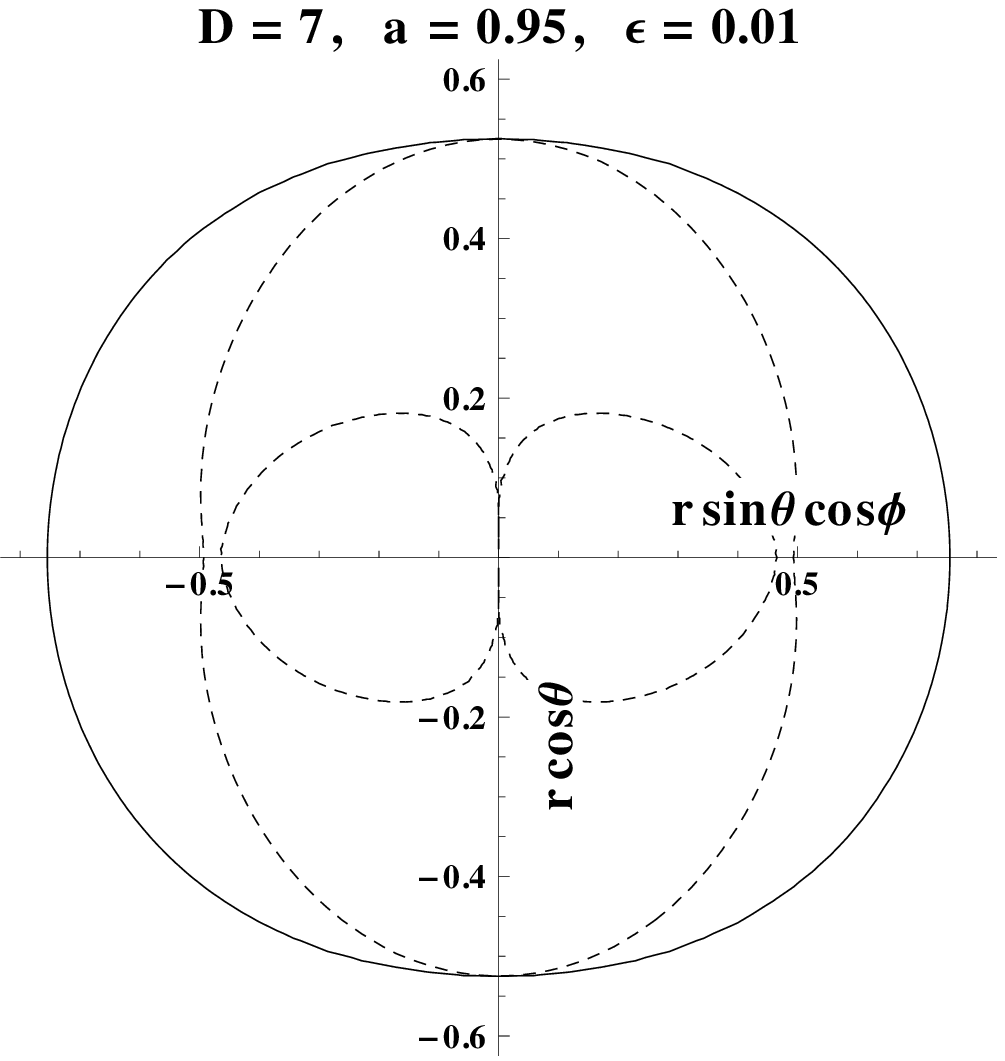}
& \includegraphics[scale=0.3]{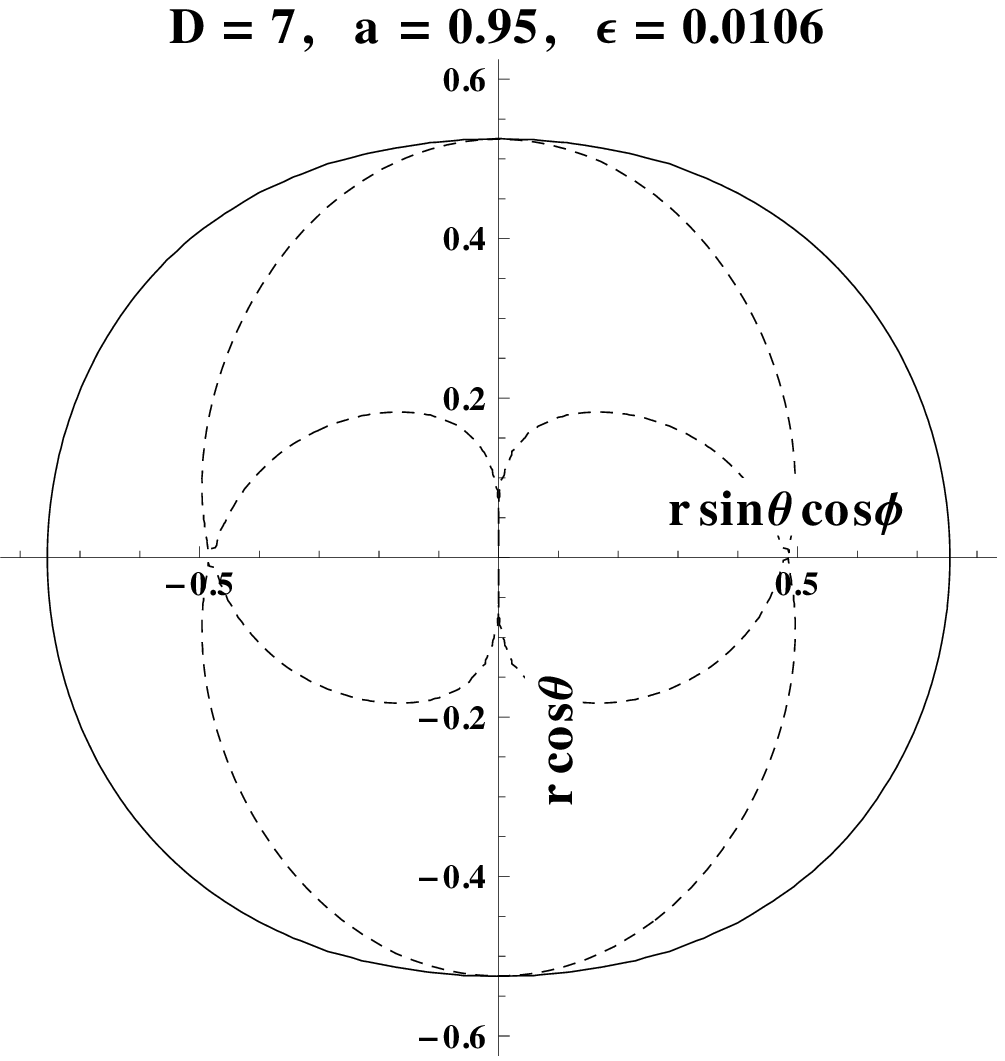}
& \includegraphics[scale=0.3]{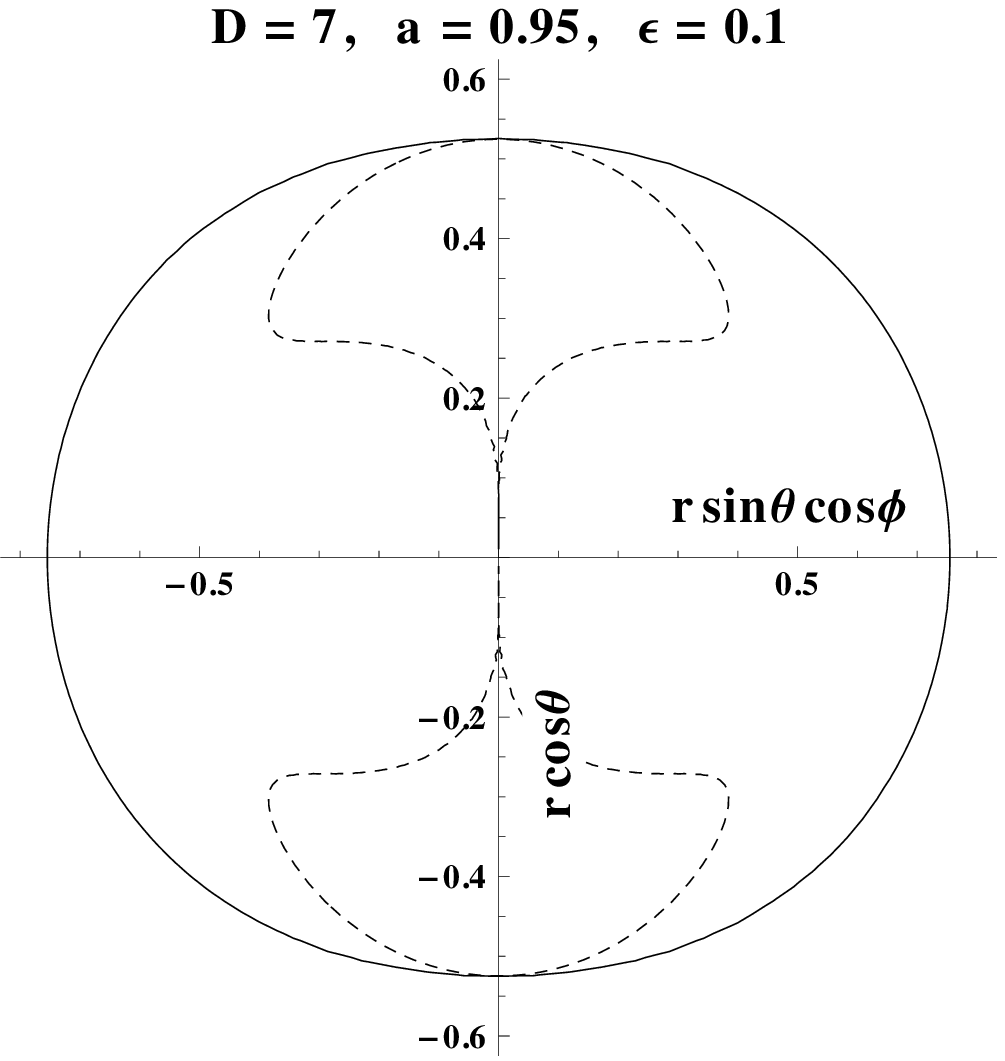}\\
\hline
\includegraphics[scale=0.3]{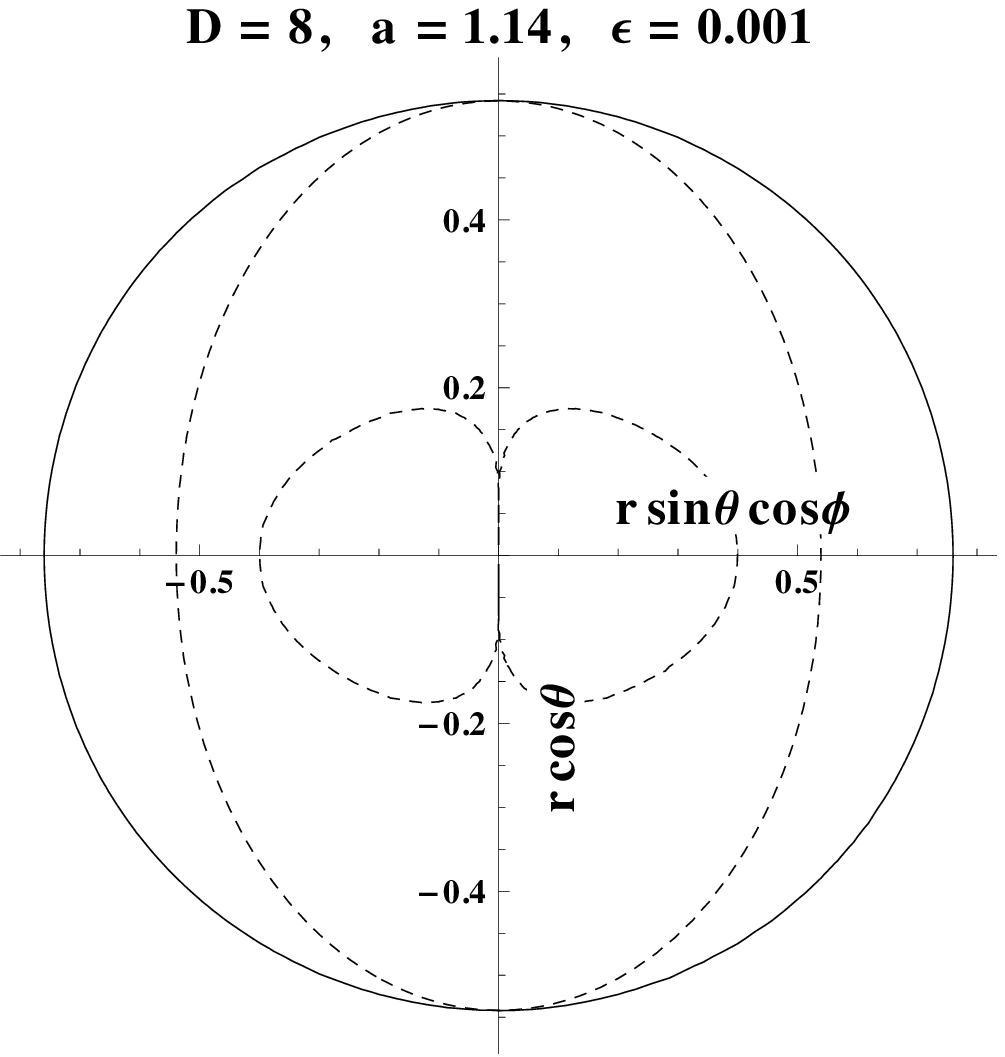}
& \includegraphics[scale=0.3]{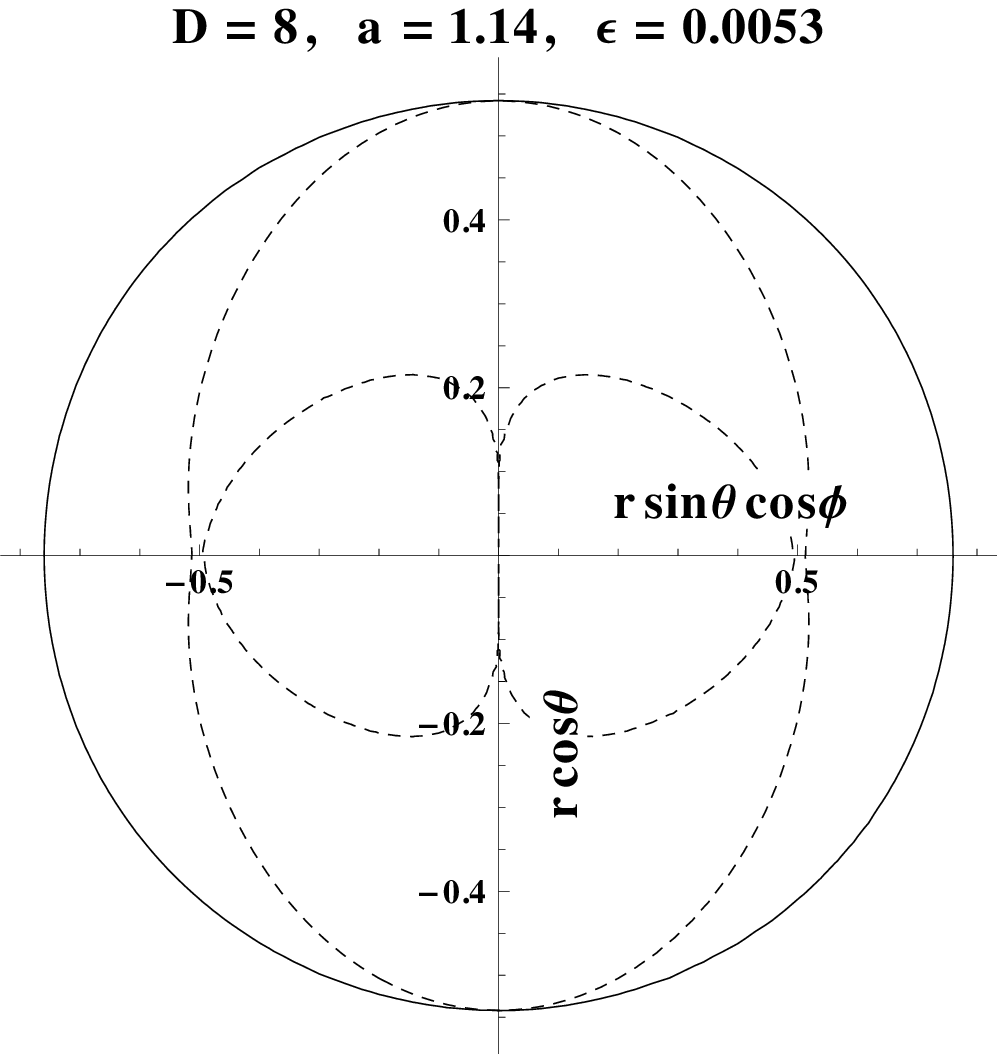}
& \includegraphics[scale=0.3]{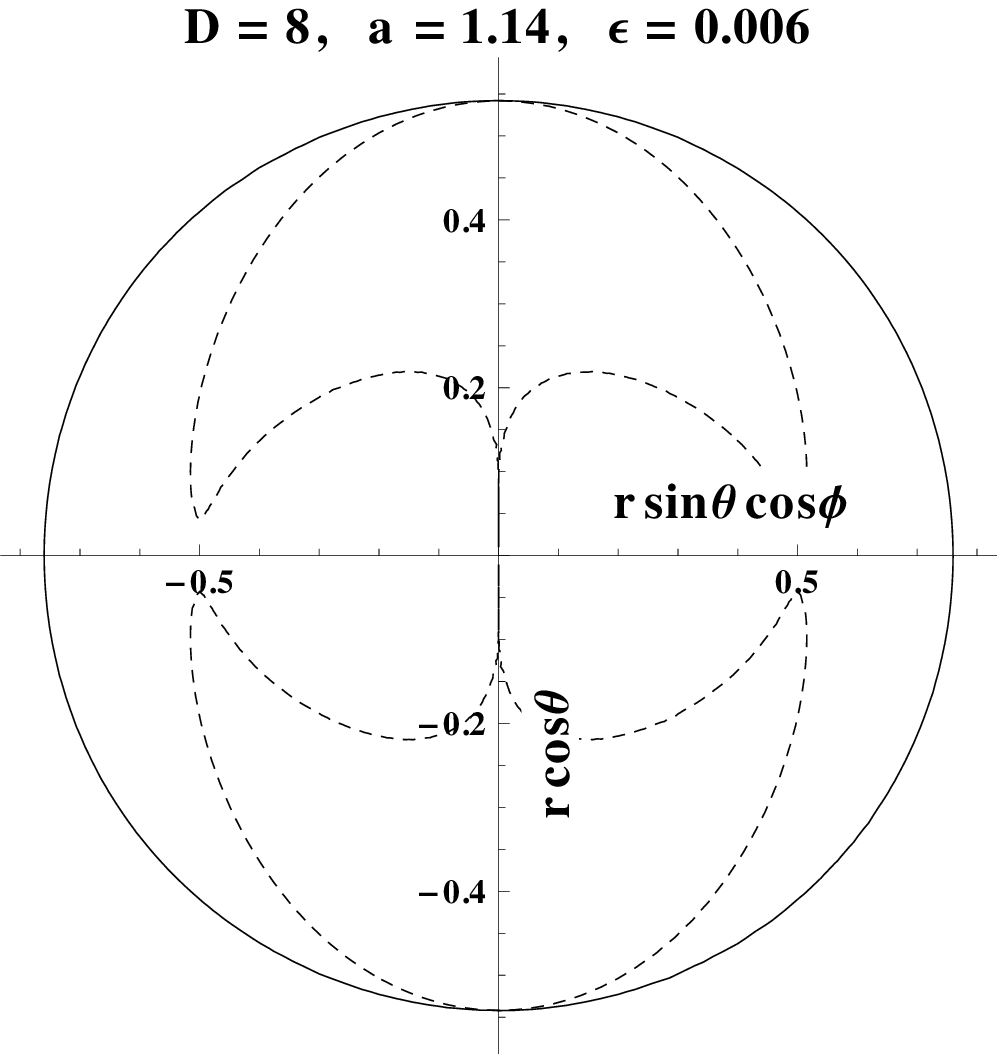}
& \includegraphics[scale=0.3]{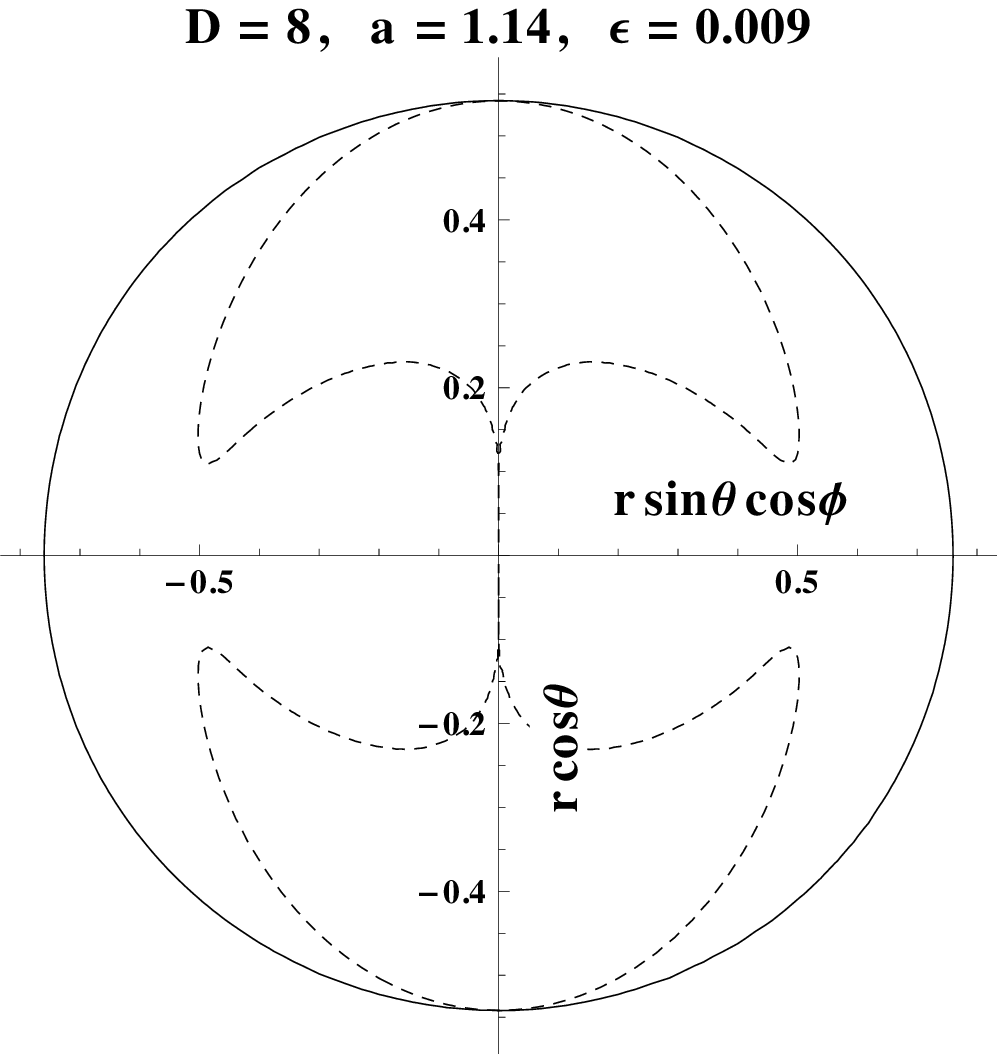}\\
\hline
  \includegraphics[scale=0.3]{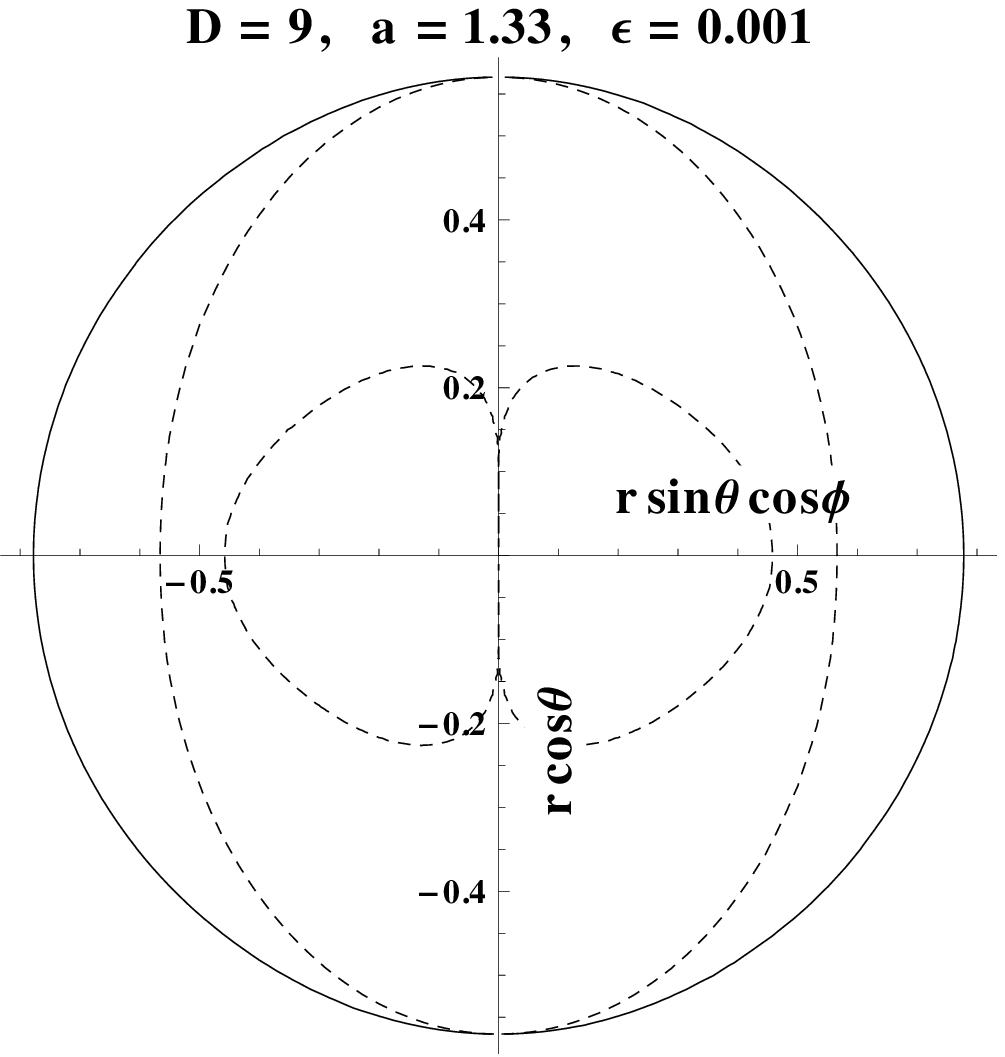}
& \includegraphics[scale=0.3]{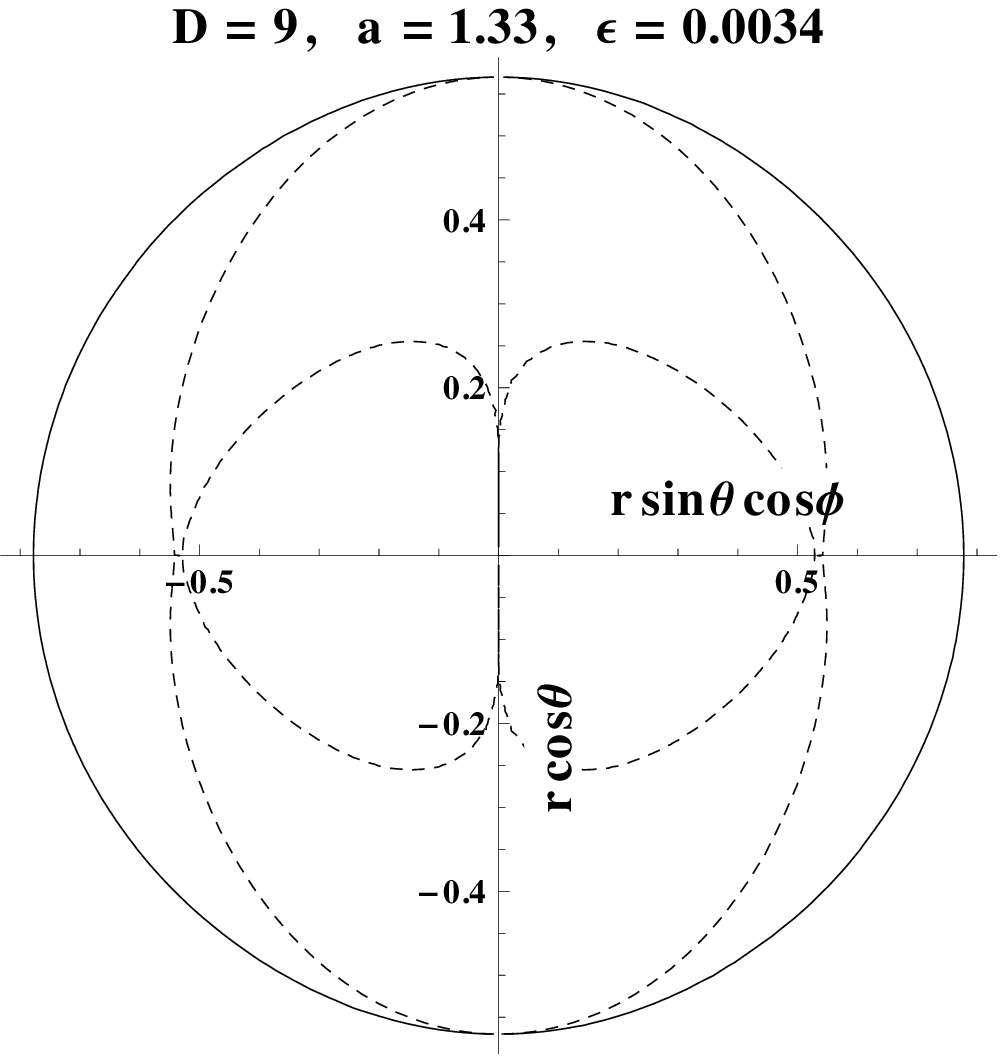}
& \includegraphics[scale=0.3]{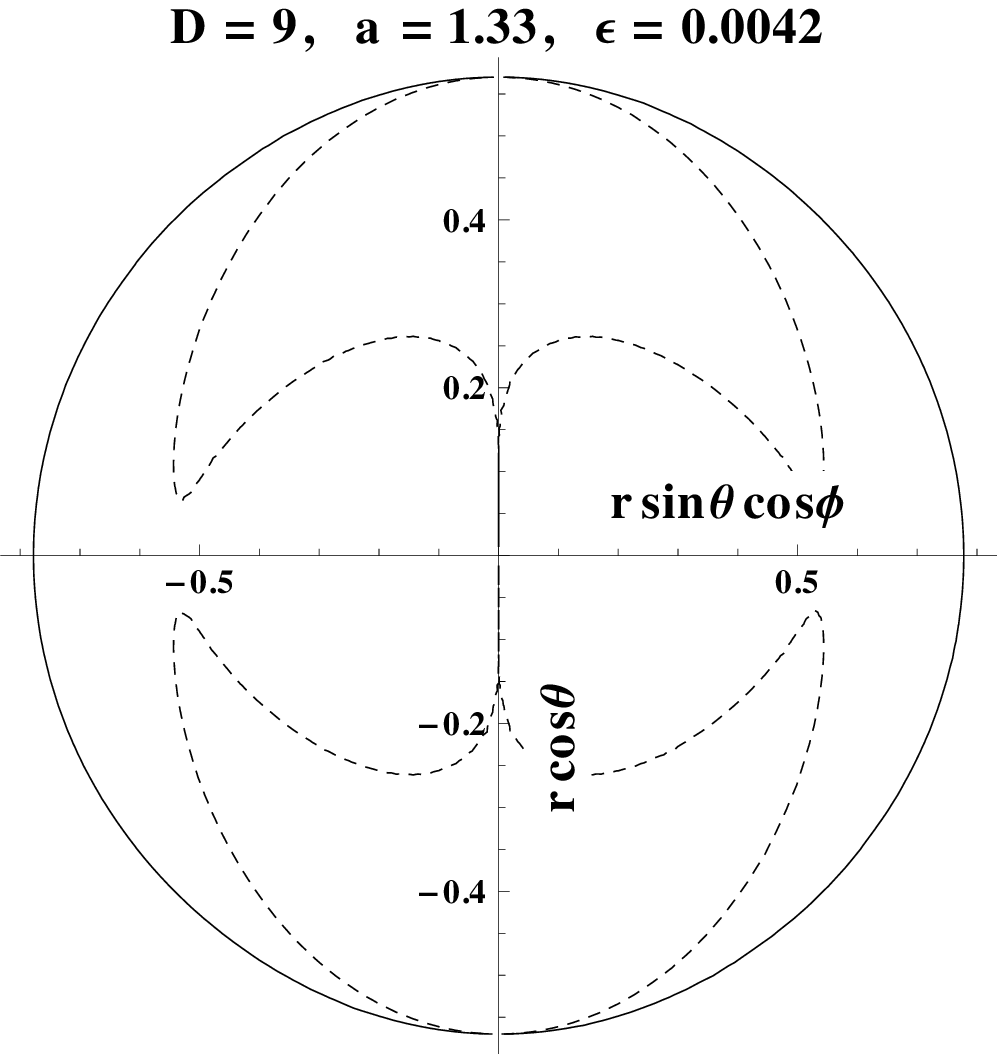}
& \includegraphics[scale=0.3]{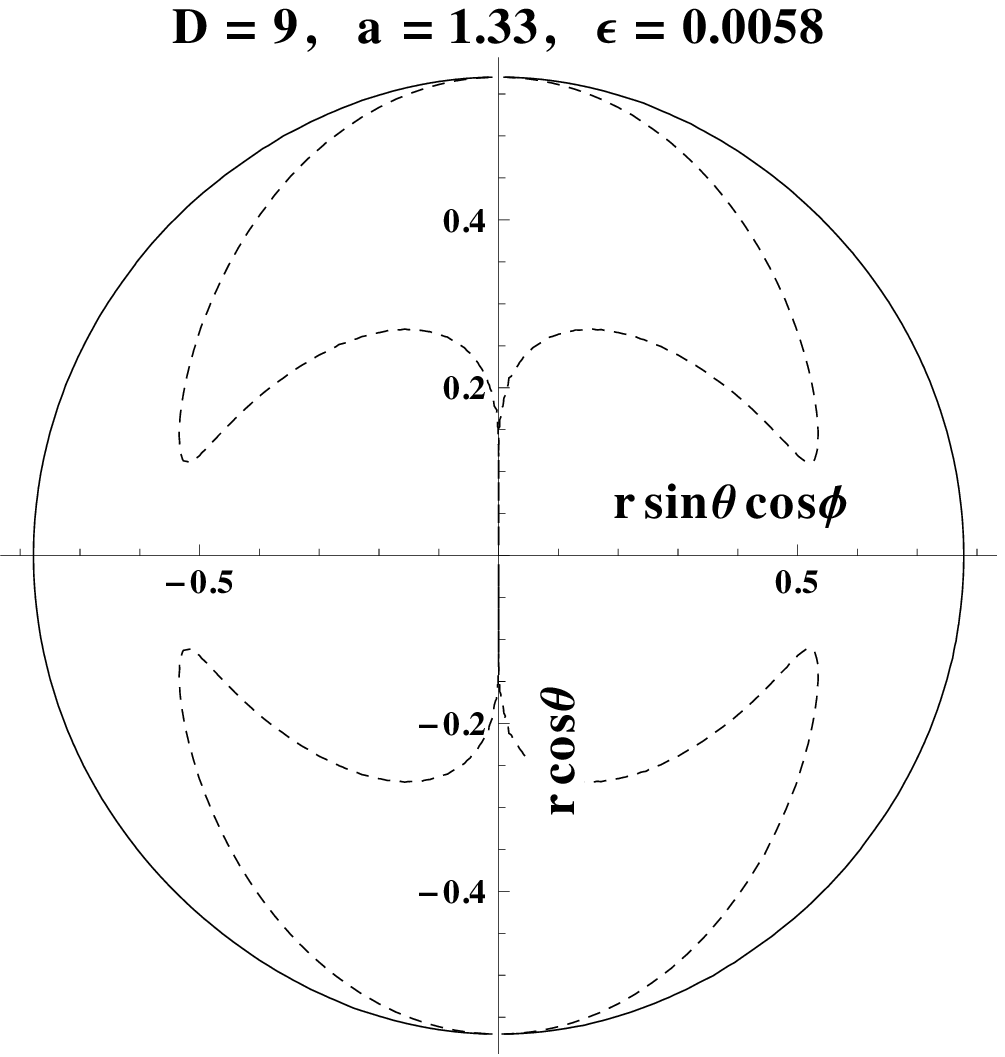}\\
  \hline
  \includegraphics[scale=0.3]{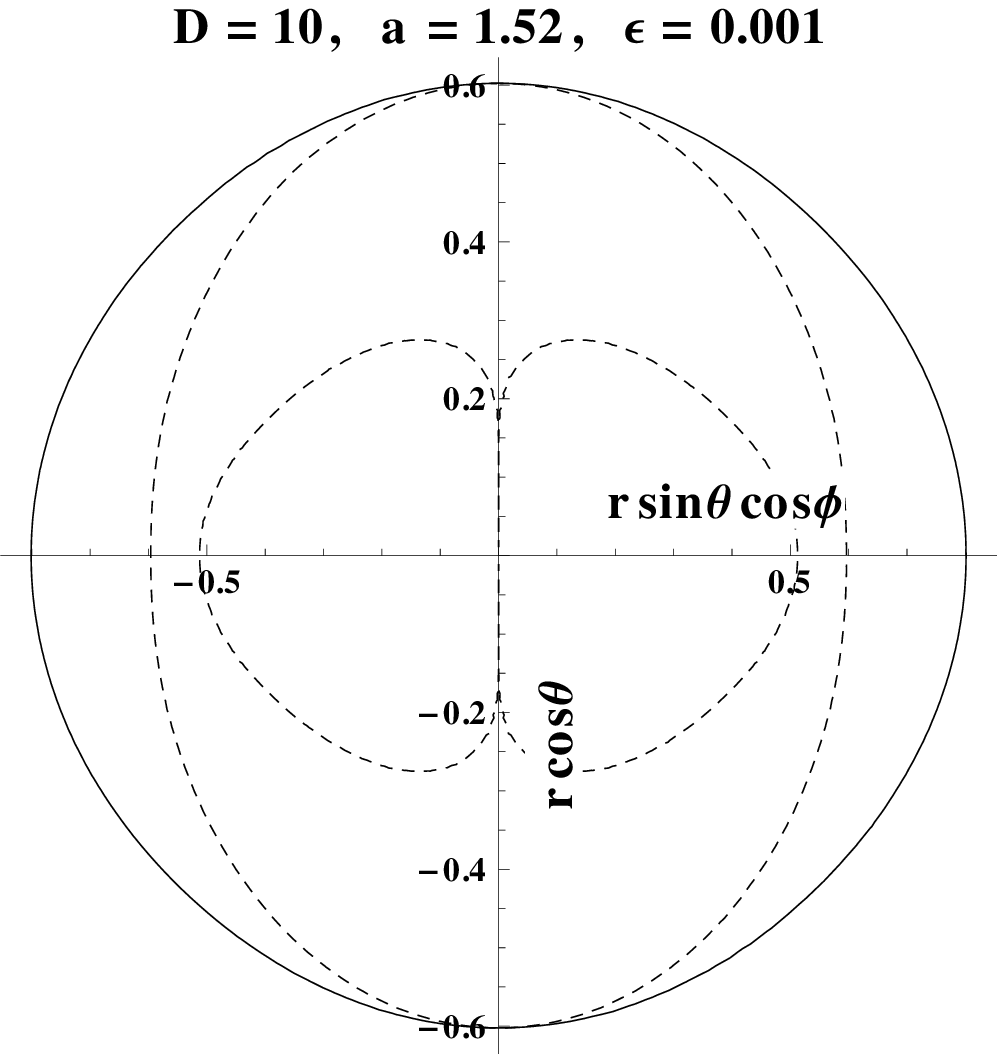}
& \includegraphics[scale=0.3]{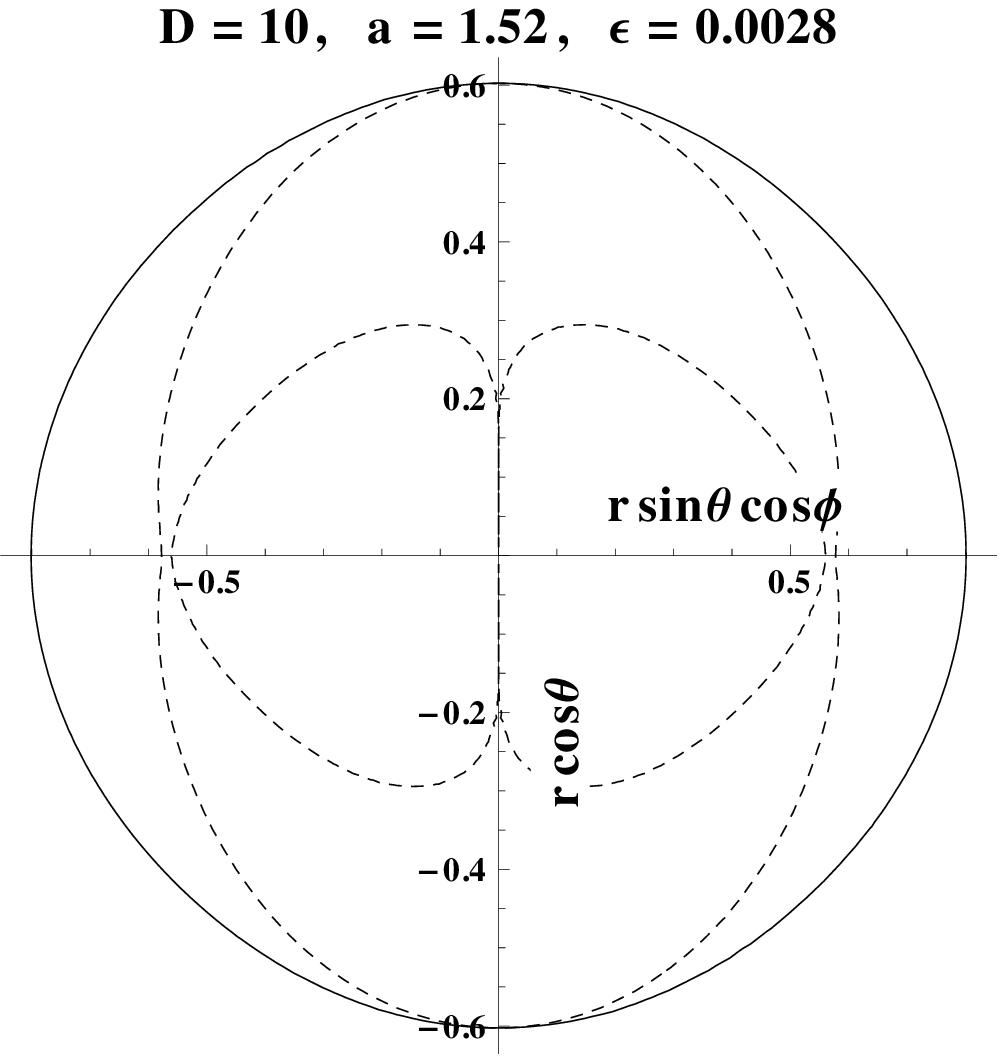}
& \includegraphics[scale=0.3]{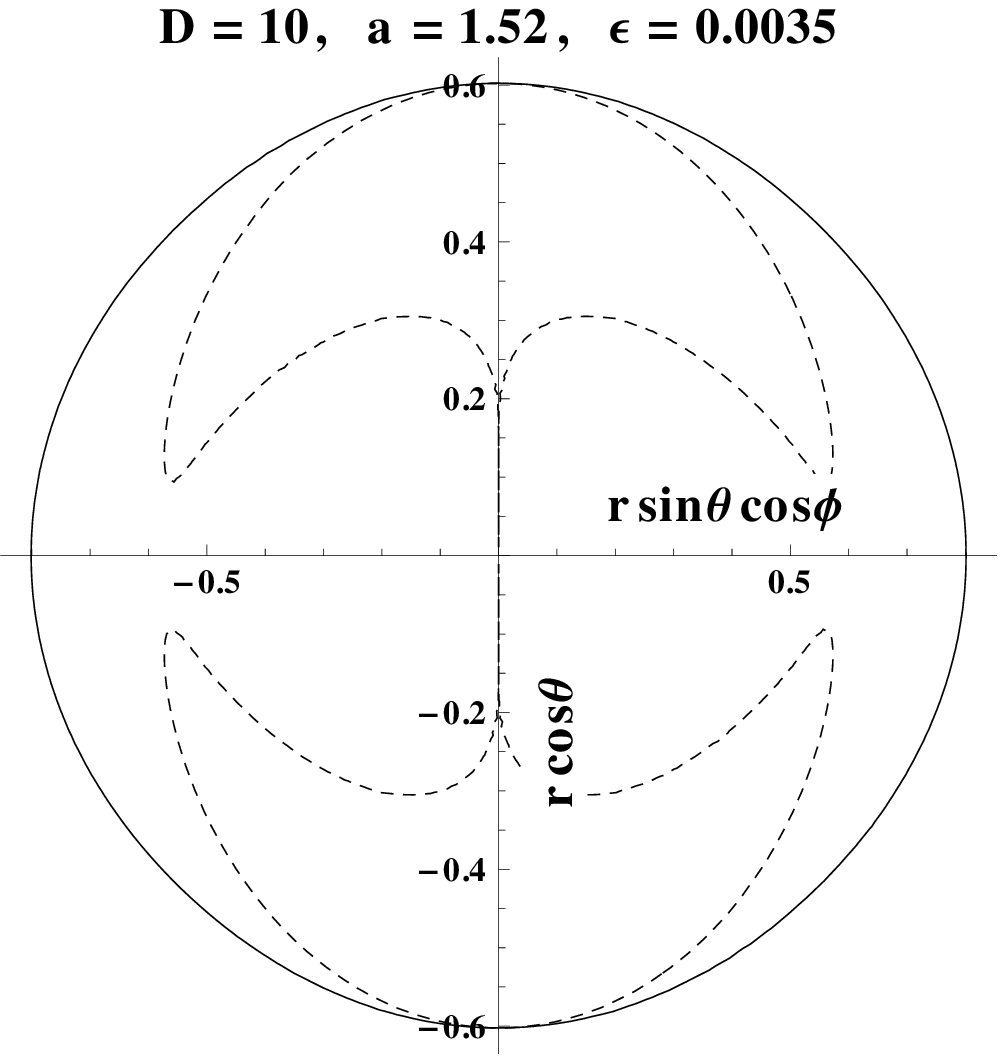}
& \includegraphics[scale=0.3]{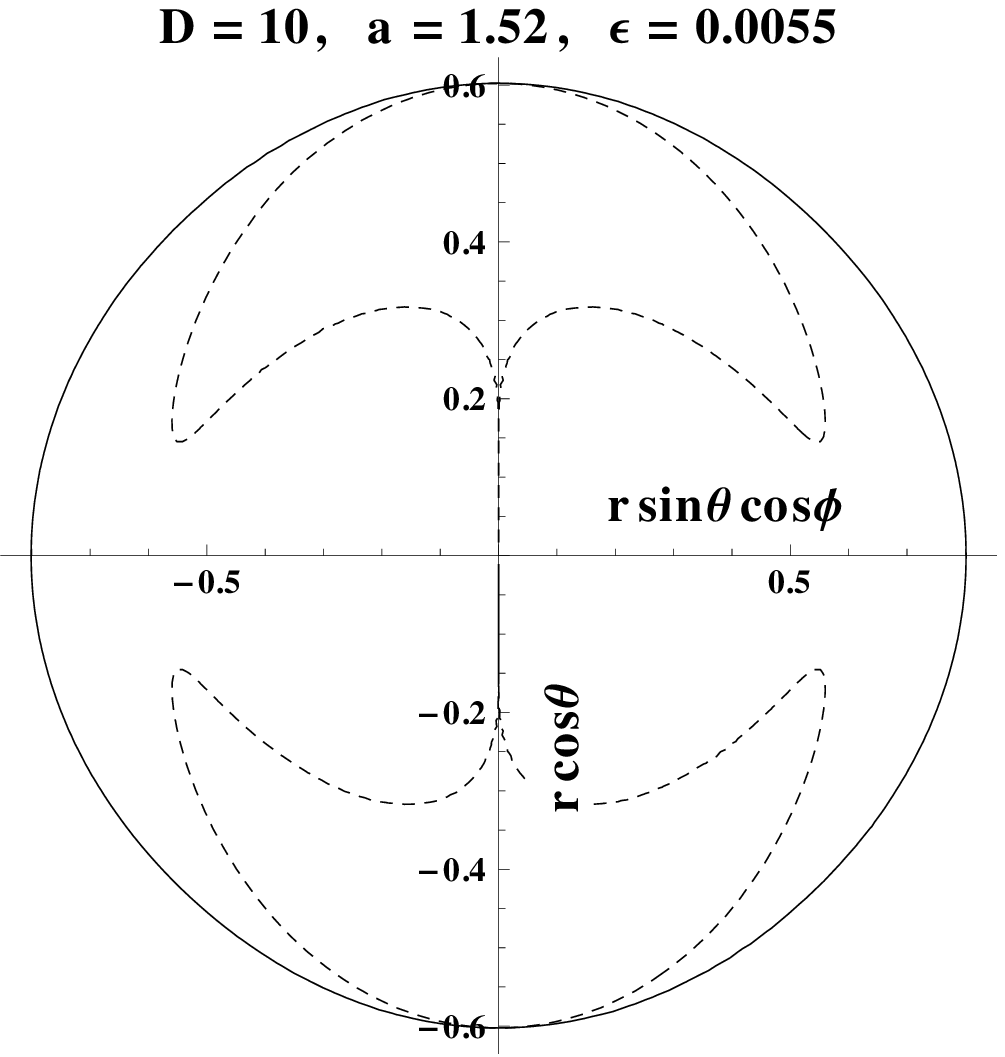}\\
  \hline
\end{tabular}
\caption{The cross section of the stationary limit surface  and event horizon and the variation
of the ergosphere for different dimensions ($D = 4,\ldots$,10) with deformation parameter
$\epsilon$ of the HD spinning non-Kerr black hole. 
The increase in the value of the deformation parameter leads to
a disconnected event horizon.}\label{ergosphere1a}
\end{figure*}

\begin{figure*}
\begin{tabular}{ | c | c | c | c | }
\hline
\includegraphics[scale=0.3]{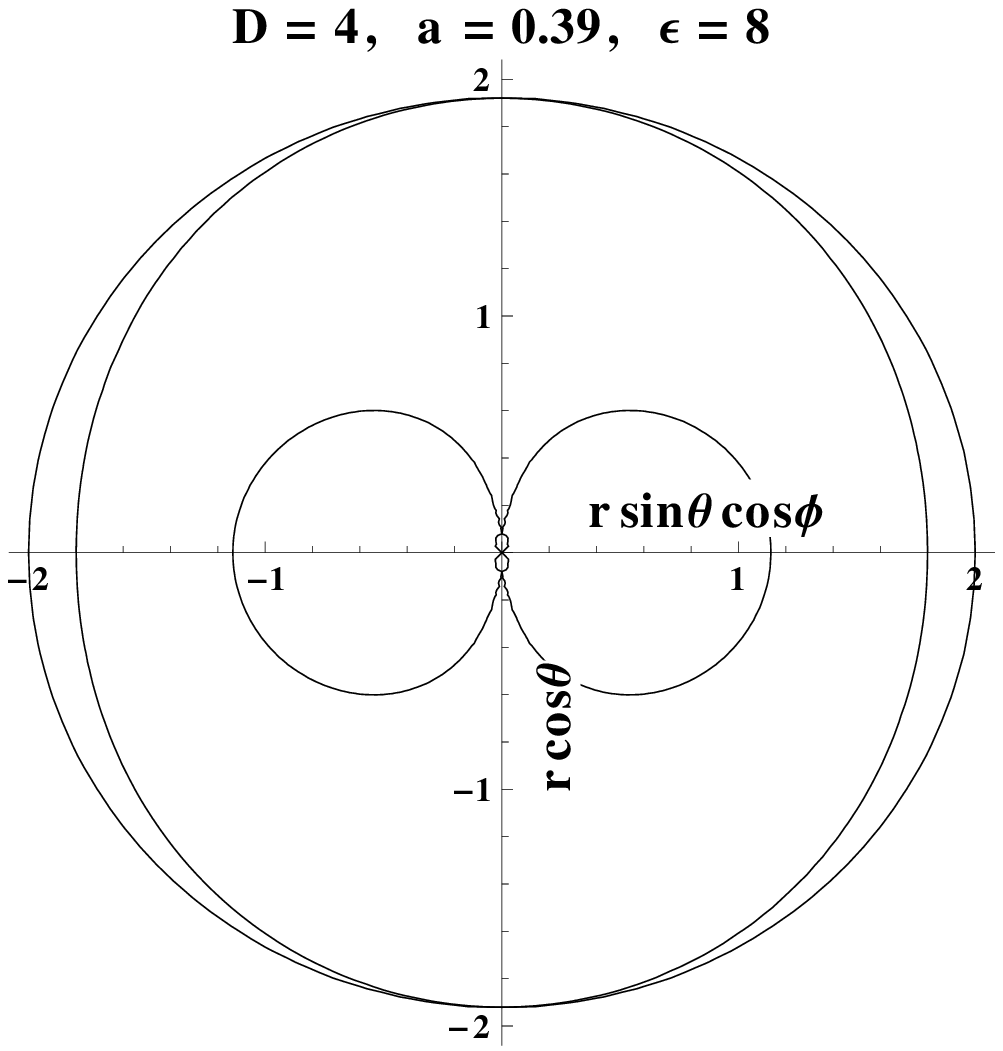}
& \includegraphics[scale=0.3]{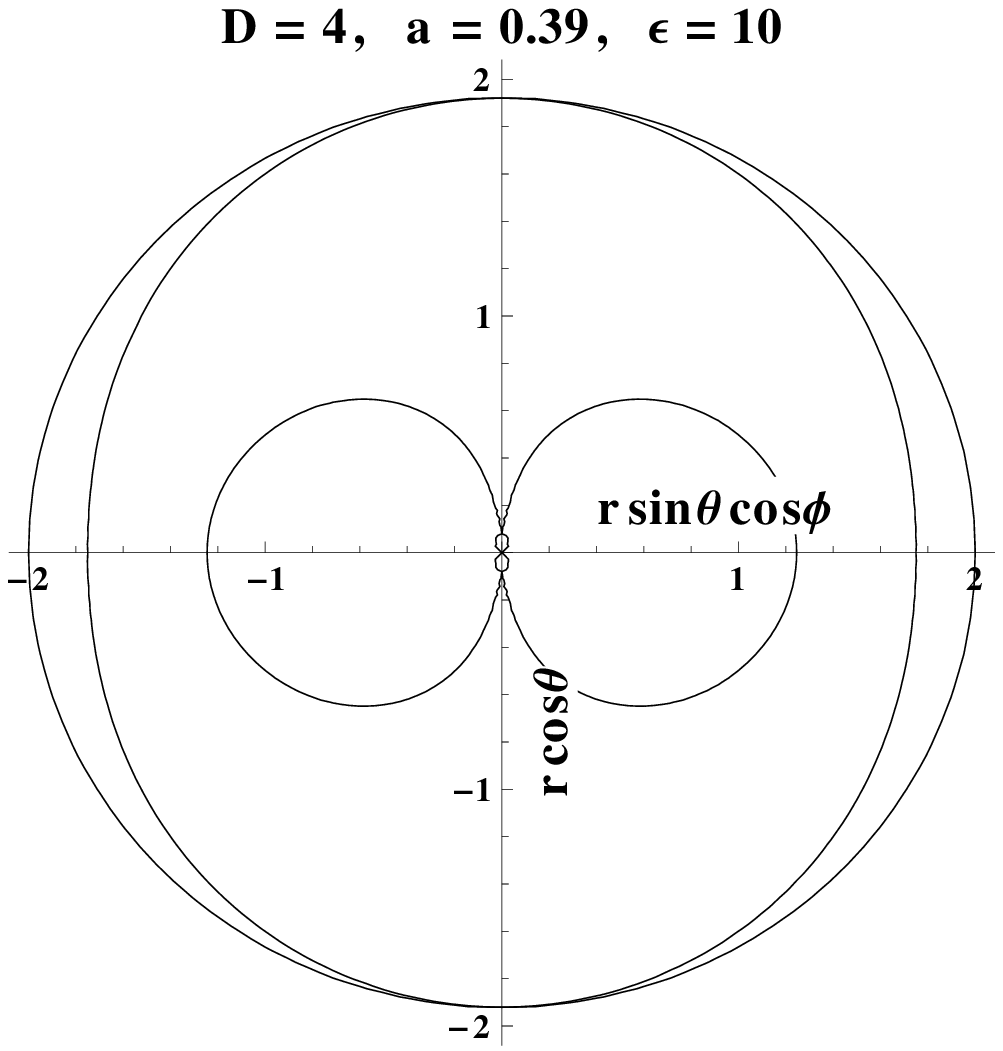}
& \includegraphics[scale=0.3]{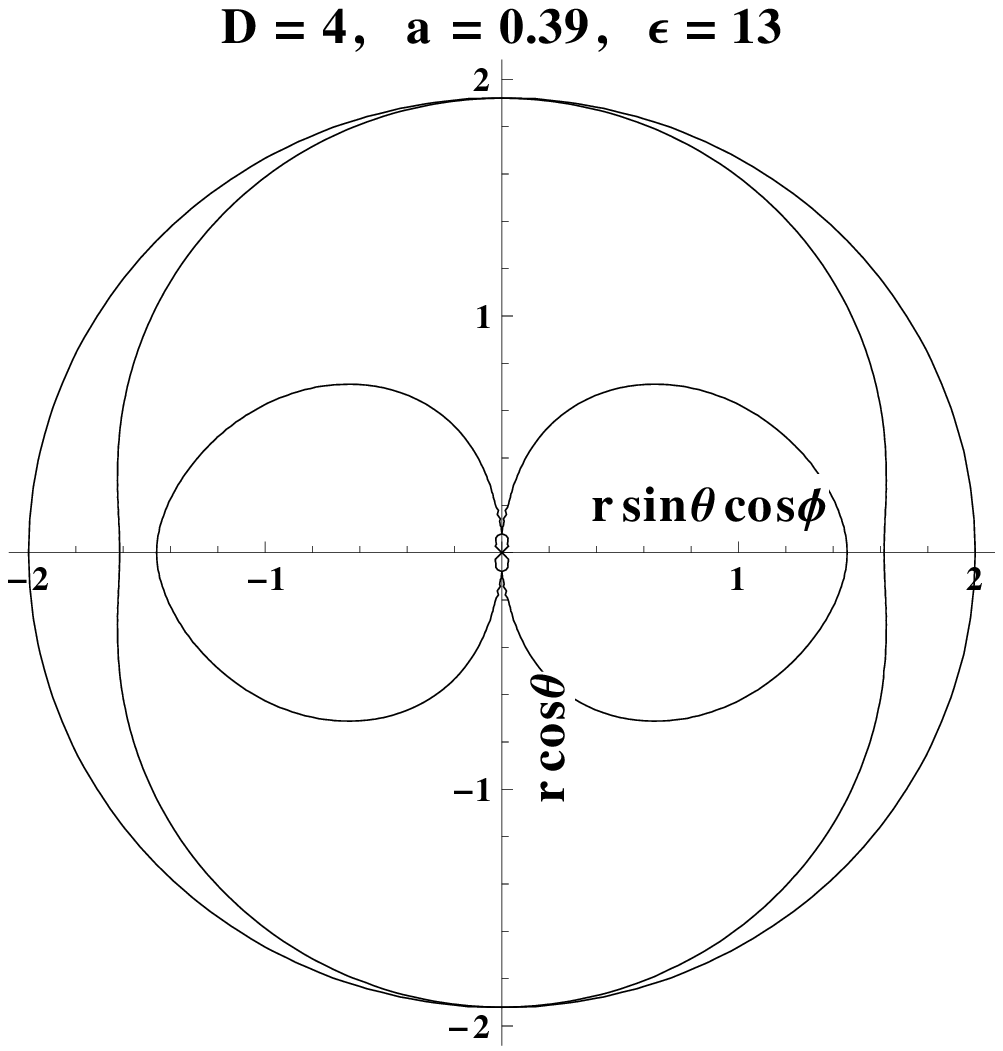}
& \includegraphics[scale=0.3]{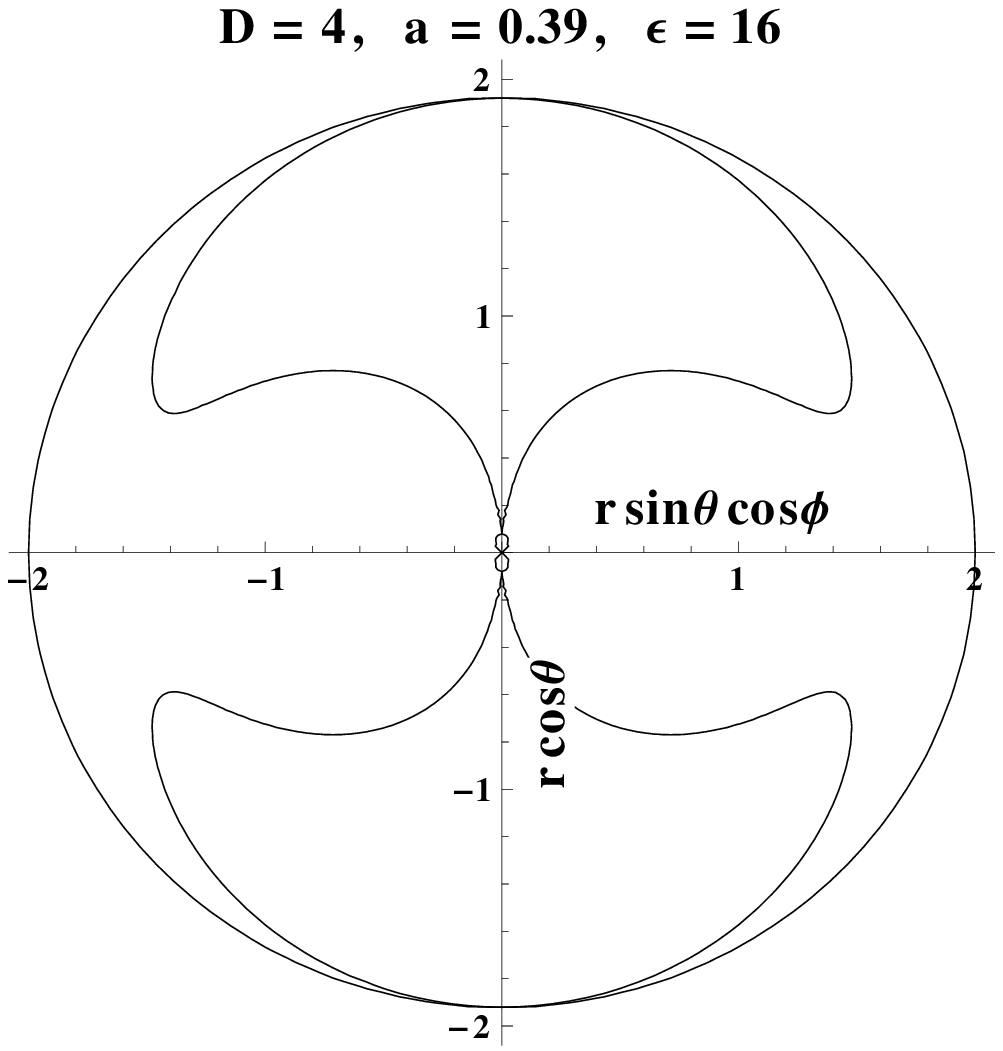}\\
\hline
\includegraphics[scale=0.3]{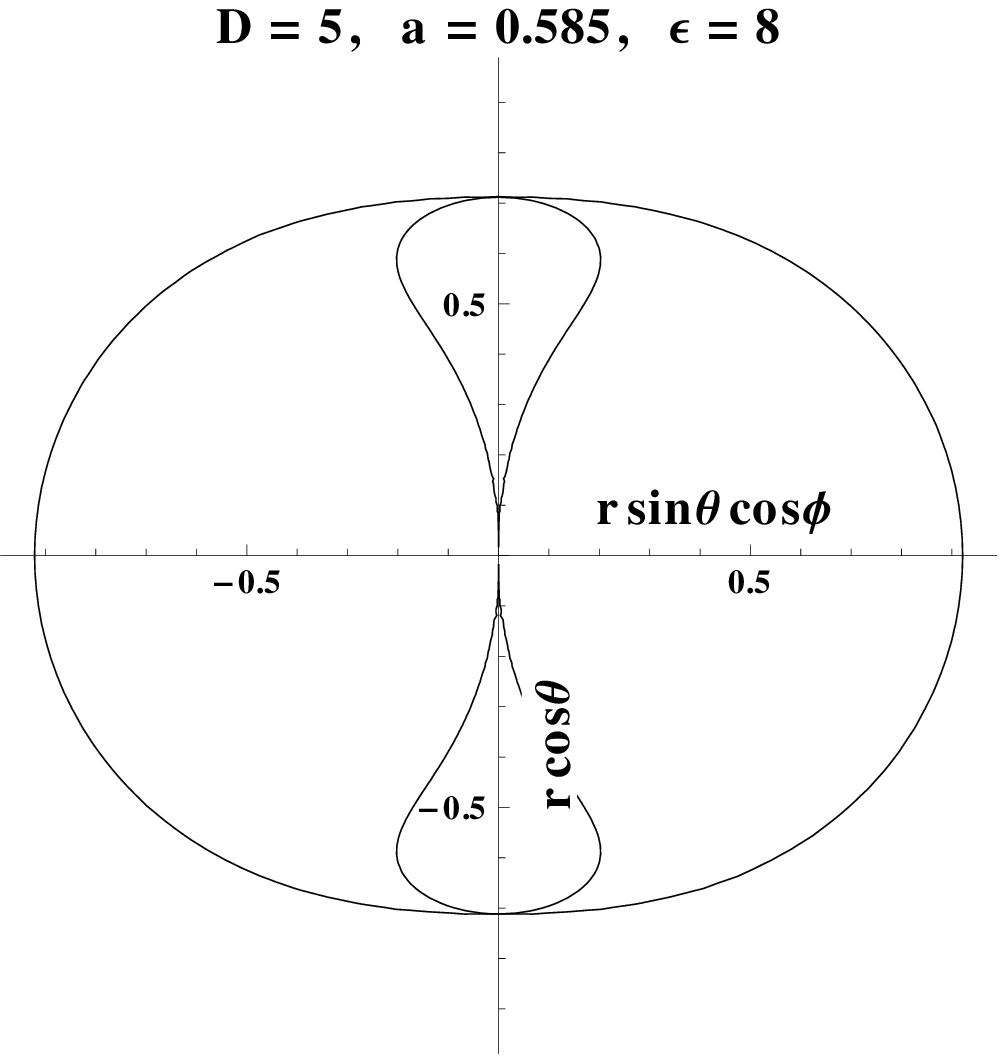}
& \includegraphics[scale=0.3]{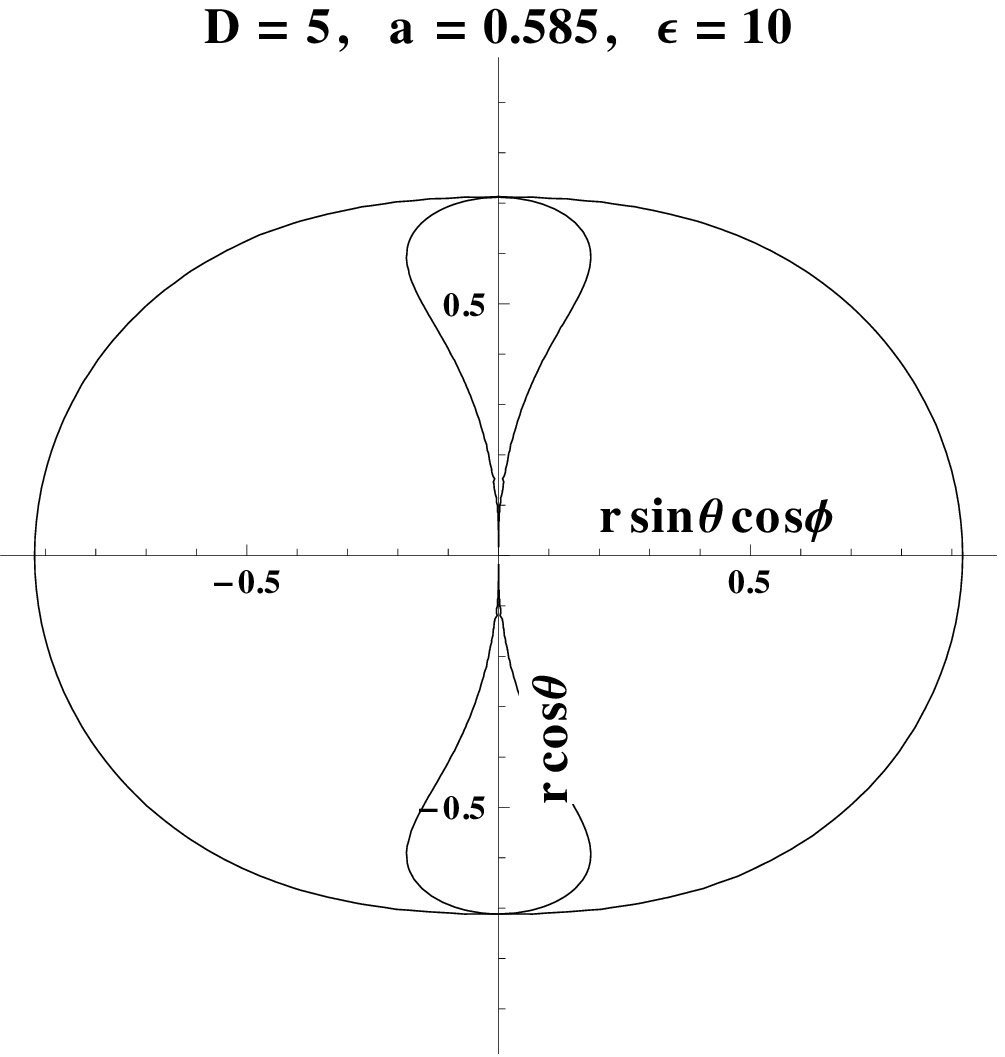}
& \includegraphics[scale=0.3]{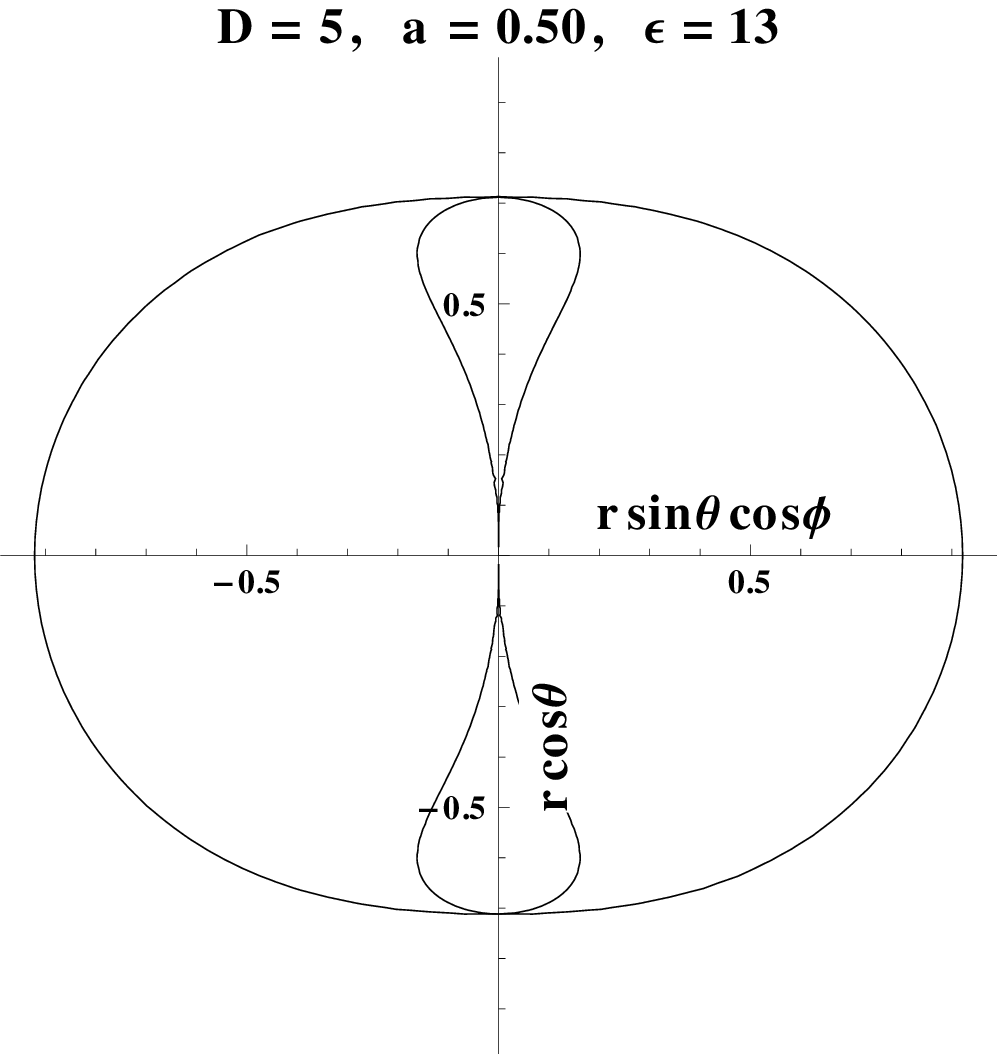}
& \includegraphics[scale=0.3]{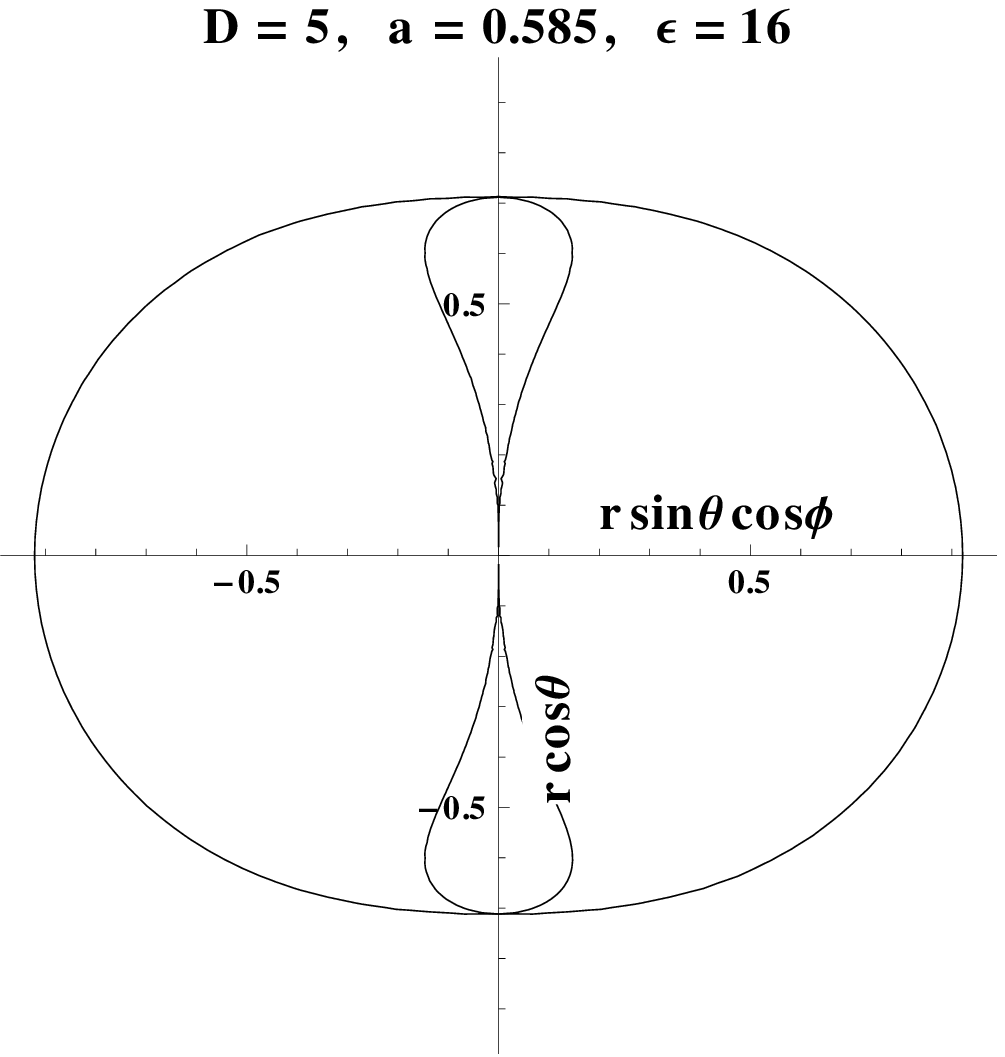}\\
\hline
\includegraphics[scale=0.3]{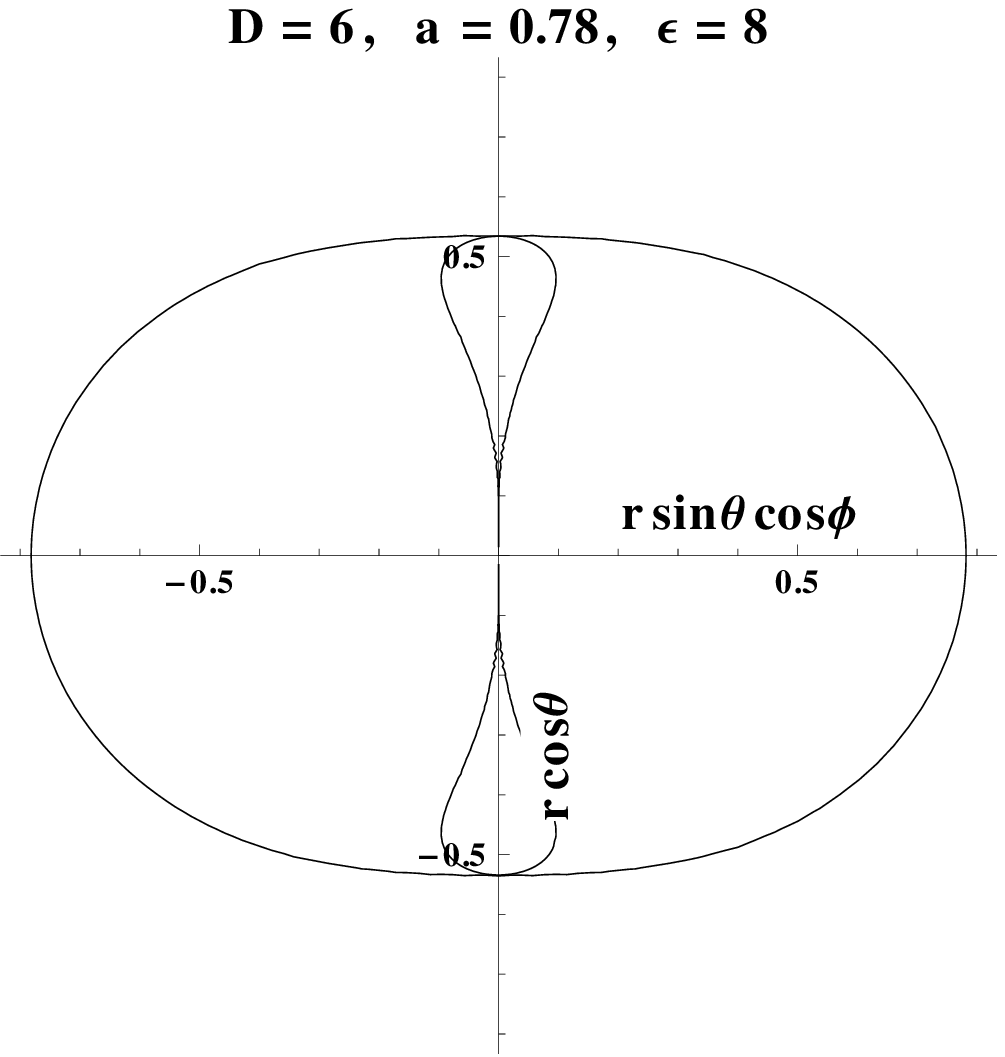}
& \includegraphics[scale=0.3]{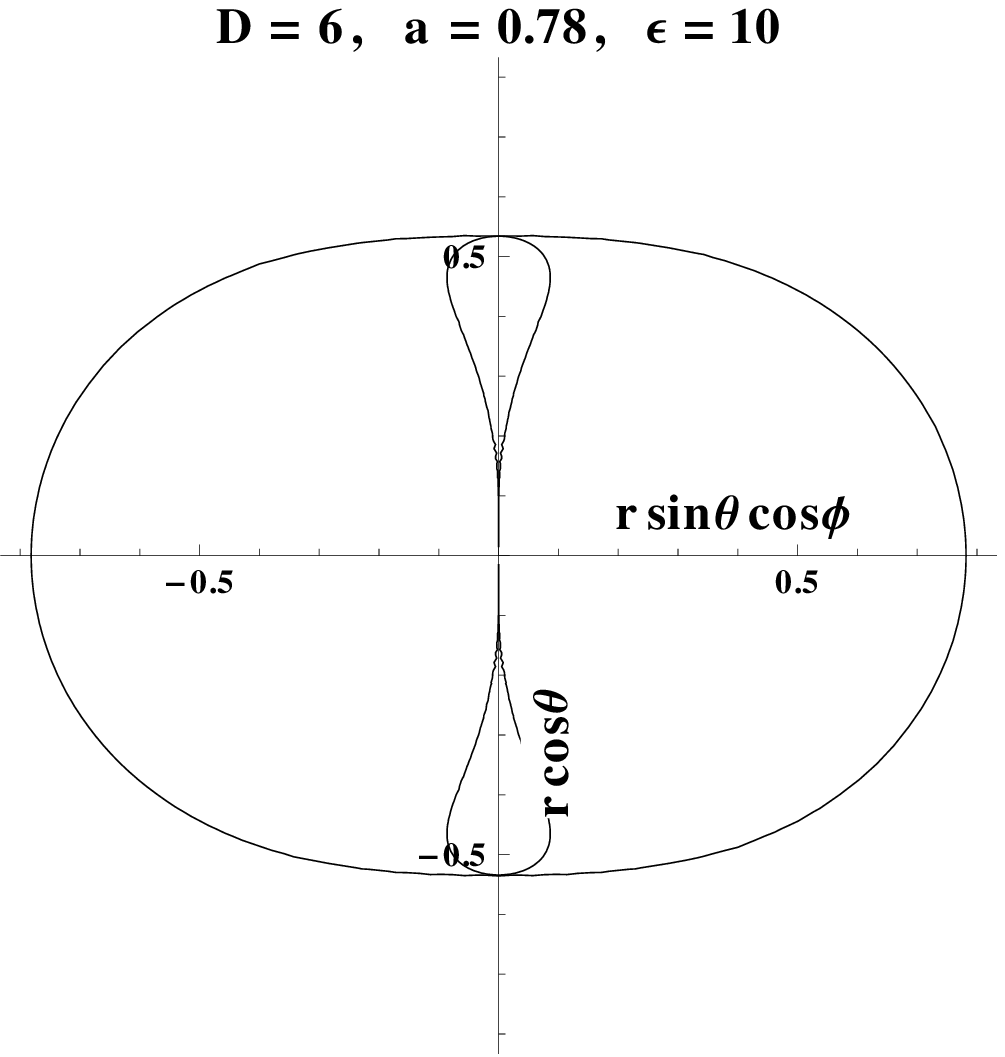}
& \includegraphics[scale=0.3]{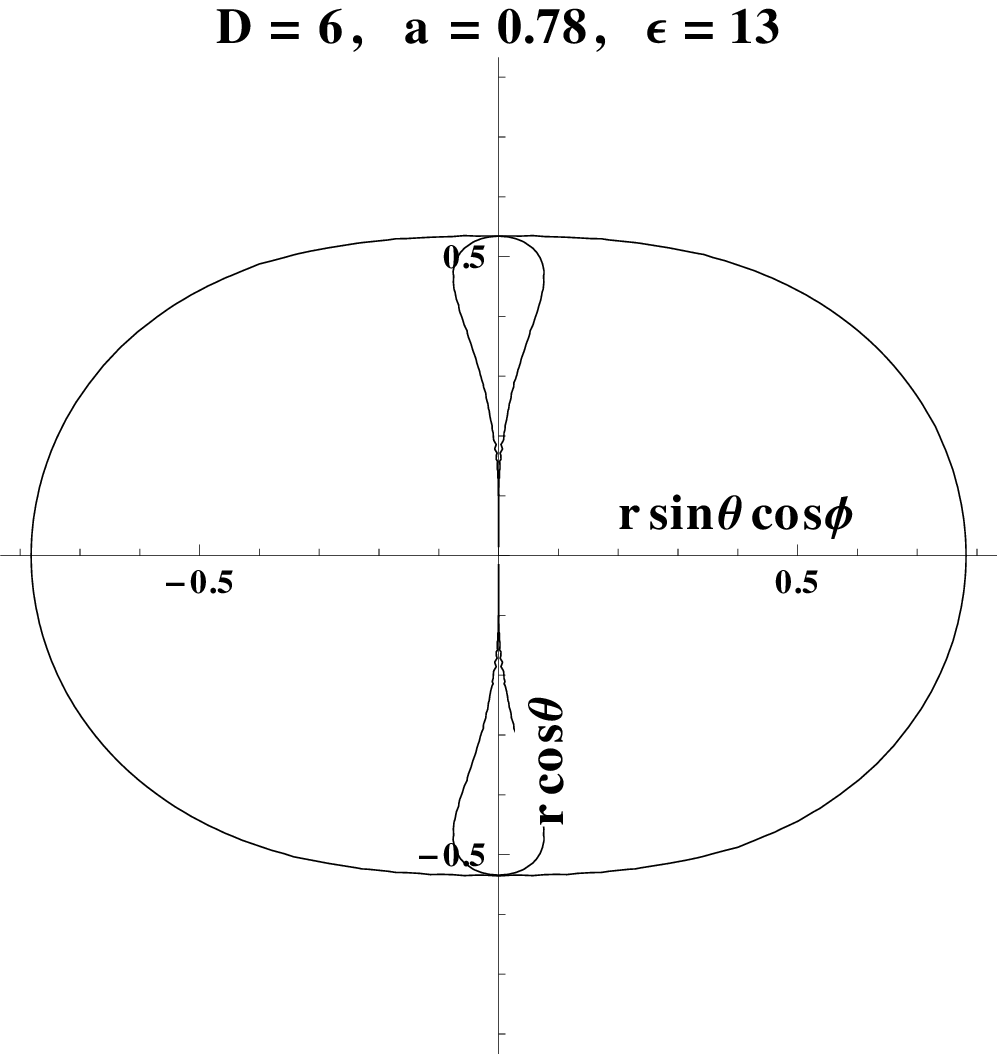}
& \includegraphics[scale=0.3]{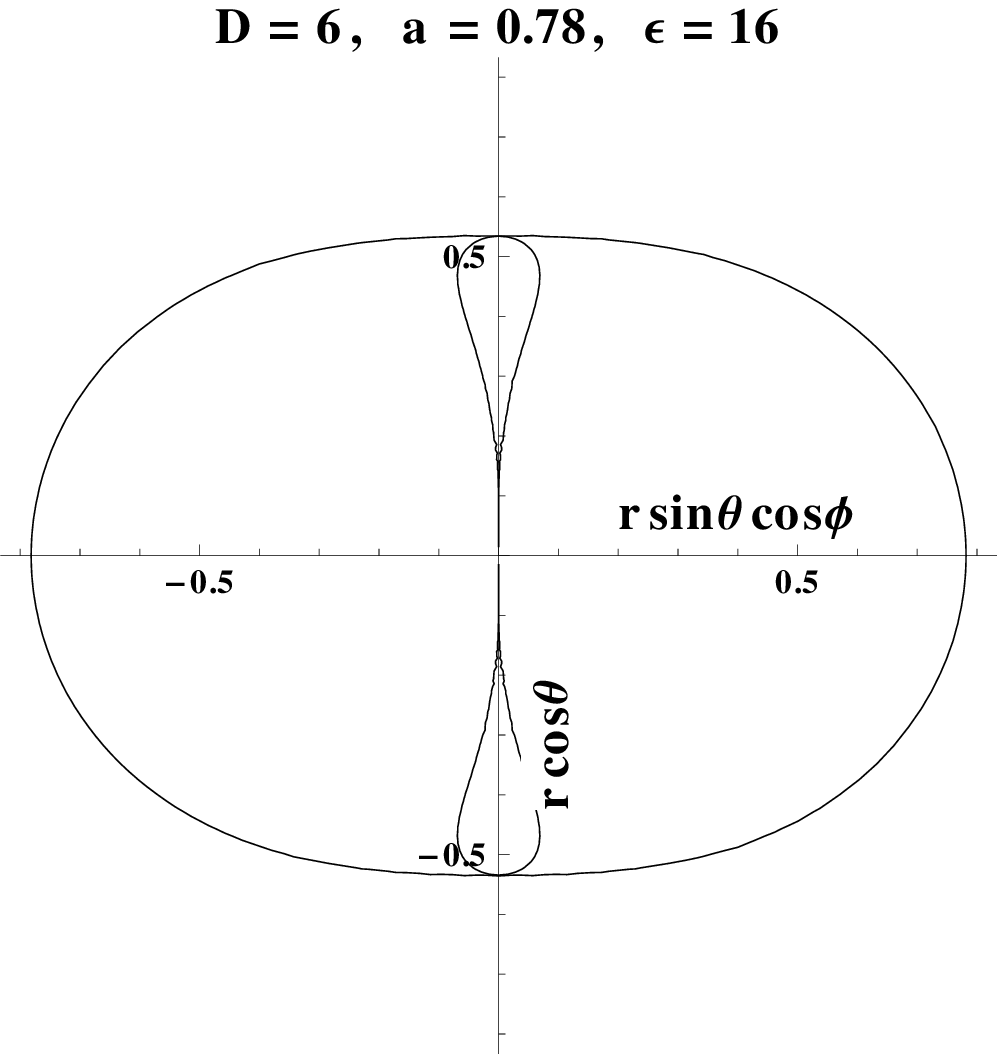}\\
\hline
\includegraphics[scale=0.3]{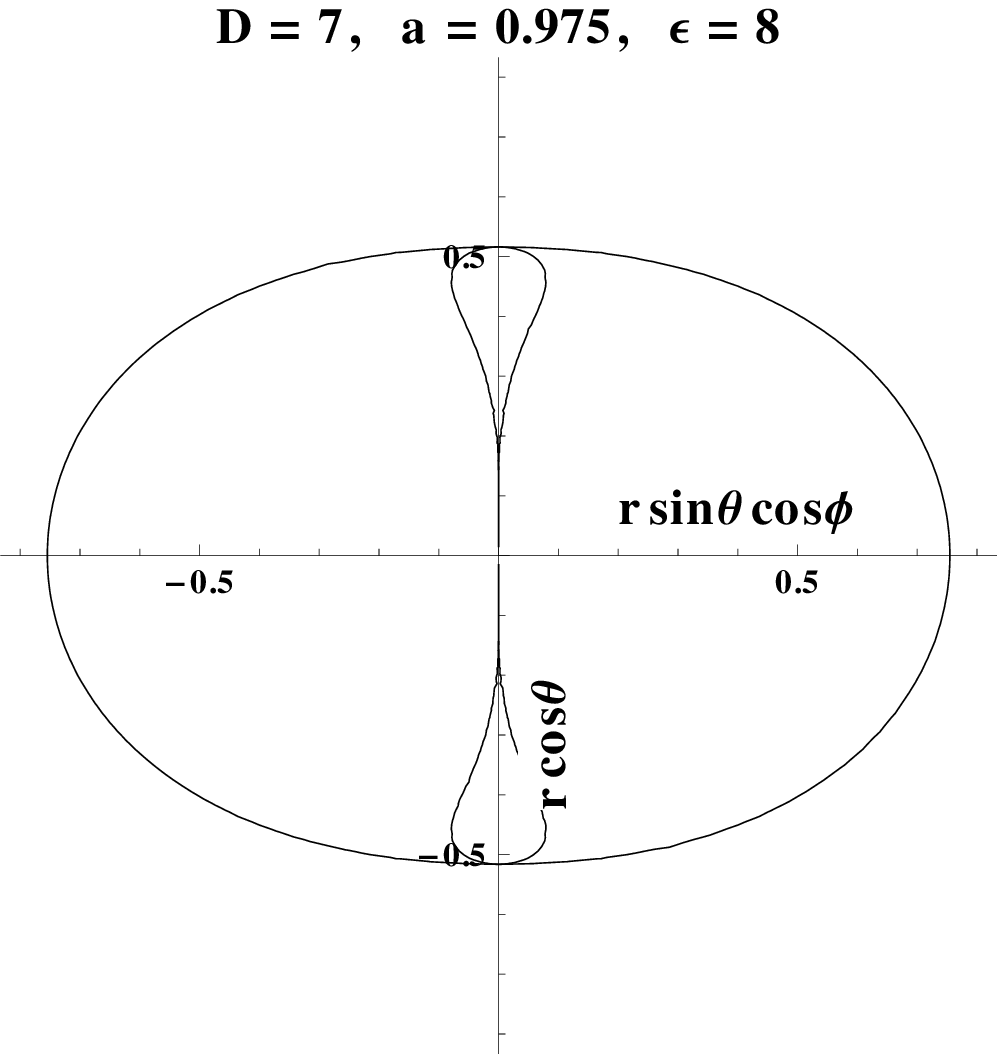}
& \includegraphics[scale=0.3]{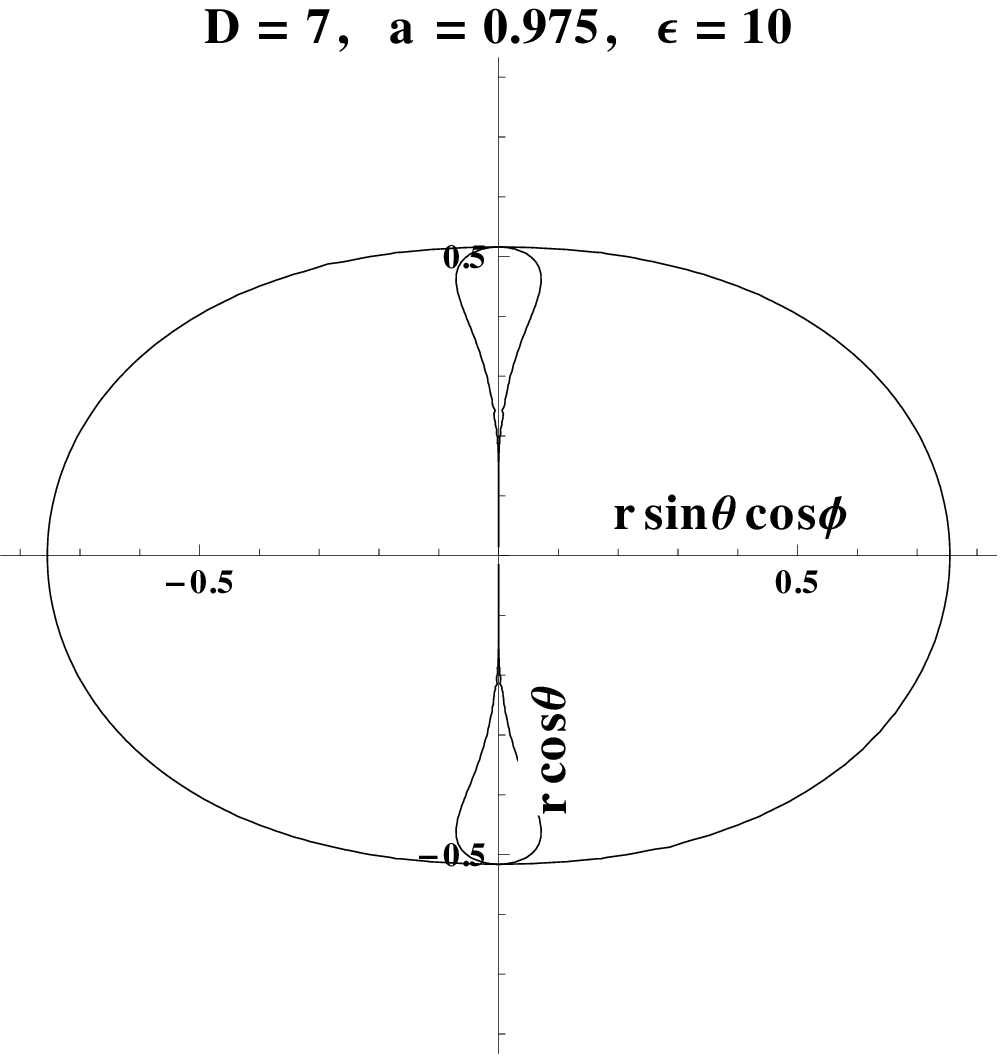}
& \includegraphics[scale=0.3]{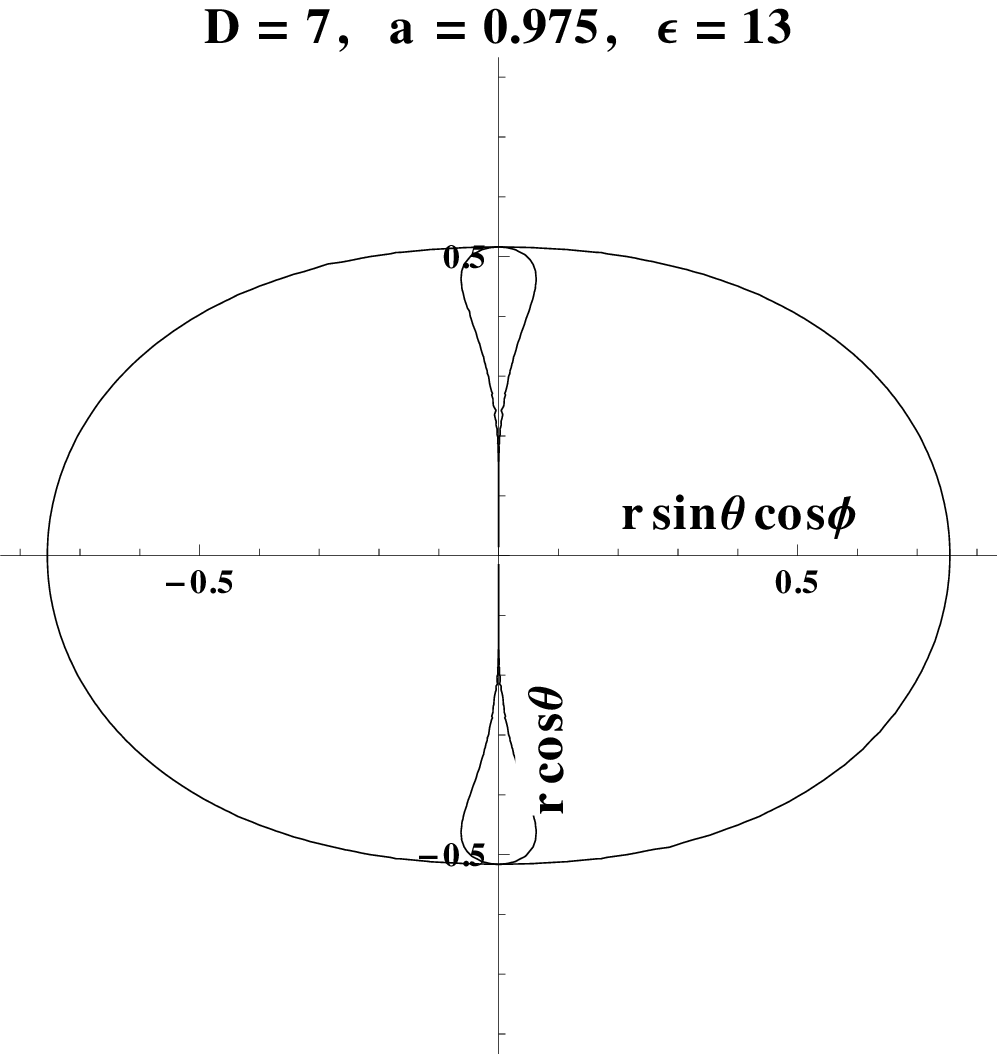}
& \includegraphics[scale=0.3]{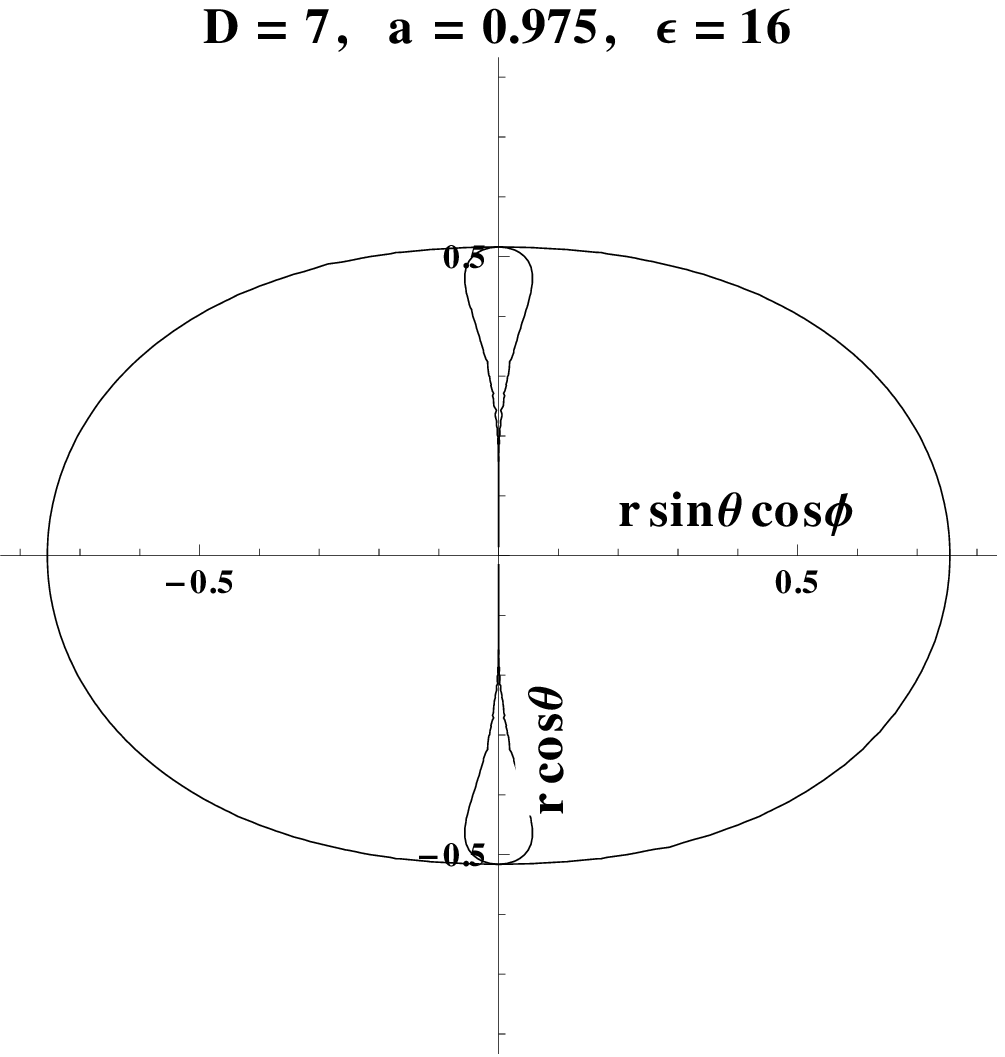}\\
\hline
\includegraphics[scale=0.3]{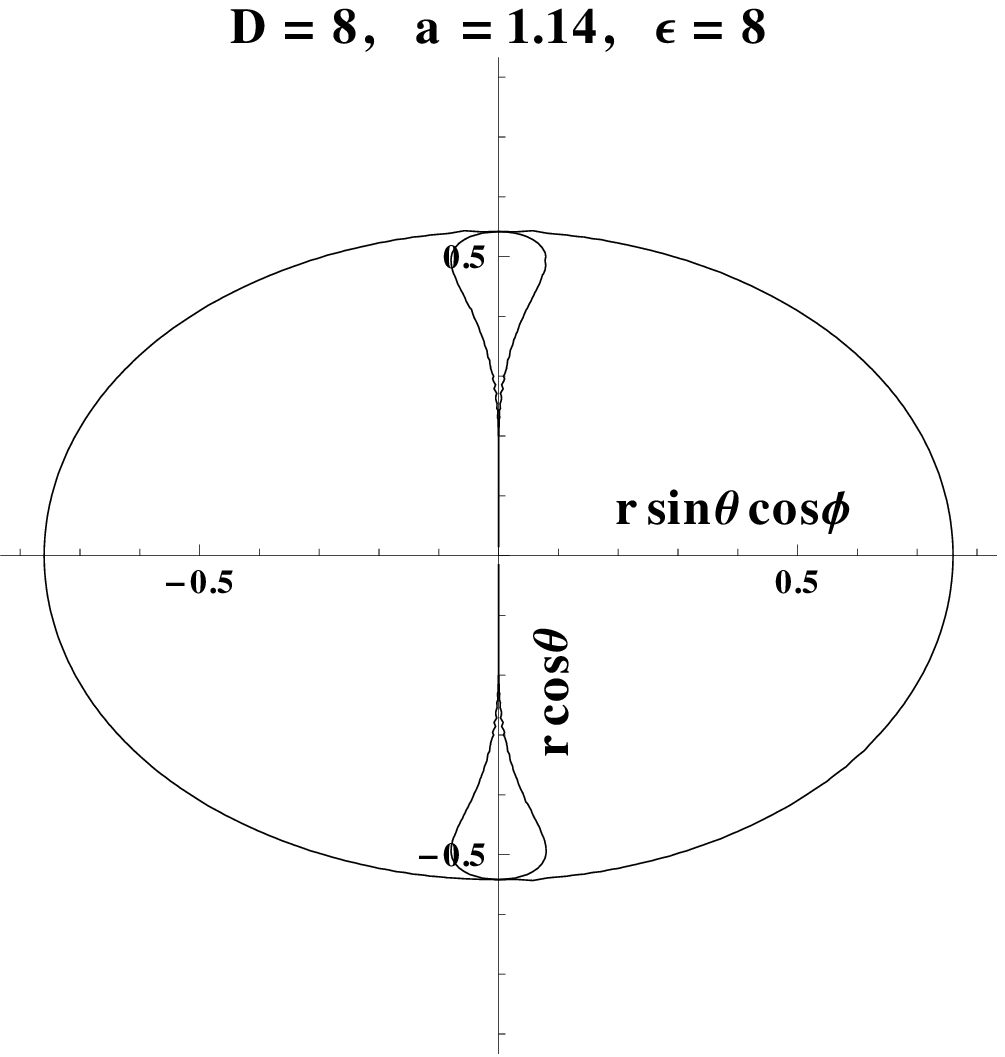}
& \includegraphics[scale=0.3]{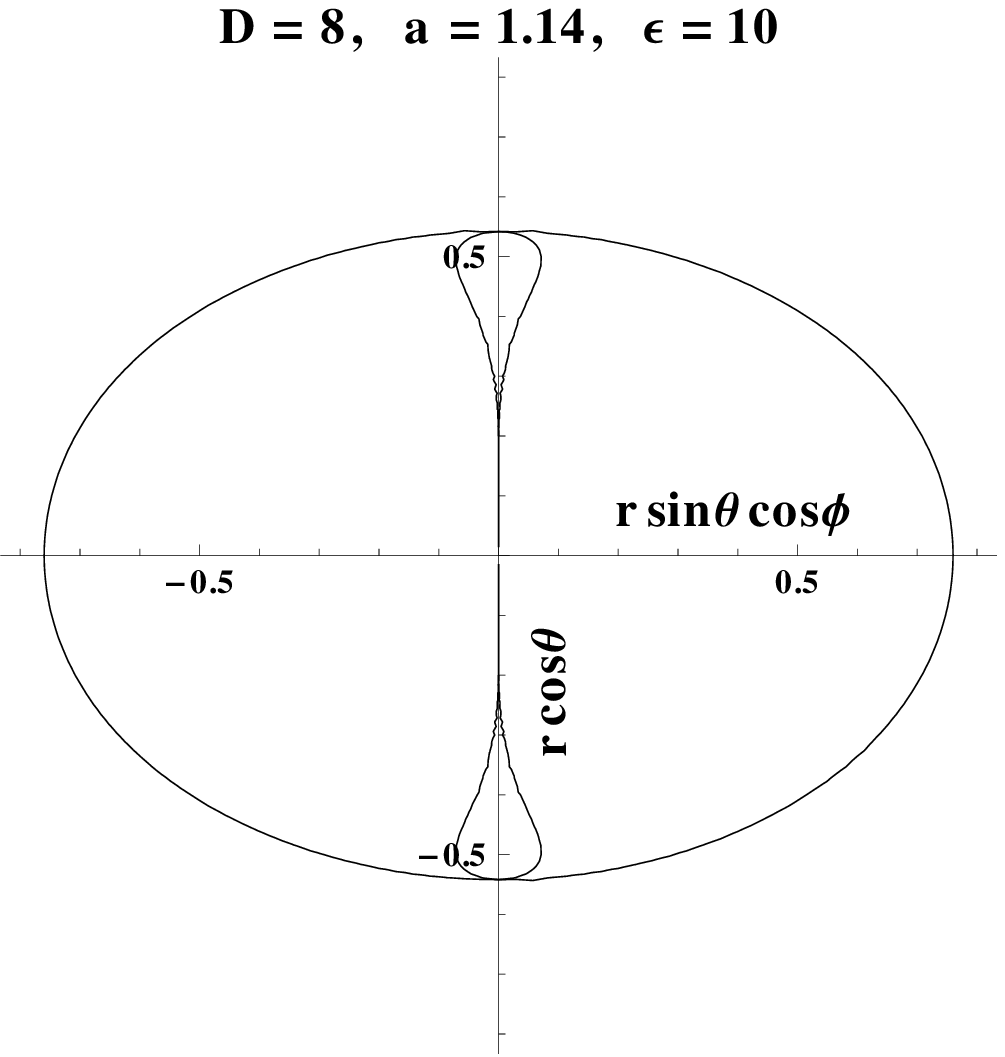}
& \includegraphics[scale=0.3]{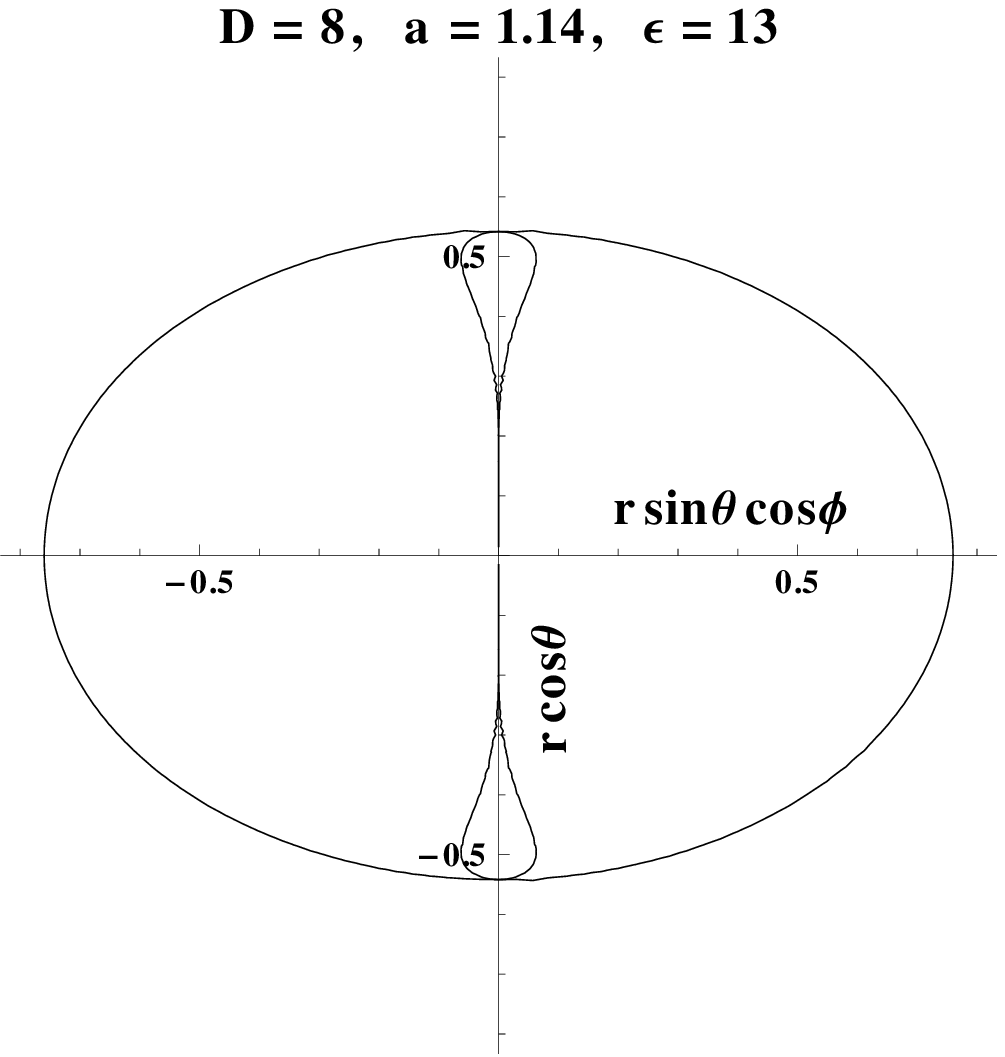}
& \includegraphics[scale=0.3]{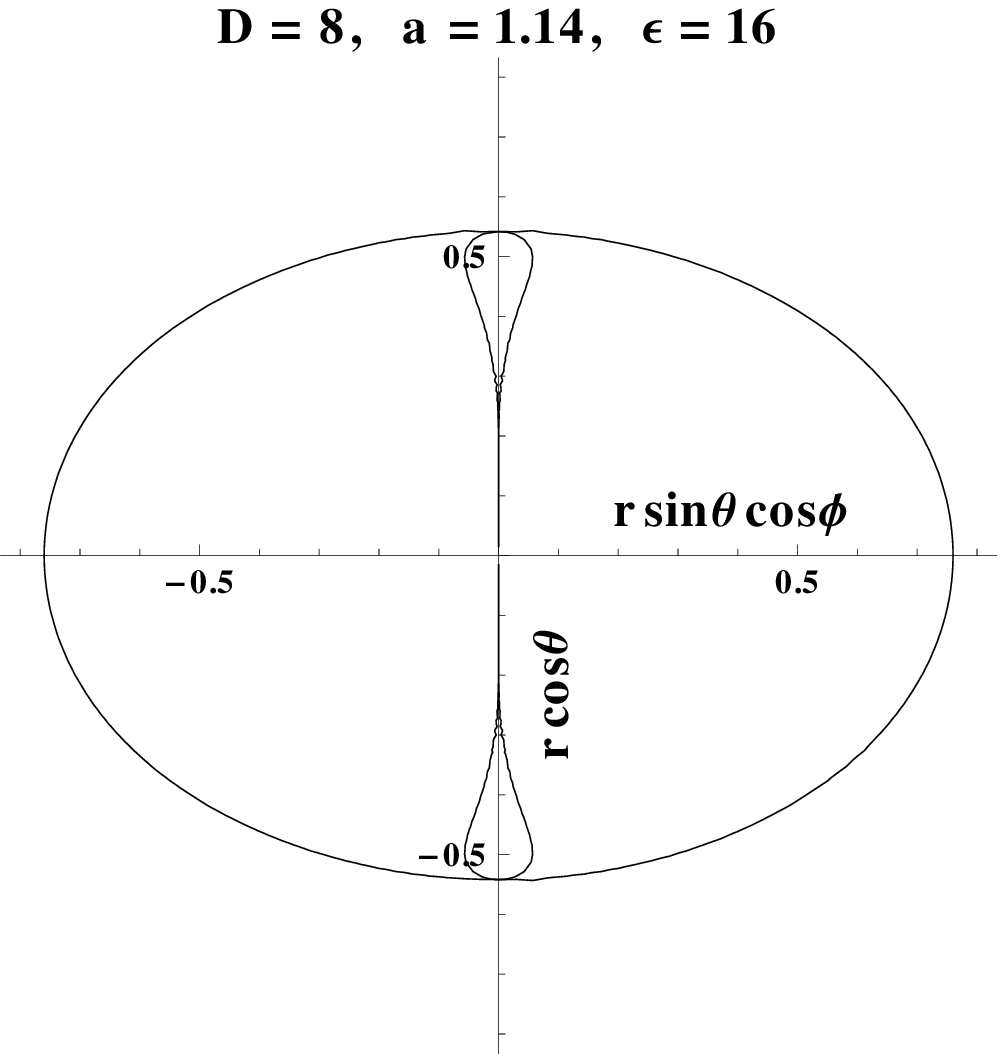}\\
\hline
\includegraphics[scale=0.3]{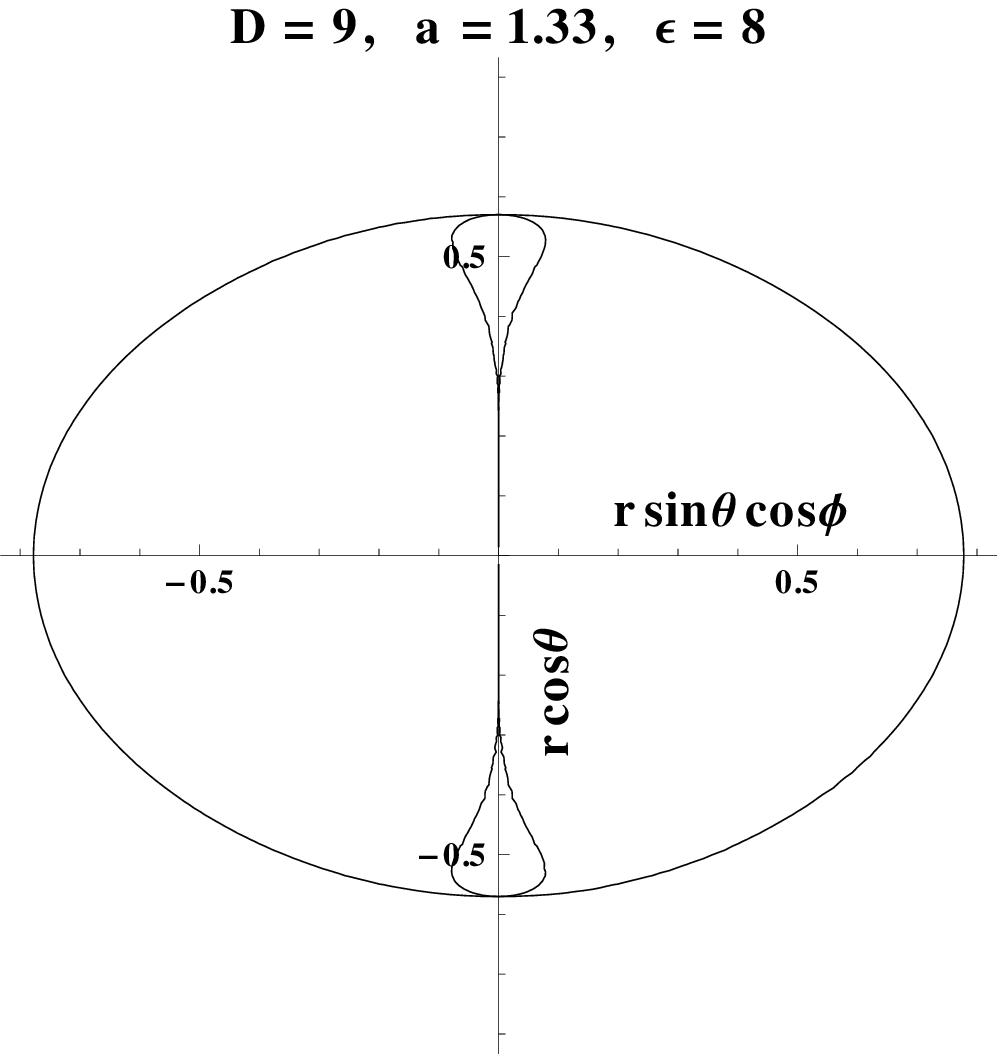}
& \includegraphics[scale=0.3]{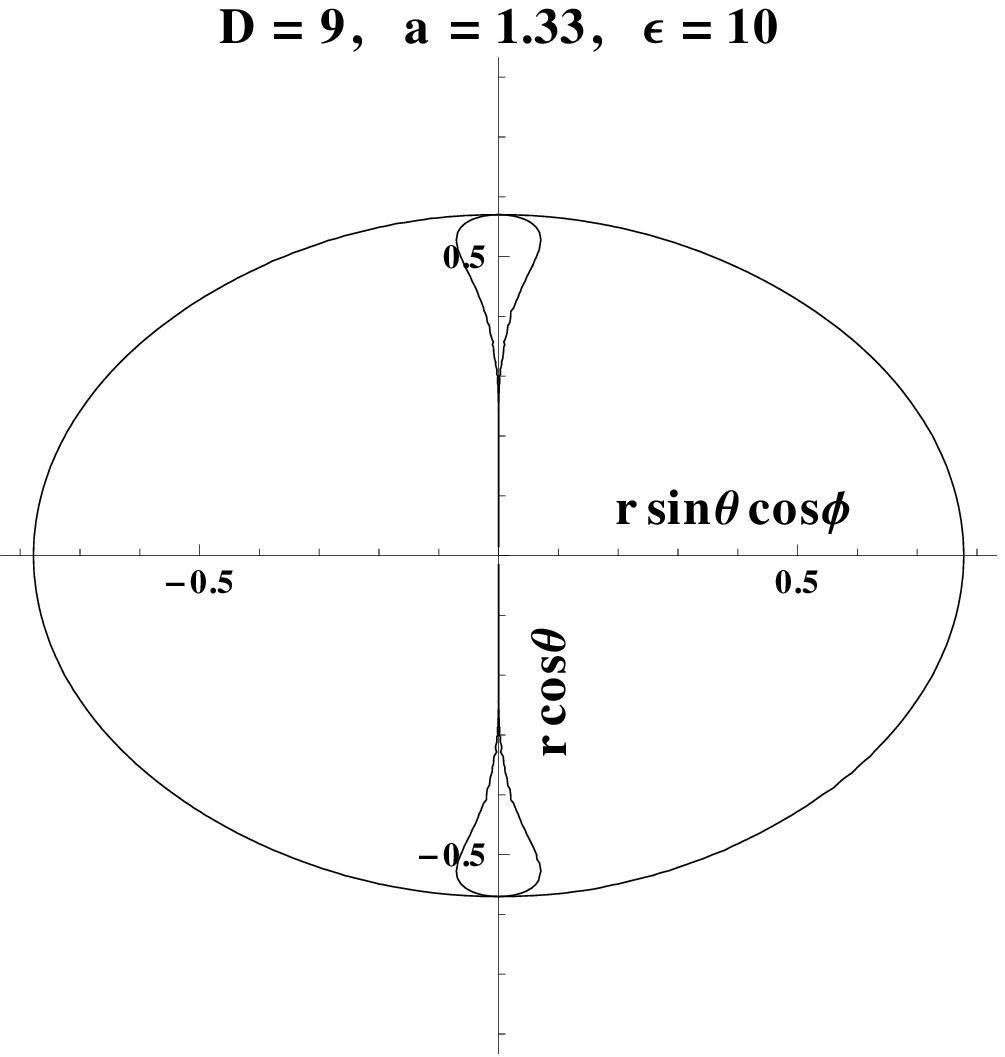}
& \includegraphics[scale=0.3]{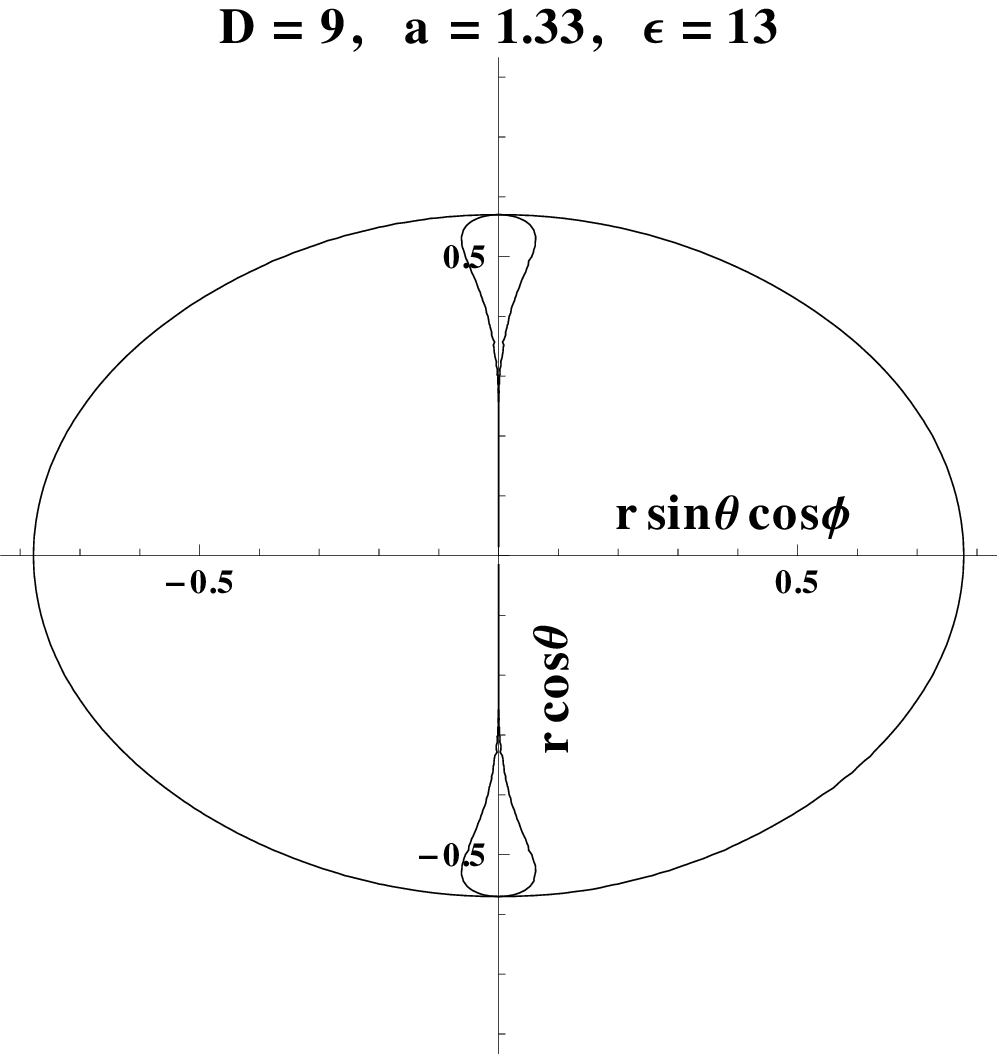}
& \includegraphics[scale=0.3]{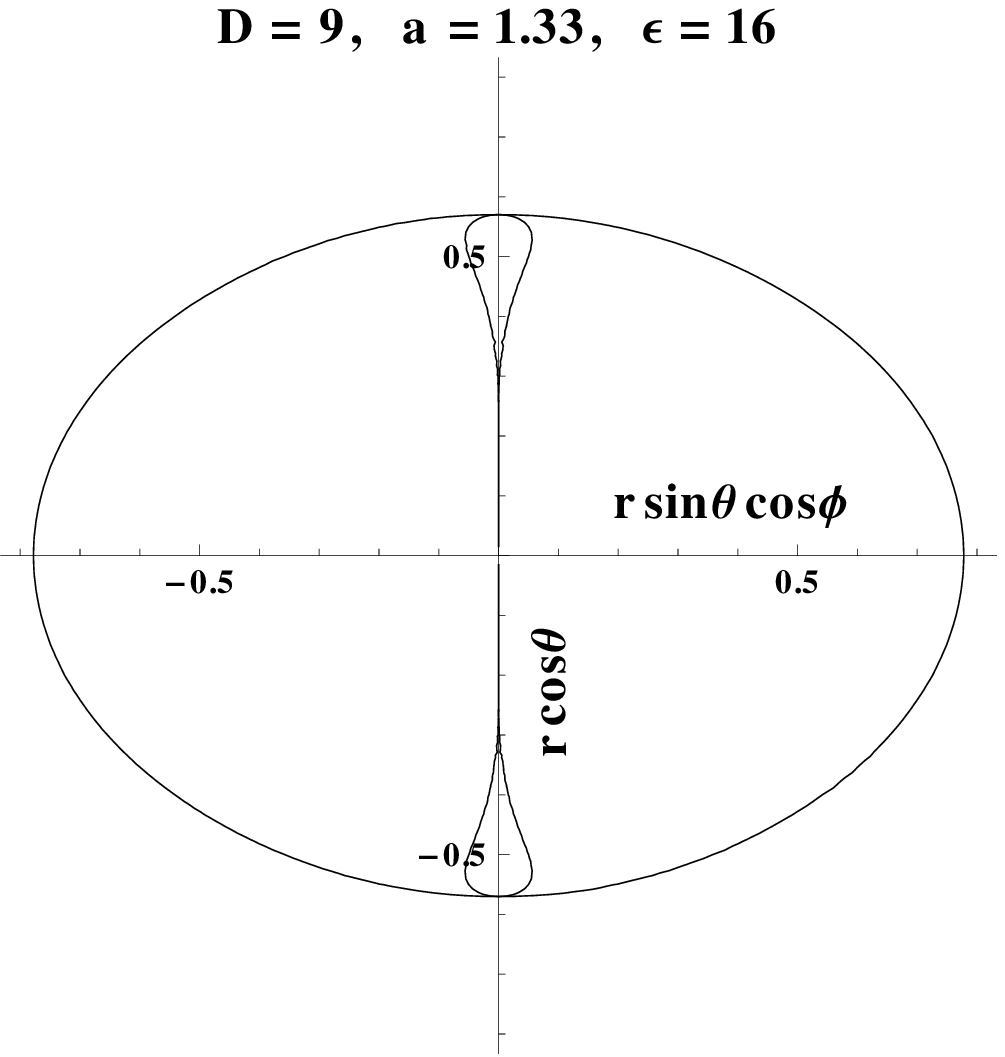}\\
\hline
\includegraphics[scale=0.3]{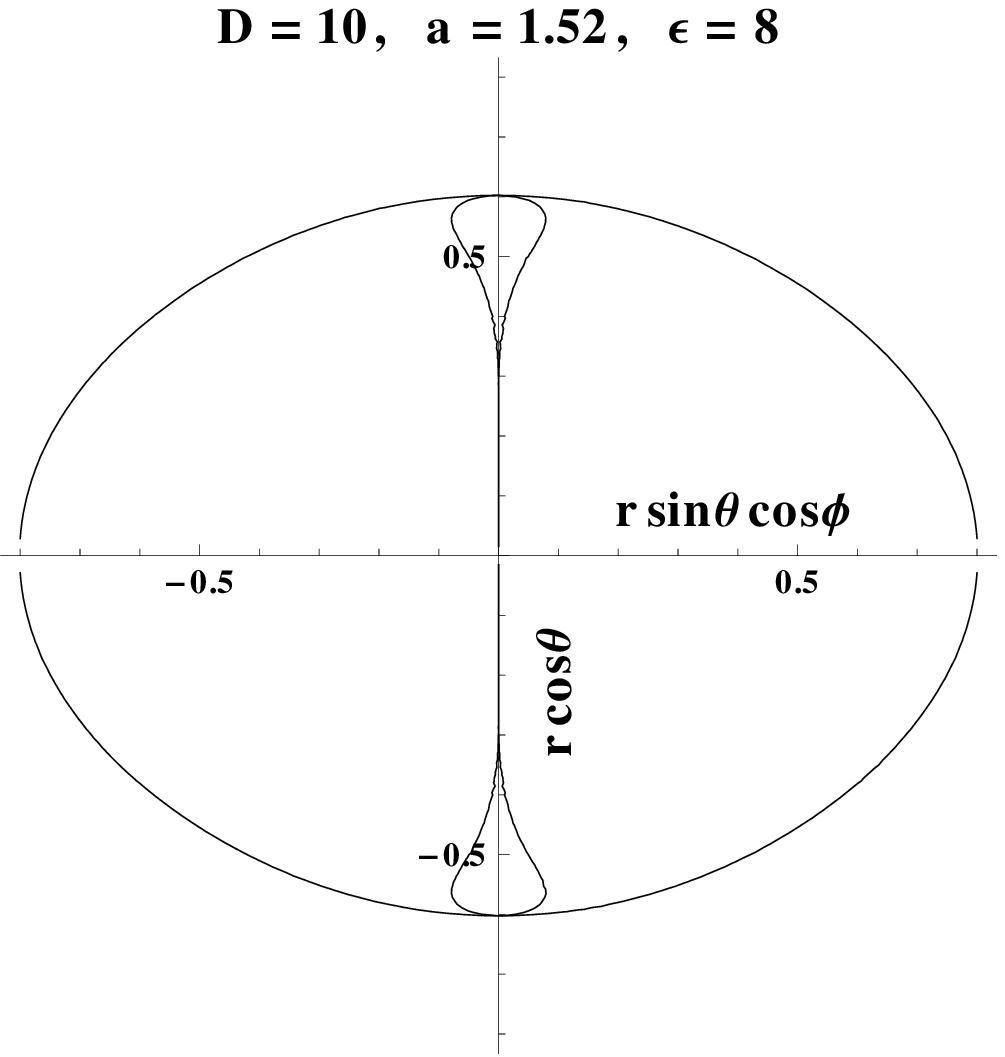}
& \includegraphics[scale=0.3]{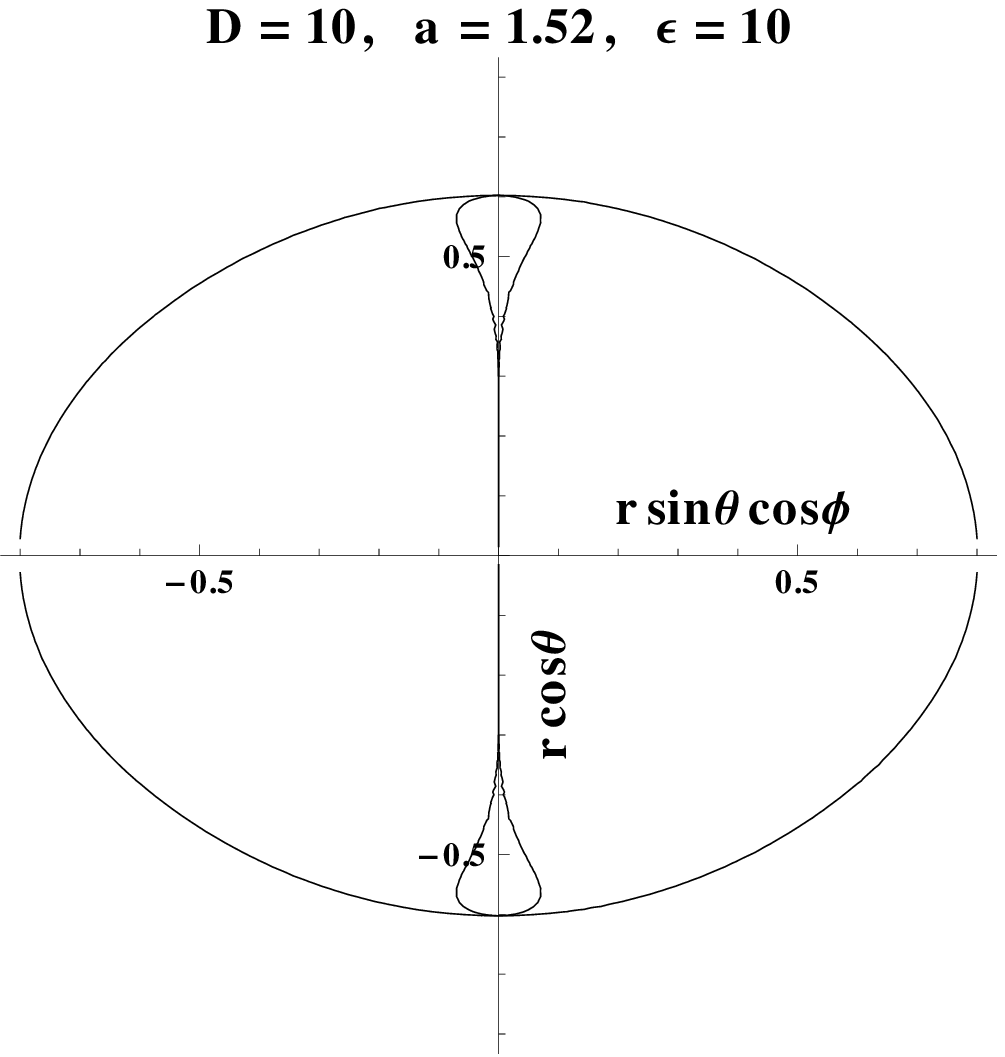}
& \includegraphics[scale=0.3]{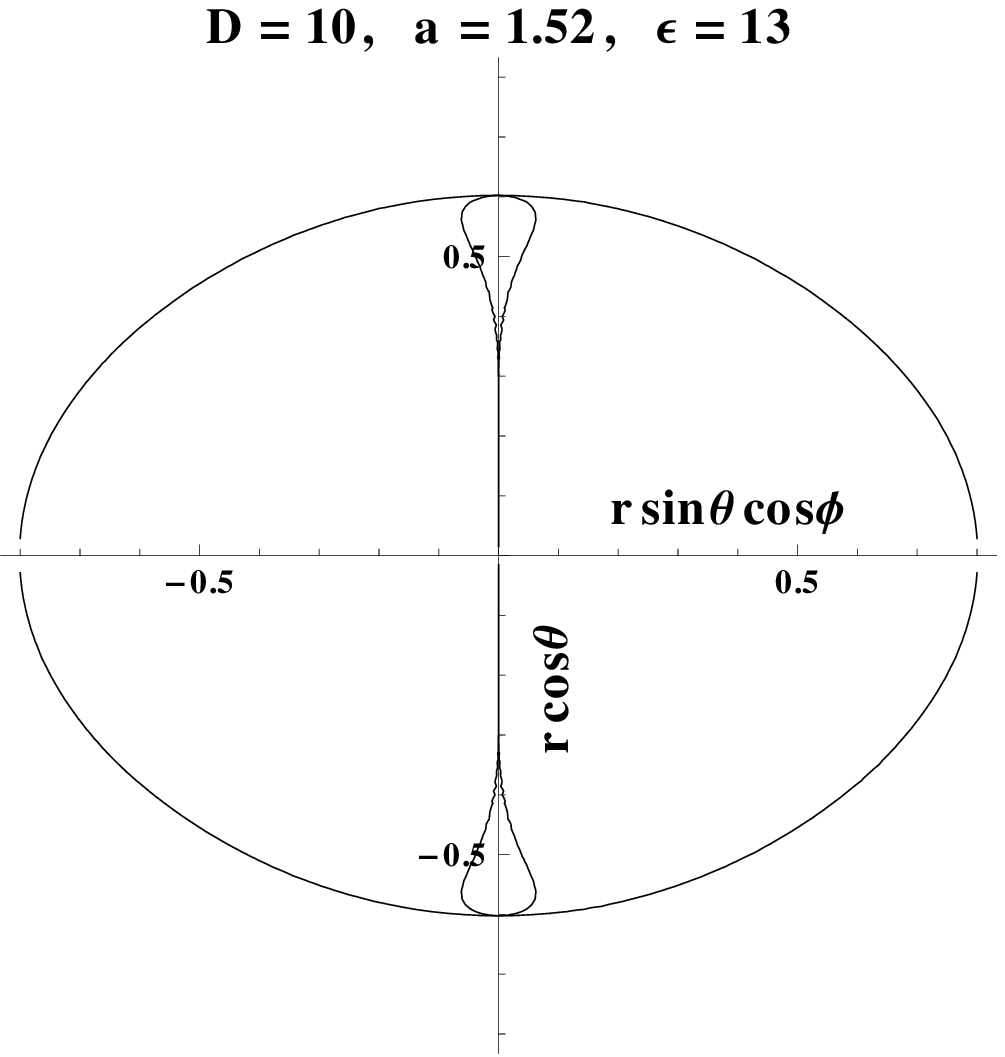}
& \includegraphics[scale=0.3]{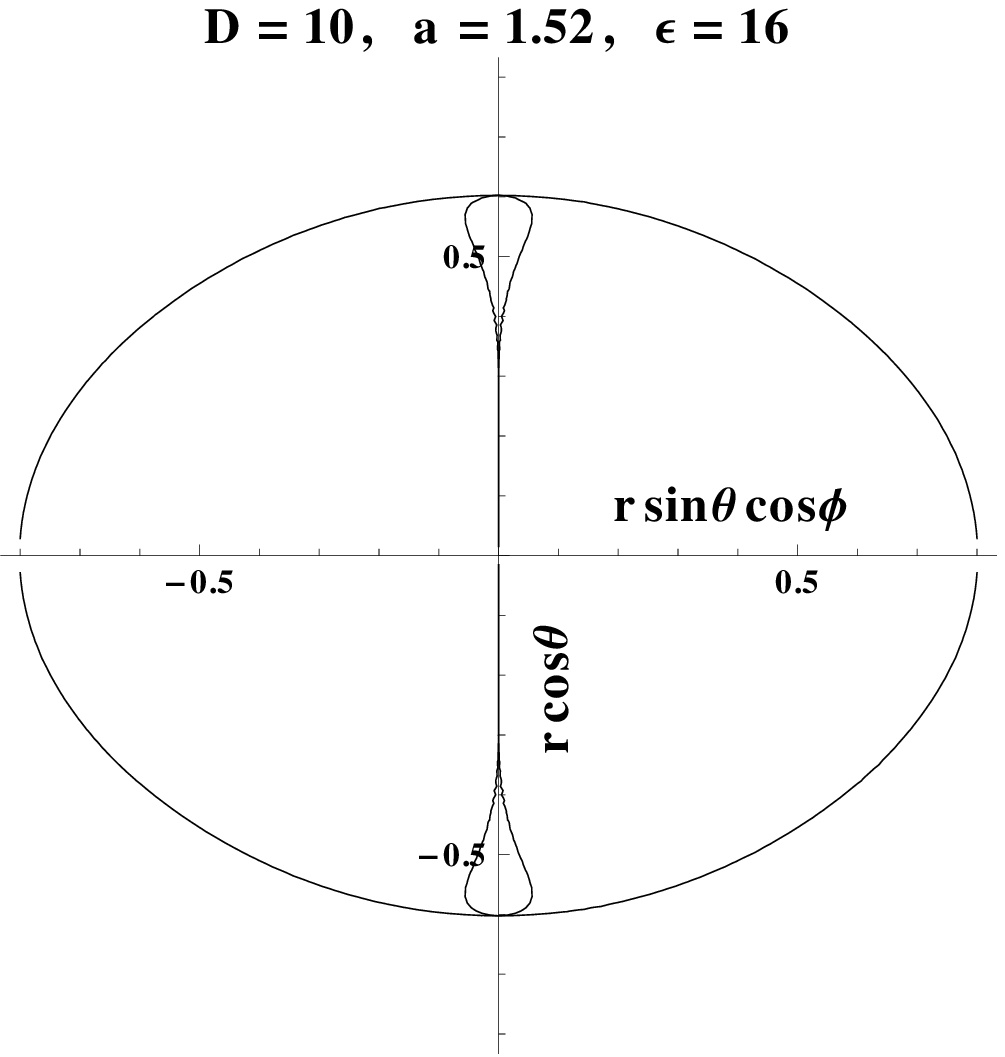}\\
\hline
\end{tabular}
\caption{The cross section of the stationary limit surface and  event horizon and the
variation of the ergosphere for different dimensions ($D$ = 4,$\ldots$,10) with four different
values of deformation parameter $\epsilon$. It shows a disconnected event horizon for higher values of deformation
parameter $\epsilon>0$.}\label{ergosphere2a}
\end{figure*}

\begin{figure*}
\begin{tabular}{ | c | c | c | }
\hline
\includegraphics[scale=0.3]{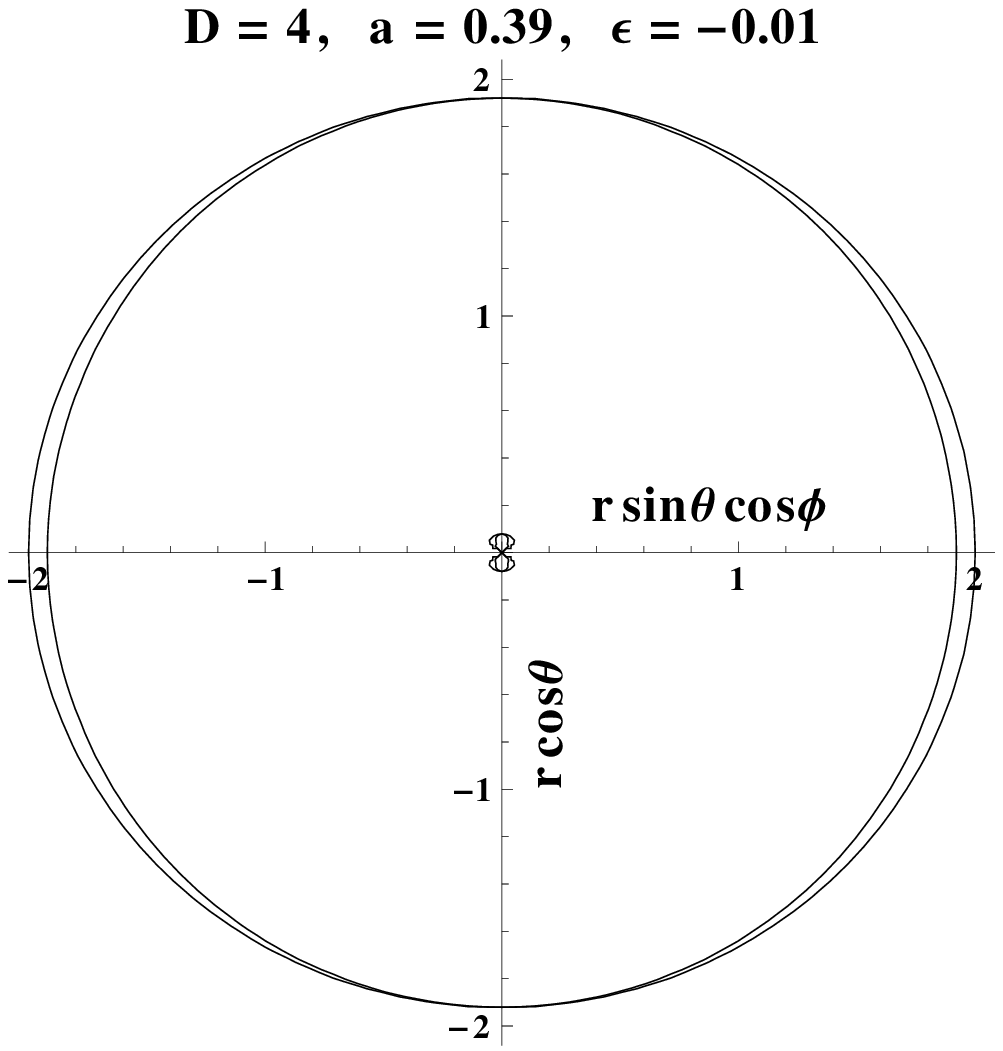}
& \includegraphics[scale=0.3]{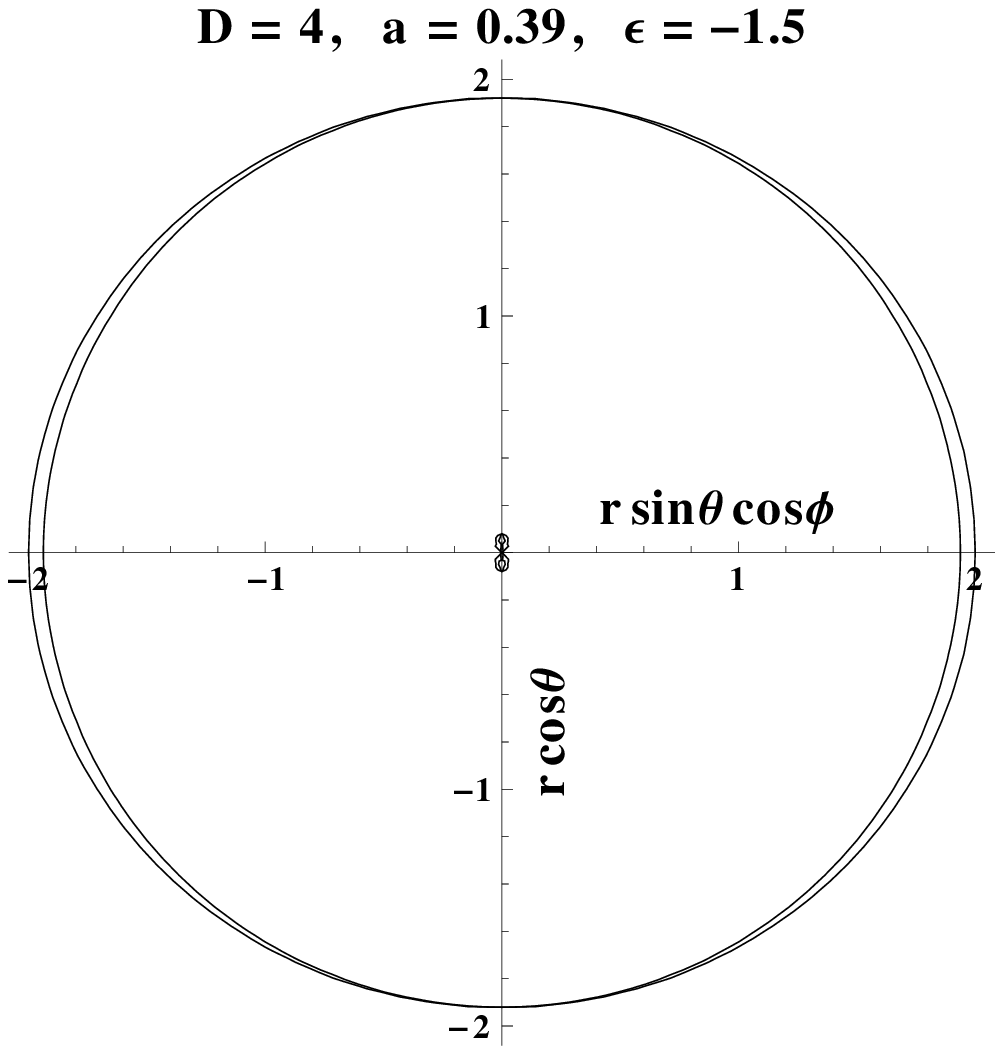}
& \includegraphics[scale=0.33]{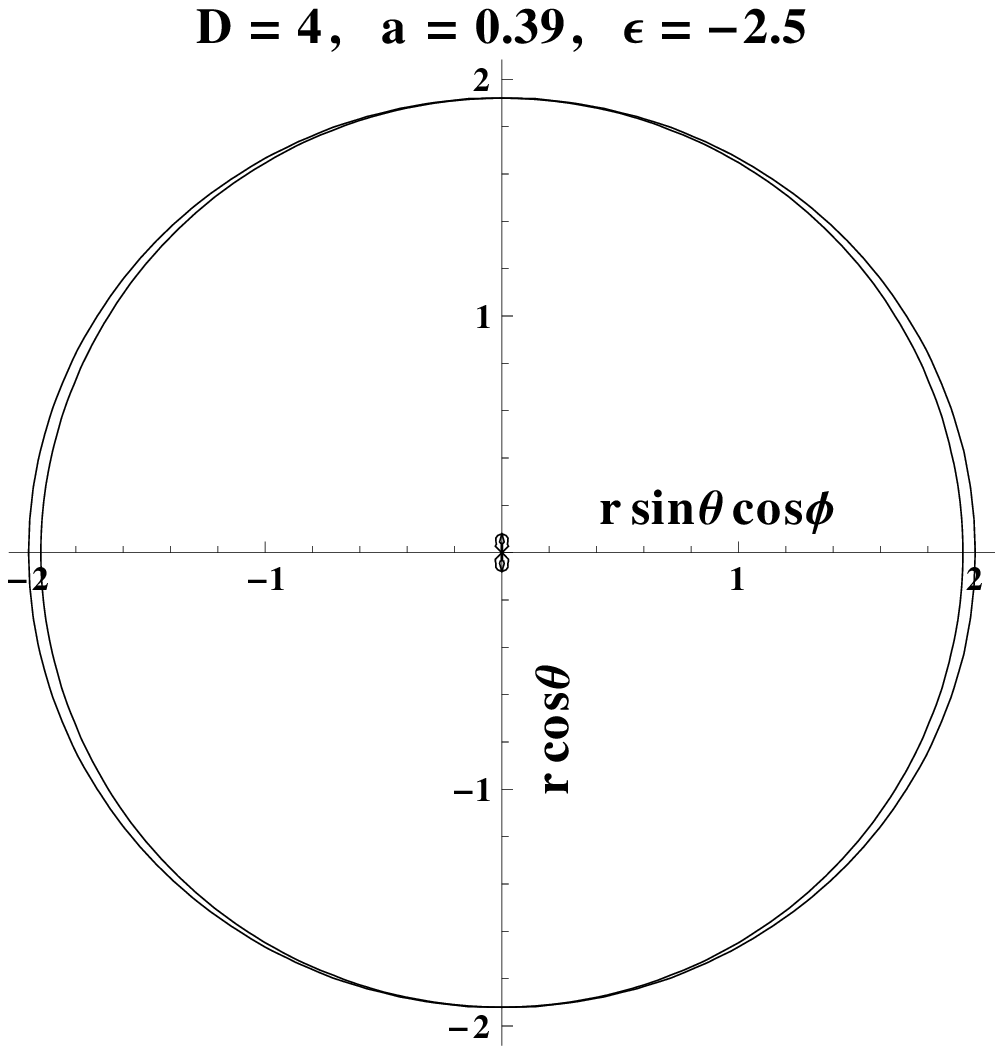}
\\
\hline
\includegraphics[scale=0.3]{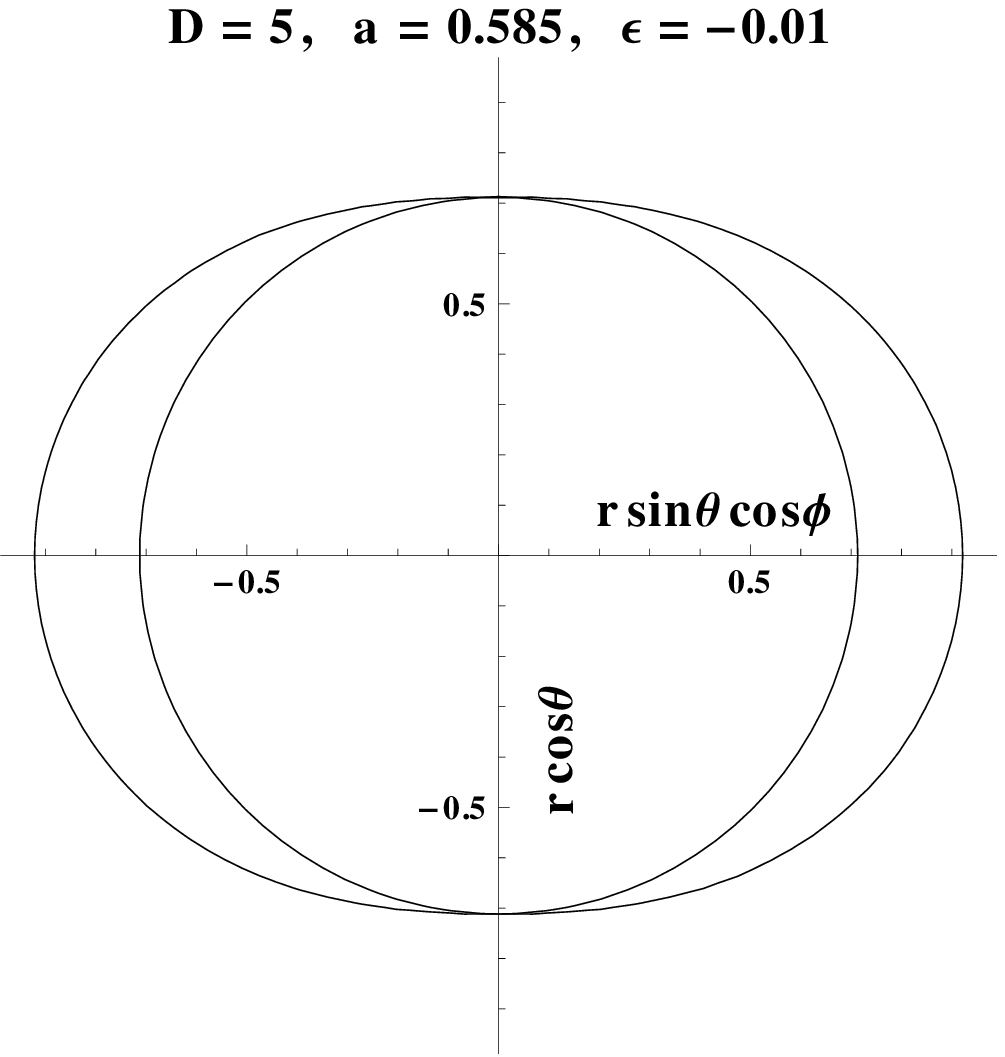}
& \includegraphics[scale=0.3]{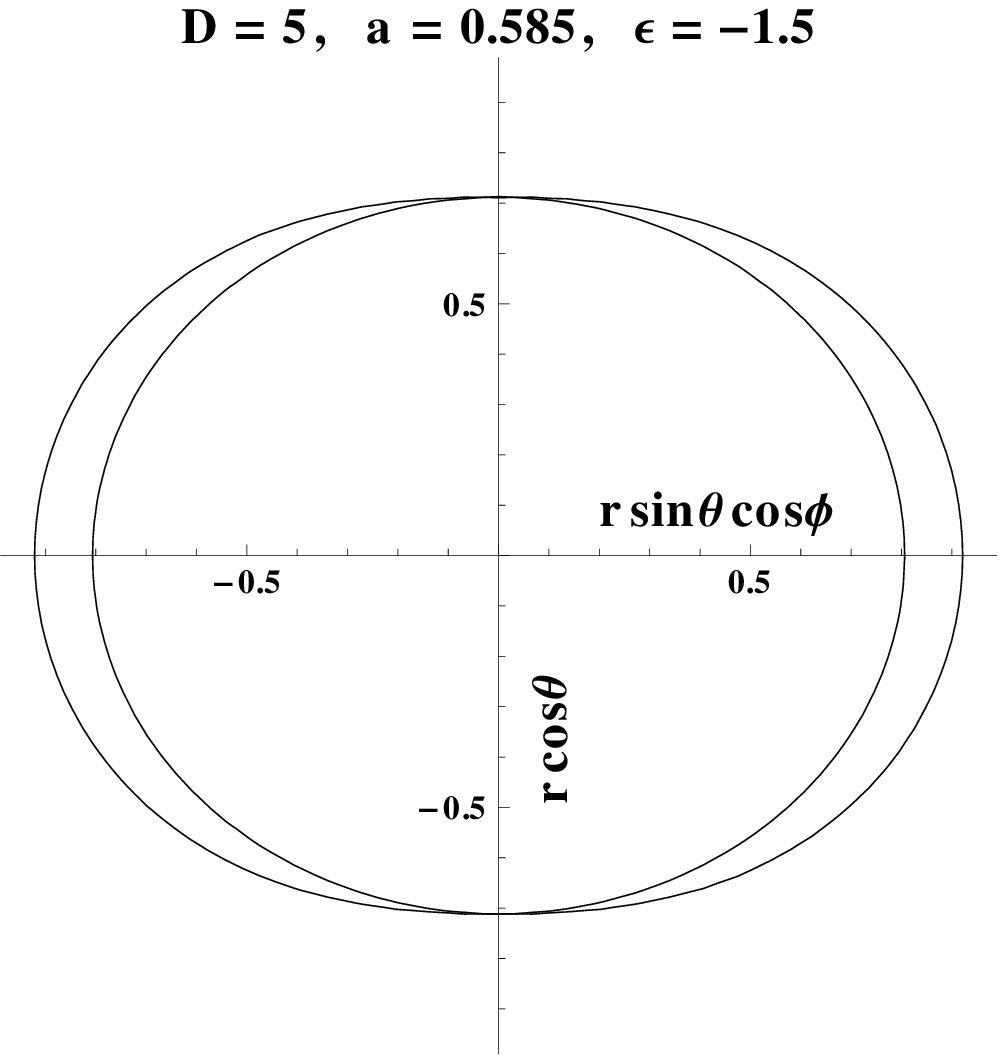}
& \includegraphics[scale=0.3]{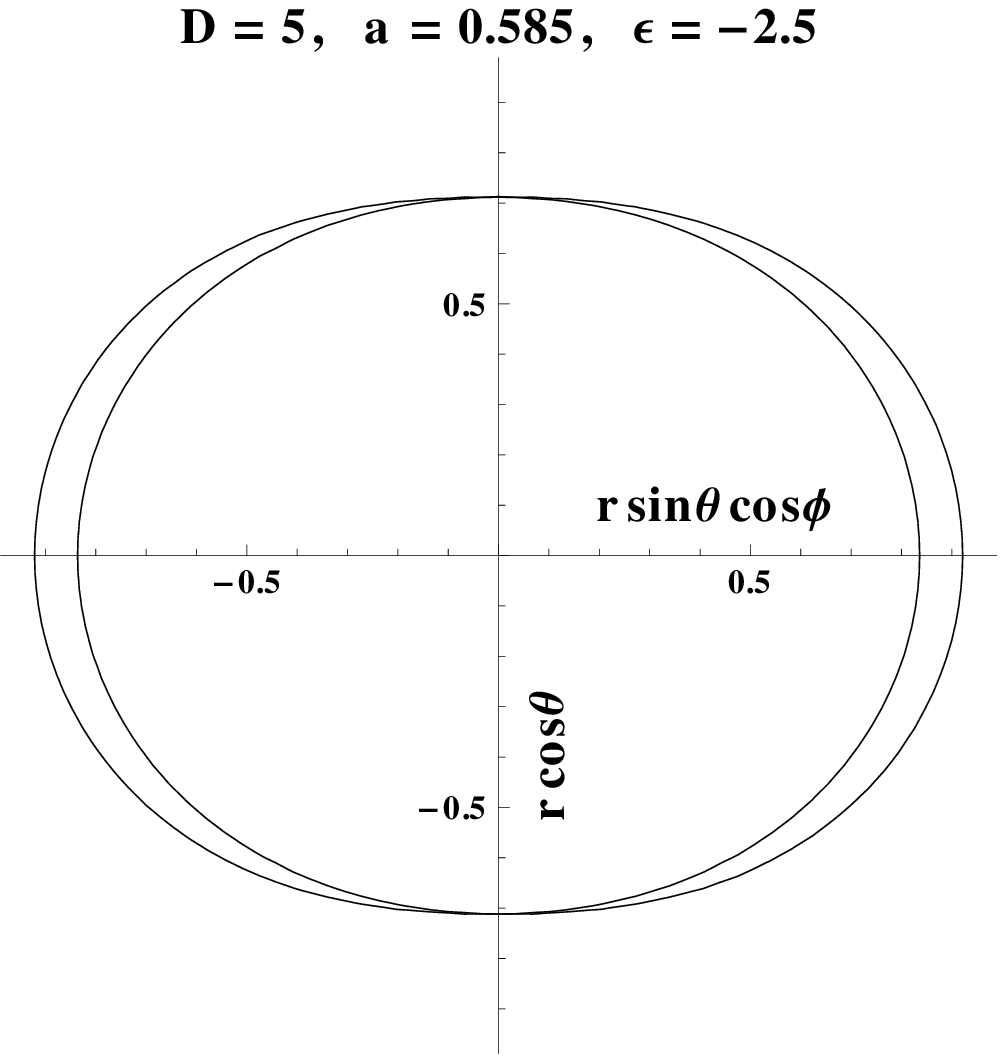}
\\
\hline
\includegraphics[scale=0.3]{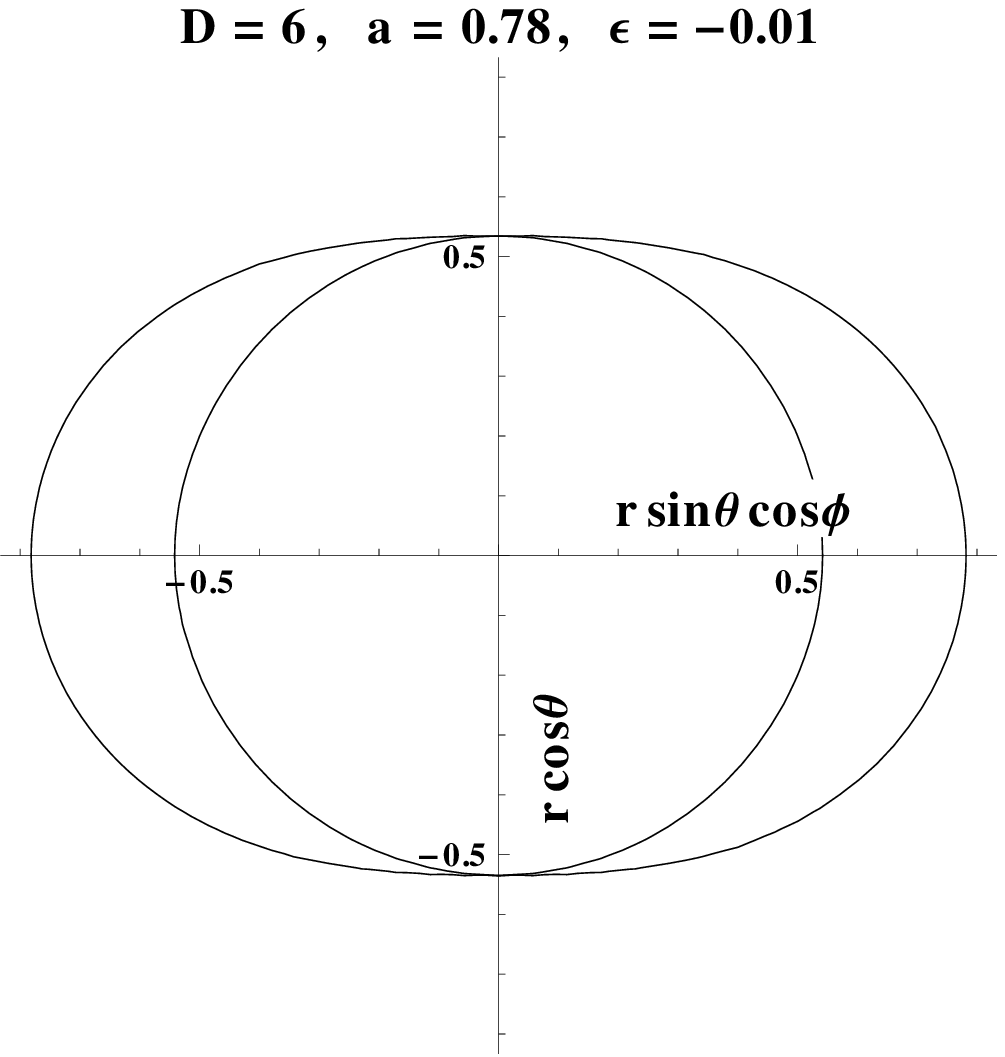}
& \includegraphics[scale=0.3]{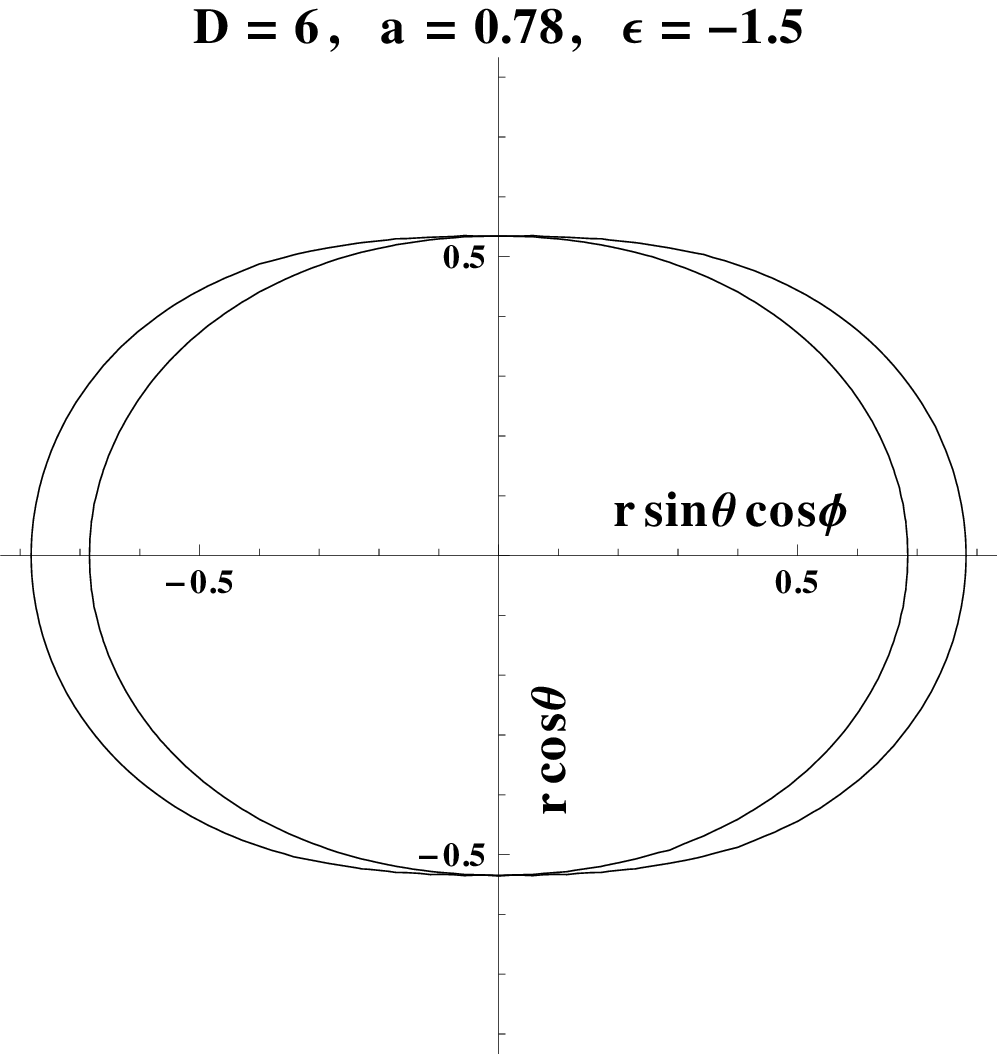}
& \includegraphics[scale=0.3]{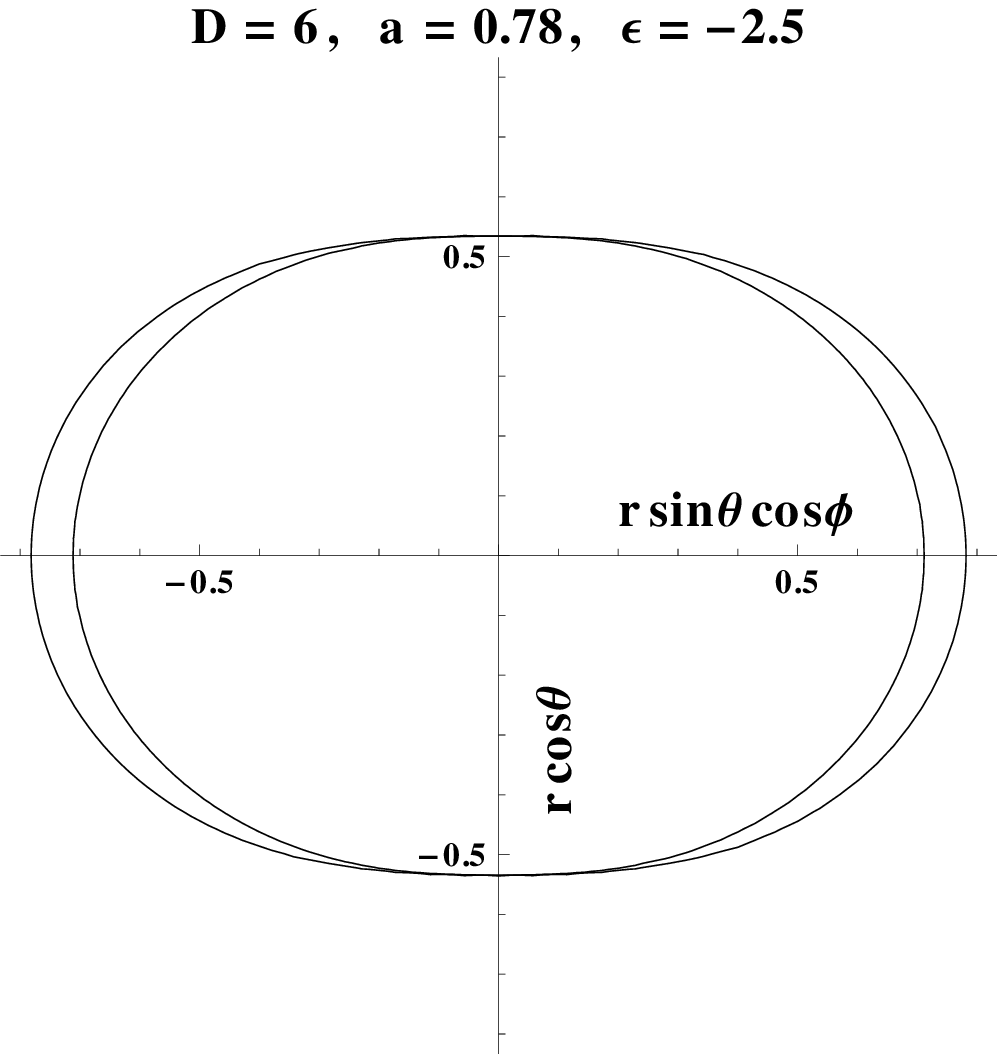}
\\
\hline
\includegraphics[scale=0.3]{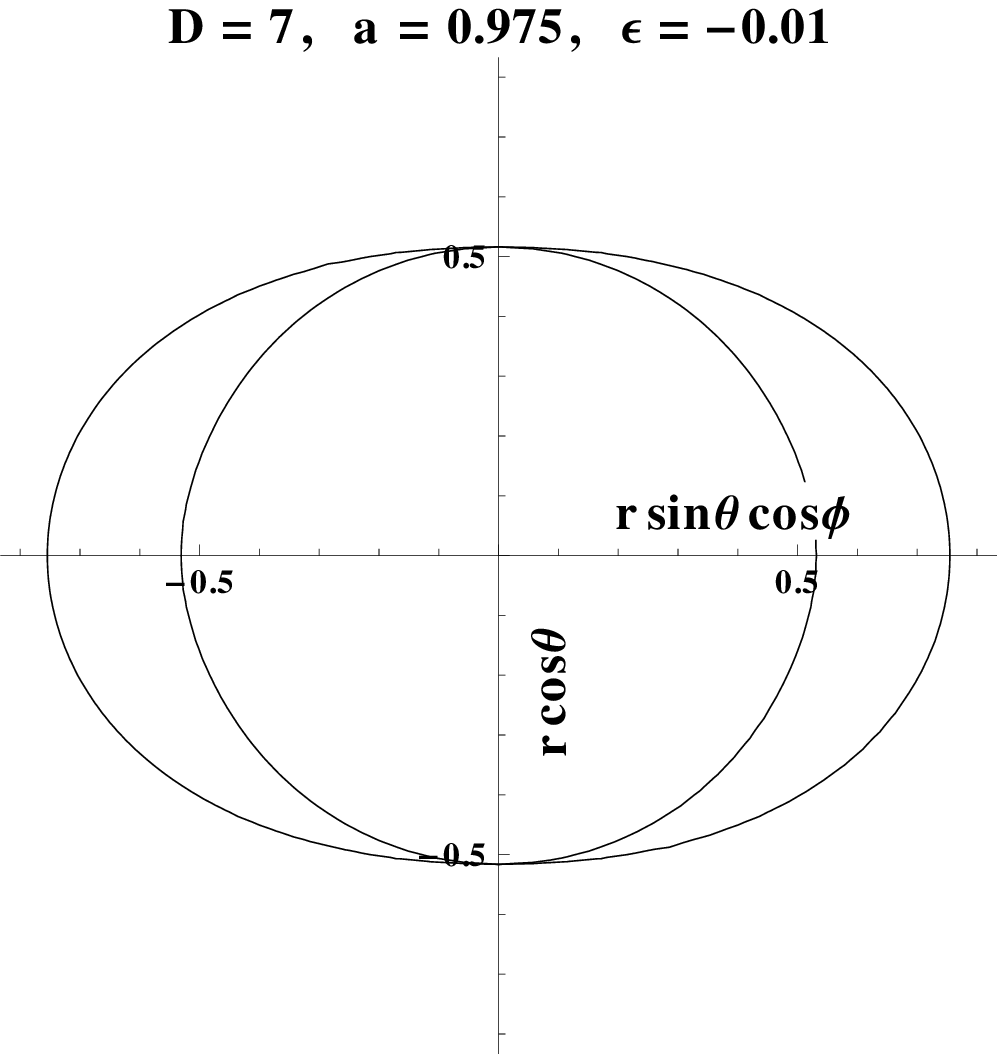}
& \includegraphics[scale=0.3]{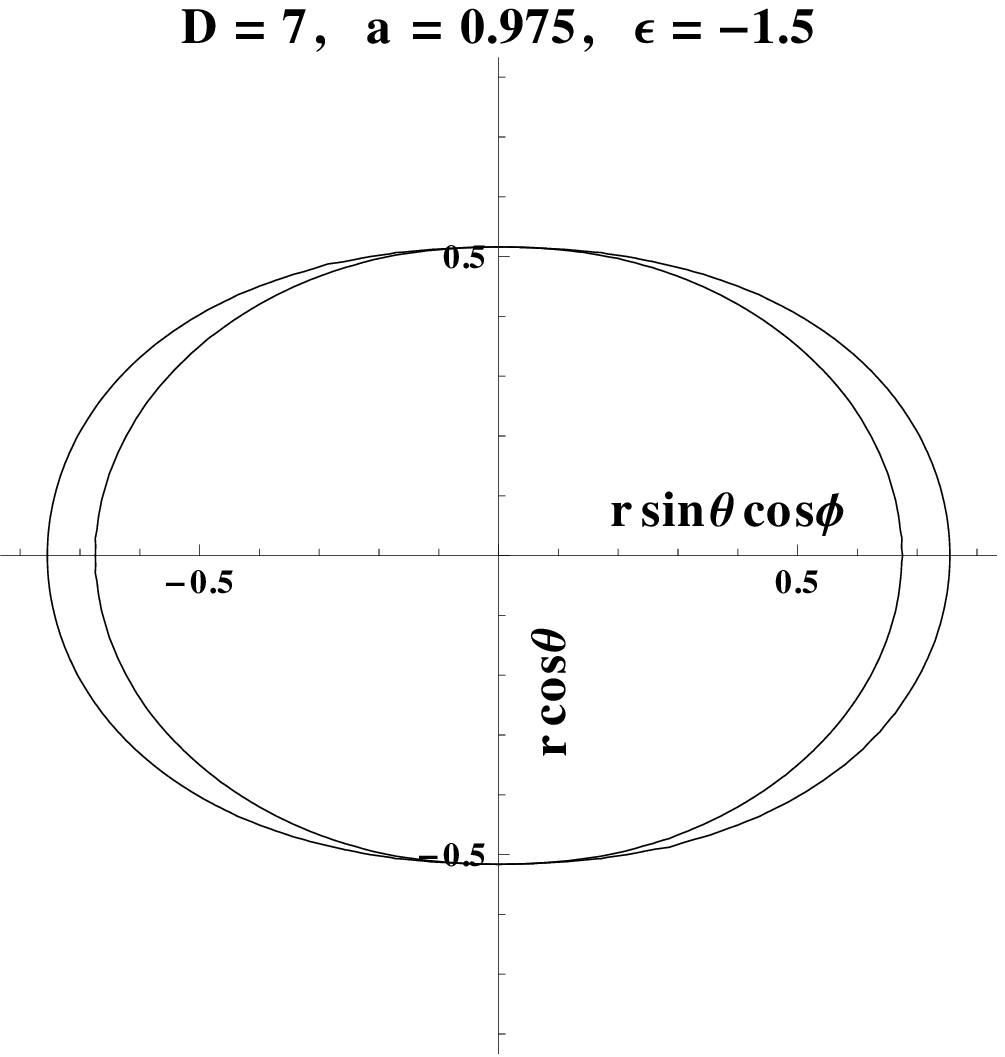}
& \includegraphics[scale=0.3]{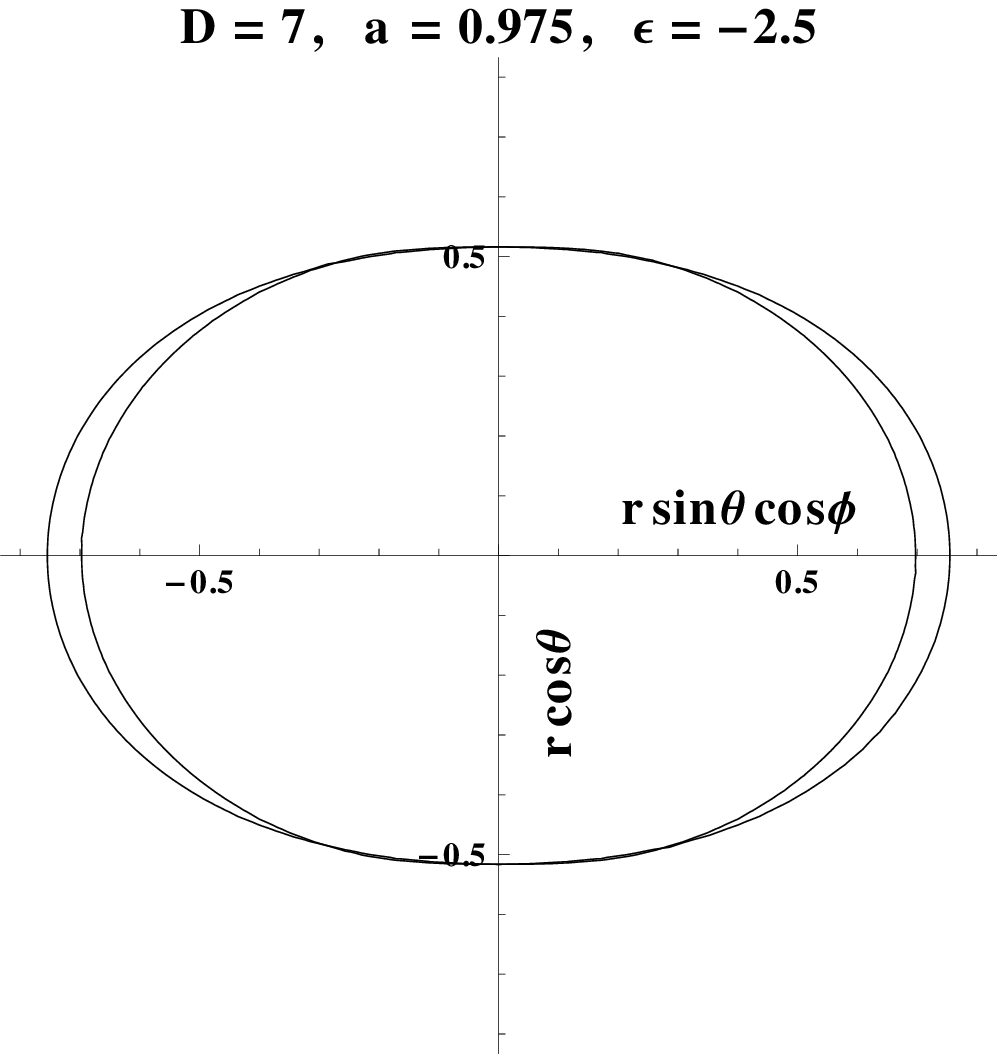}
\\
\hline
\includegraphics[scale=0.3]{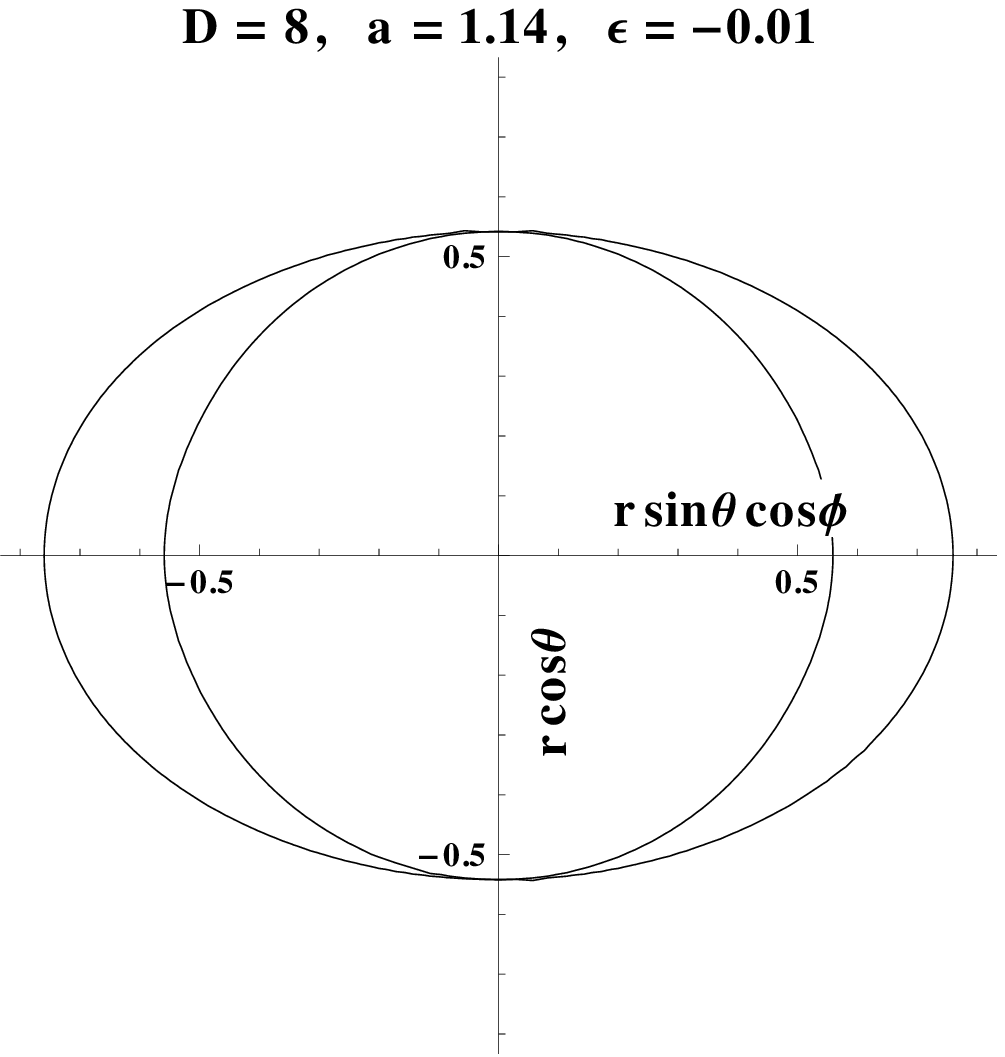}
& \includegraphics[scale=0.3]{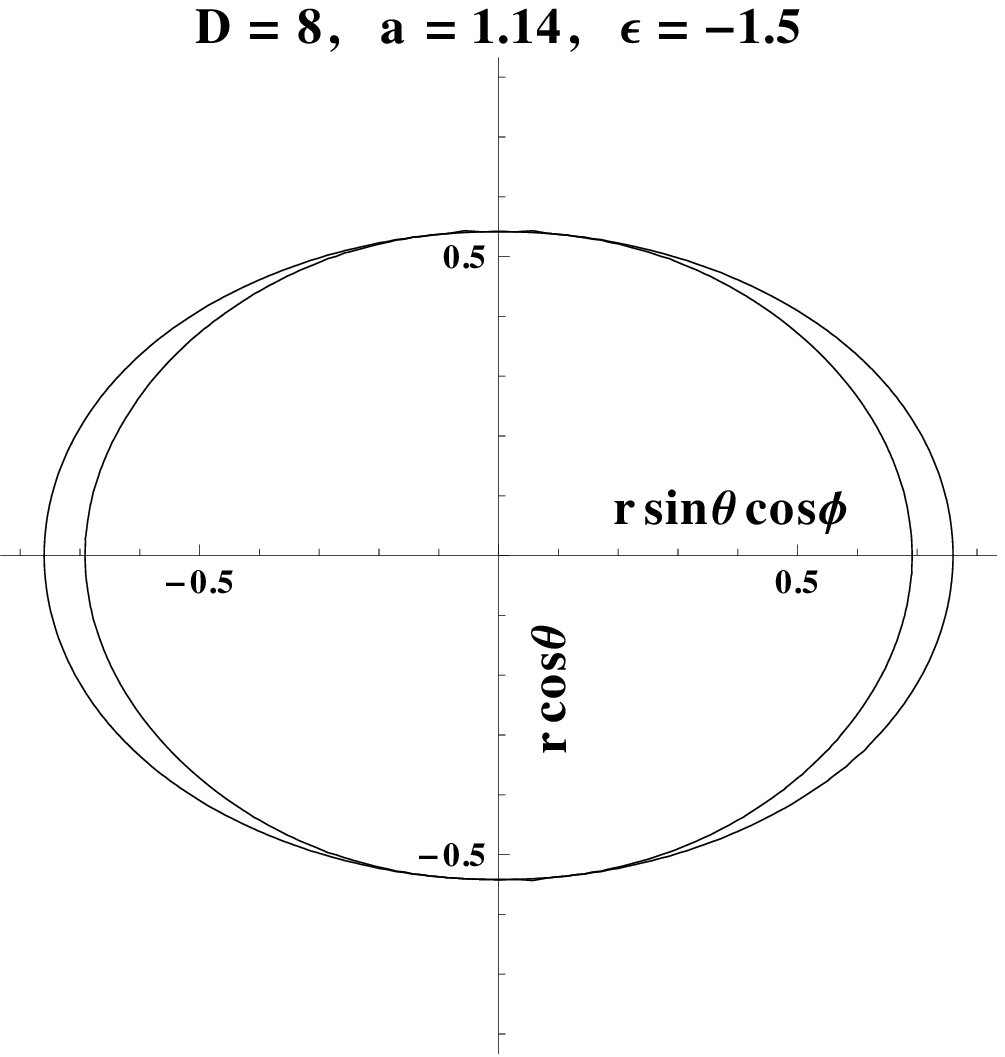}
& \includegraphics[scale=0.3]{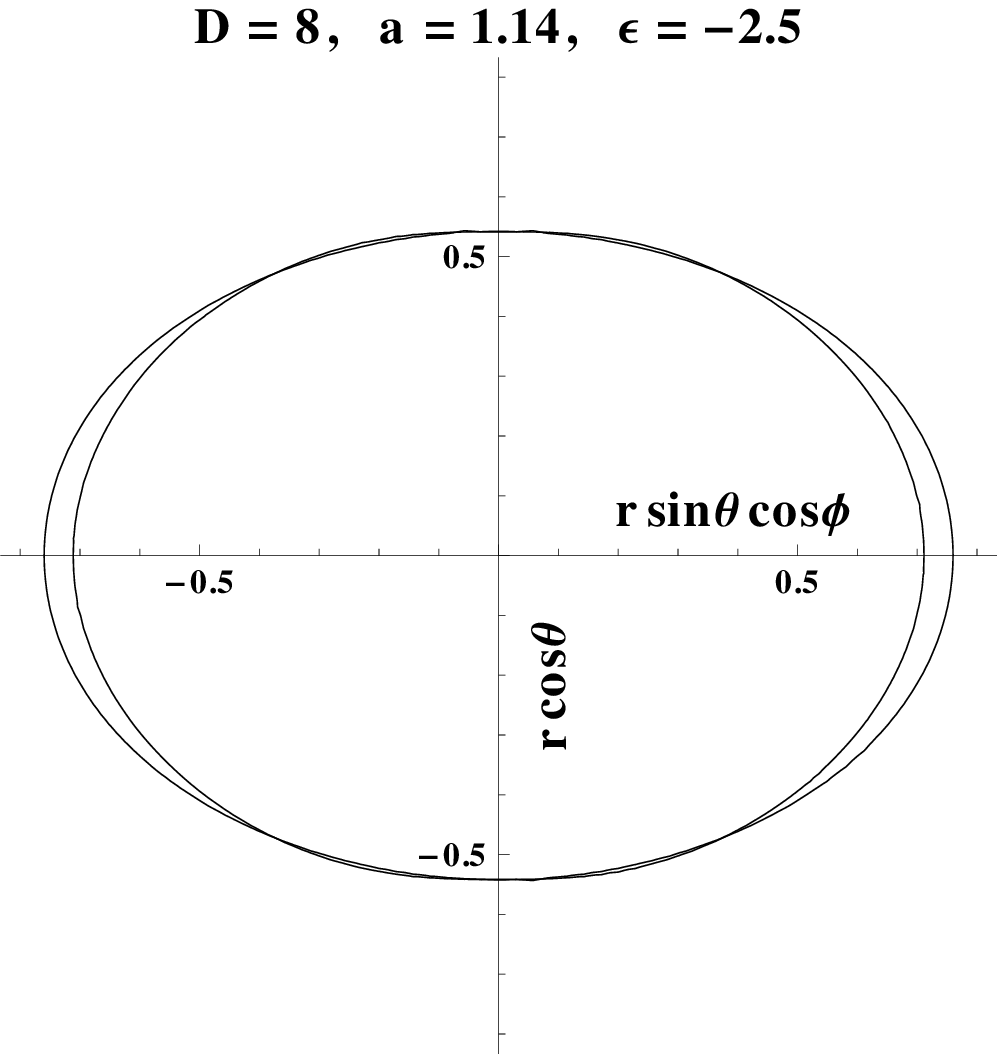}
\\
\hline
\includegraphics[scale=0.3]{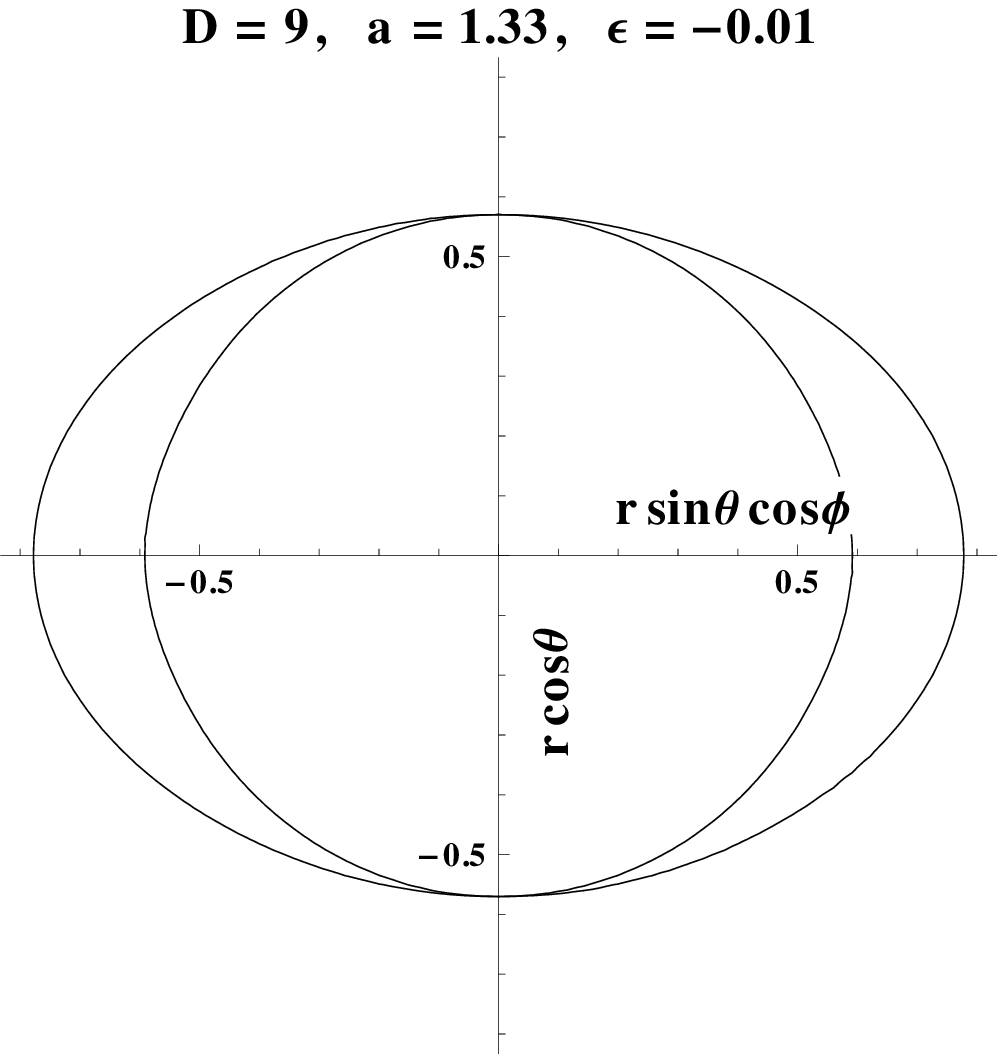}
& \includegraphics[scale=0.3]{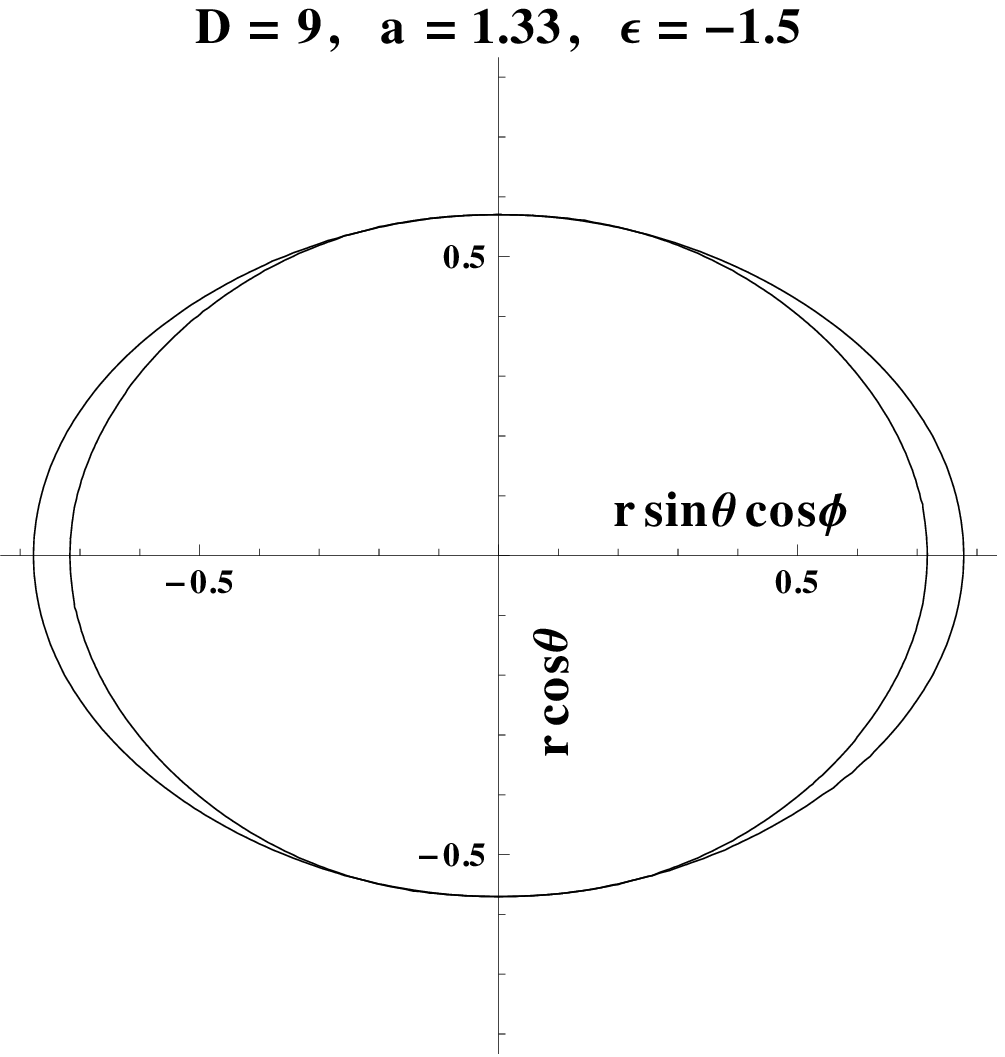}
& \includegraphics[scale=0.3]{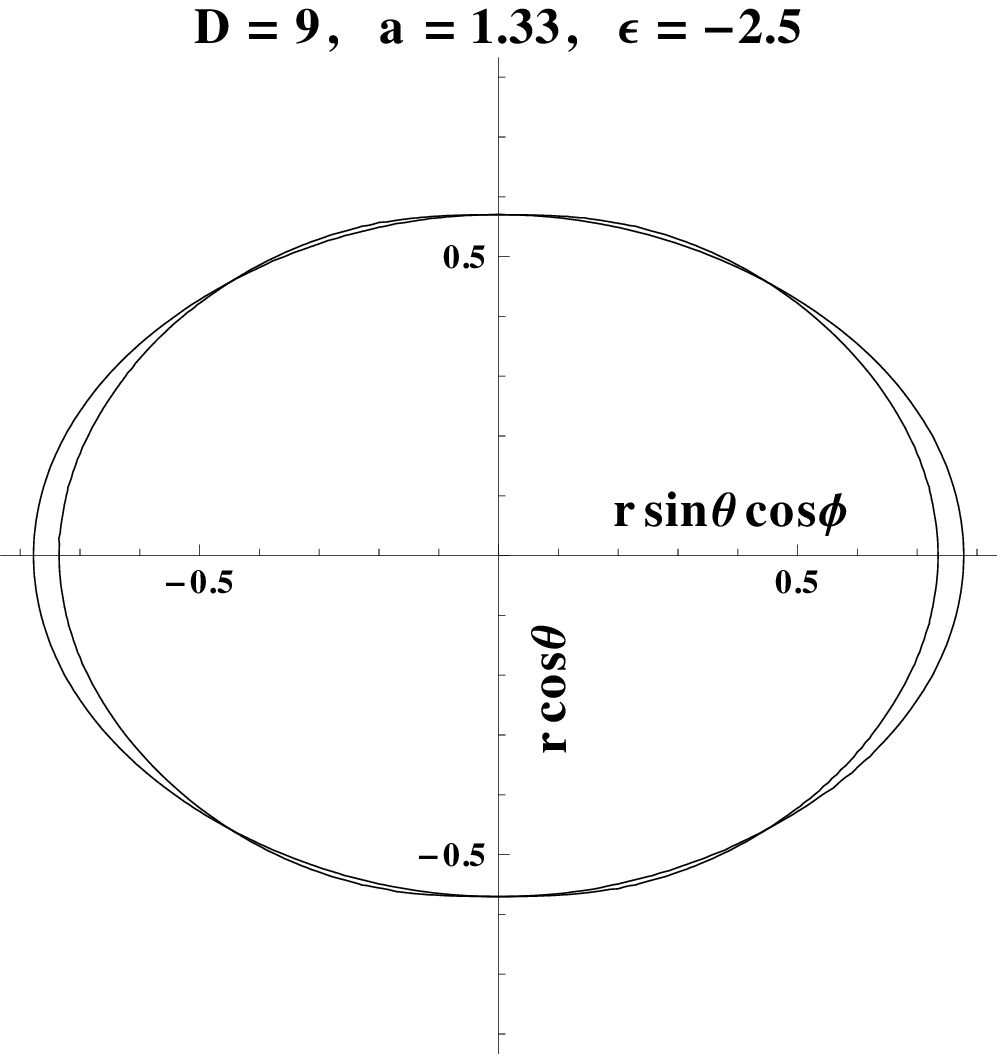}
\\
\hline
\includegraphics[scale=0.3]{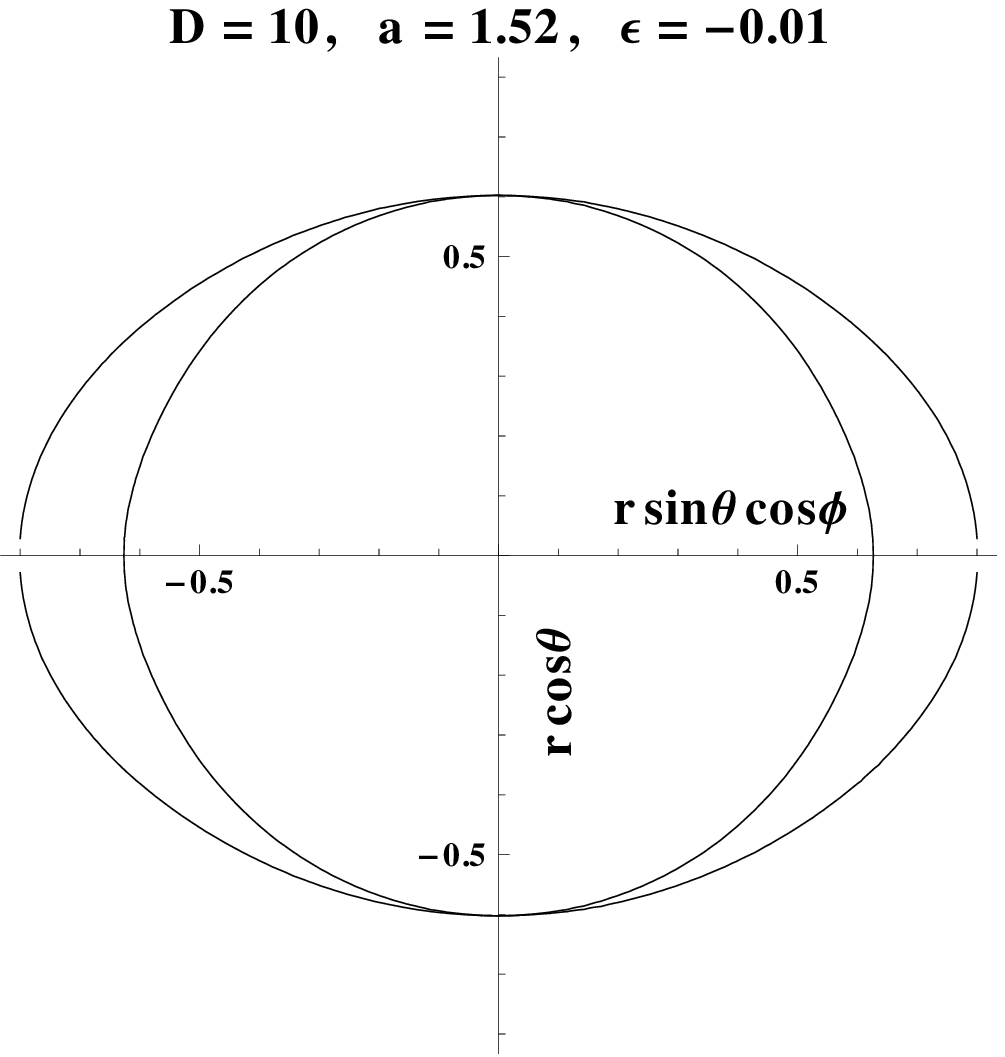}
& \includegraphics[scale=0.3]{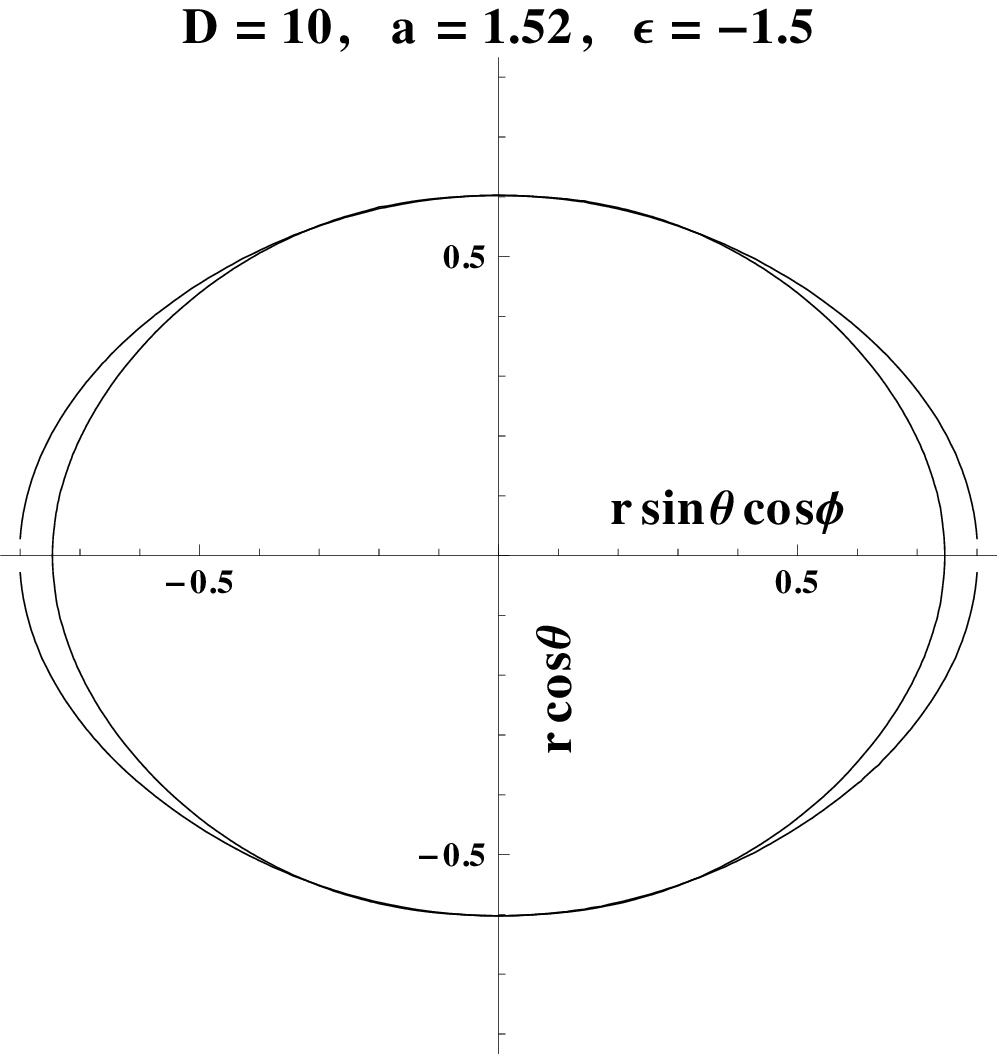}
& \includegraphics[scale=0.3]{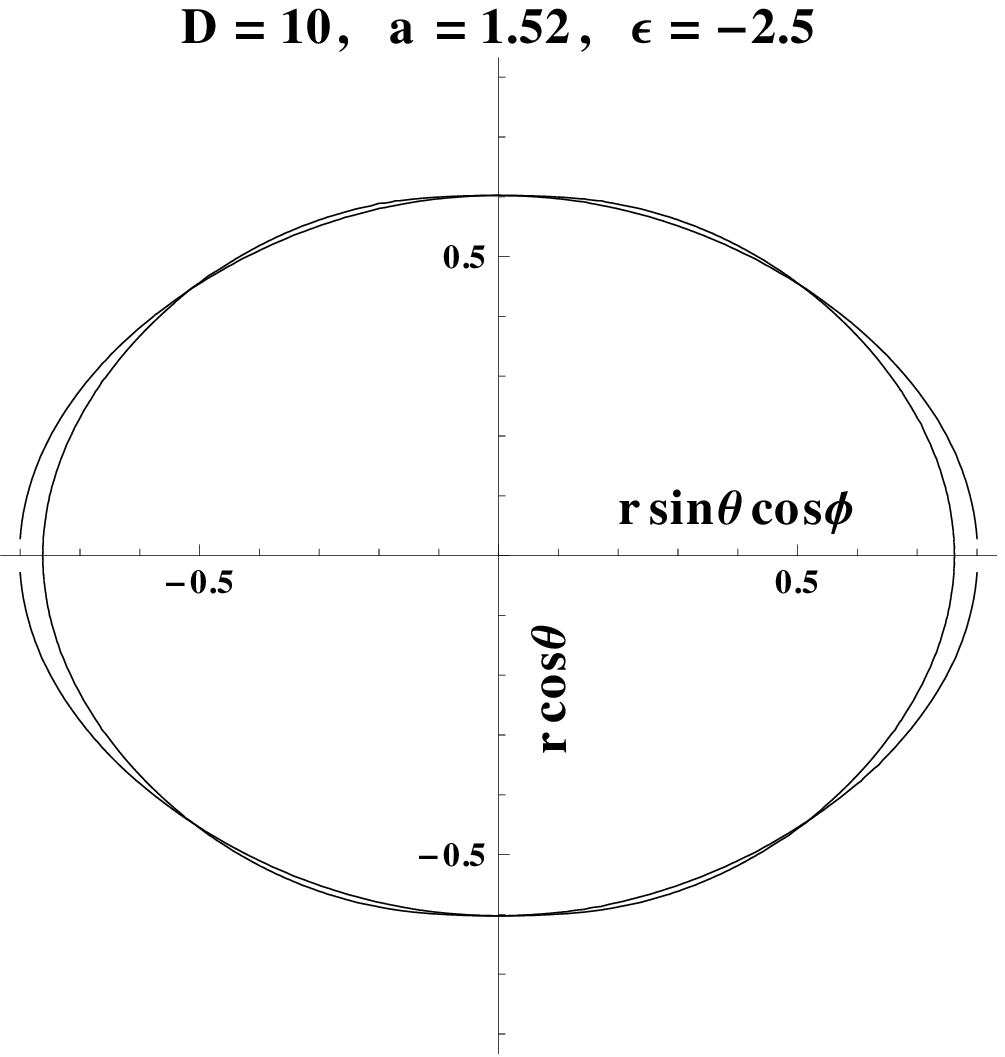}
\\
\hline
\end{tabular}
\caption{The cross section of the stationary limit surface and the event horizon and the variation
of the ergosphere for different dimensions ($D$ = 4,$\ldots$,10) with negative values of
deformation parameter $\epsilon$. For $\epsilon<0$ the ergosphere 
region shrinks with an increase in the
magnitude of $\epsilon$.}\label{ergosphere3a}
\end{figure*}

The non-Kerr black hole becomes more prolate than the Kerr black hole for the case
$a<M$, and the size of the ergosphere increases with the increase in
value of the deformation parameter $\epsilon$. We wish to bring out
the effect of the extra dimension on the ergosphere.  The ergospheres
in various cases are shown  in
Figs.~\ref{ergosphere1a}-\ref{ergosphere2a}, which are polar
plots of Eqs. (\ref{tls}) and (\ref{eh}).  How the deformation
parameter $\epsilon$ and $D$ affect the size of the ergosphere is
demonstrated in these figures.  We note that the relative shape of
the ergosphere becomes more prolate, thereby increasing the area of the
ergosphere with rotation parameter $a$; i.e., the faster the black hole rotates,
the more the ergosphere grows. The area of ergospheres also grows with
an increase in the dimension $D$.  The positive value of the deformation
parameter $\epsilon>0$ also facilitates the increase of the area of
the ergosphere in higher dimensions, but it slightly decreases in seven dimensions (7D).
However, we get disconnected horizons for the high values of
$\epsilon>0$ and $a$.  In Fig.~\ref{ergosphere3a}, we plot
the variation in size of the ergosphere for the negatives values of the
deformation parameter $\epsilon<0$. It turns out the increase in
negative values of the deformation parameter $\epsilon$ leads to
shrinkage of the size of the ergosphere.

The term turning point $(\epsilon_{tp})$ corresponding to the upper limiting value of the deformation
parameter corresponds to the largest positive root of $(\partial \epsilon/ \partial r)=0$ with 
the help of Eq. (\ref{eh}). Following \cite{lcj}, the value of the deformation parameter for the turning
point $(\epsilon_{tp})$ can be obtained from Eq. (\ref{eh}) as
\begin{equation}
\epsilon_{tp} = -\frac{8 \Delta \Sigma^{2} r^{3N-4}}{\mu^{3} a^{2} \sin^{2}{\theta}}\Bigg|_{r=r_{tp}}.
\end{equation}

\begin{table}
\begin{center}
\caption{The value of the  deformation parameter at the turning
point $\epsilon_{tp}$ in the HD non-Kerr black hole for different values of
spin parameter $a$. \label{etp1}}
\begin{tabular}{ cc | cc  }
\hline\hline
\multicolumn{2}{c|}{$D = 4$}  & \multicolumn{2}{c}{$D = 5$}  \\
\hline            
$a$ & $\epsilon_{tp}$ & $a$ & $\epsilon_{tp}$\\
\hline
0.05 &  1044.48              & 0.075 & 123.98            \\
0.10 &   258.06              & 0.150 &  25.58            \\
0.15 &   112.44              & 0.225 &  11.06            \\
0.20 &    61.49              & 0.300 &   5.08            \\
0.25 &    37.93              & 0.375 &   2.46            \\
0.30 &    25.16              & 0.450 &   1.18            \\
0.35 &    17.479             & 0.525 &\;\;\;\;\;   5.40 $\times$ 10$^{-1}$\\
0.40 &    12.51              & 0.600 &\;\;\;\;\;   2.18 $\times$ 10$^{-1}$\\
0.45 &     9.14              & 0.675 &\;\;\;\;\;   7.23 $\times$ 10$^{-1}$\\
0.50 &     6.75              & 0.750 &\;\;\;\;\;   1.64 $\times$ 10$^{-1}$\\
\hline  
\end{tabular}
\end{center}
\end{table}\vspace{0.5cm}

The $r_{tp}$ is the largest positive root of $(\partial \epsilon/ \partial r)=0$. 
The value of the deformation parameter at the intersection of the three 
surfaces $\Delta+a^{2}h\sin^{2}{\theta}=0$, $1+h=0$, and
$\left[1-\mu/(r^{N-2}\Sigma)\right]=0$  is denoted by $\epsilon_{ip}$. The
allowed value of the deformation parameter $\epsilon$ should lie
within the range $\epsilon_{ip}\leq\epsilon\leq\epsilon_{tp}$ in the
ergosphere. The event horizon exists only when the deformation parameter $\epsilon$
lies between $\epsilon_{ip}$ and  $\epsilon_{tp}$; when the value
of the deformation parameter $\epsilon\geq\epsilon_{tp}$ and
$\epsilon<\epsilon_{ip}$, no event horizon exists \cite{lcj}. The
value of the deformation parameter $\epsilon_{ip}$ at the
intersection point remains constant in all the dimensions, i.e.,
$\epsilon_{ip}=-8$, whereas from Tables~ \ref{etp1} and \ref{etp2} we
conclude that the value of the deformation parameter at the turning
point $\epsilon_{tp}$ decreases as the value of the spin parameter
$a$ increases in each dimension.

\begin{table}[H]
\begin{center}
\caption{The value of the deformation parameter at the turning point
$\epsilon_{tp}$ in the HD non-Kerr black hole for different values of spin
parameter $a$. \label{etp2}}
\begin{tabular}{ cc | cc }
\hline\hline
 \multicolumn{2}{c|}{$D = 6$}  & \multicolumn{2}{c}{D = 7} \\
\hline            
 $a$ & $\epsilon_{tp}$   & $a$ & $\epsilon_{tp}$\\
\hline
 0.10     &  54.01              & 0.125 & 33.12           \\
 0.20     &  11.46              & 0.250 & 6.50            \\
 0.30     &   3.86              & 0.375 & 1.94            \\
 0.40     &   1.46              & 0.500   & $\;\;\;\;$6.33 $\times$ 10$^{-1}$\\
 0.50     &   $\;\;\;\;$5.53 $\times$ 10$^{-1}$  & 0.625 & $\;\;\;\;$2.07 $\times$ 10$^{-1}$\\
 0.60     &   $\;\;\;\;$1.98 $\times$ 10$^{-1}$  & 0.750  & $\;\;\;\;$ 6.64 $\times$ 10$^{-2}$\\
 0.70     &   $\;\;\;\;$6.58 $\times$ 10$^{-2}$  & 0.875 & $\;\;\;\;$ 2.11 $\times$ 10$^{-2}$\\
 0.80     &   $\;\;\;\;$1.99 $\times$ 10$^{-2}$  & 1.000  & $\;\;\;\;$6.79 $\times$ 10$^{-3}$\\
 0.90     &   $\;\;\;\;$5.57 $\times$ 10$^{-3}$  & 1.125 & $\;\;\;\;\;$2.25 $\times$ 10$^{-3}$\\
 1.0      &   $\;\;\;\;$1.47 $\times$ 10$^{-3}$  & 1.250 & $\;\;\;\;\;$7.79 $\times$ 10$^{-4}$\\
\hline  
\end{tabular}
\end{center}
\end{table}
We shall next show how this ergosphere can be used to extract energy
from the black hole, i.e., by throwing in particles with suitable parameters,
so they attain negative energy relative to an asymptotic observer.
We also explicitly study the effect of extra dimensions and the
deformation parameter on the energy extraction process.

\section{Energy extraction from HD Black hole}
The Penrose process theoretically suggested by Penrose
\cite{pen} is that the energy can be extracted from a spinning black hole.
This is made possible because of the existence of the ergosphere,  where it
is possible to have timelike or a null trajectory with negative
energy. Penrose considered an infalling particle disintegration in
the ergosphere of a Kerr black hole.
One of the particles produced in this
process might be thrown into a negative energy (with respect to
infinity) orbit, while the other one will have an energy  larger
than that of the infalling one. The particle with a negative energy will be swallowed by a
black hole while the other one will escape to infinity with a net gain in energy.
 The energy excess arises eventually from the rotational energy of the black hole. Here  we apply
the Penrose process to the spinning HD non-Kerr black hole. We now consider the
trajectory of such a negative energy particle in $D$ dimensions. The
equation of motion of such a particle can   be derived from the
Lagrangian $\mathcal{L}$,
\begin{equation}\label{lag}
\mathcal{L} = \frac{m}{2} g_{ij} \dot{x}^i \dot{x}^j,
\end{equation}
where an overdot denotes the derivative with respect to affine parameter
$\tau/m$ ($\tau$ being the proper time). Since the metric
(\ref{metric0}) is stationary and axis symmetric,  the motion of a
test particle with $D$-momentum $p^{i}$ is described by its rest mass
$m$, the total energy $E$ (as measured from $\infty$) is
${\partial\mathcal{L}}/{\partial\dot{t}} = p_{t} = -m E,$ and the
component of angular momentum is
${\partial\mathcal{L}}/{\partial\dot{\phi}} = p_{\phi} = m L.$ These
expressions on using the metric (\ref{metric0}) become
\begin{eqnarray}\label{E}
m E =\left[1+h\right]\left[ \left( 1-\frac{\mu}{r^{N-2}\Sigma}\right)\dot{t}
+\frac{\mu}{r^{N-2}\Sigma}a\sin^2\theta\;\dot{\phi}\right],
\end{eqnarray}
and
\begin{eqnarray}\label{L}
m L &=& -\left[1+h\right]\frac{\mu}{r^{N-2}\Sigma}a \sin^2\theta\hspace{0.1cm}\dot{t} \nonumber \\
&+& \left[\left(r^2+a^2+\frac{\mu}{r^{N-2}\Sigma}a^2\sin^2\theta\right)\sin^2\theta\right. \nonumber \\
&&+\left.ha^2\sin^4\theta\left(1+\frac{\mu}{r^{N-2}\Sigma}\right)\right]\dot{\phi}.
\end{eqnarray}
The conservation equation for the particles rest mass $p^{j}p_{j}=-m^{2}$ gives
\begin{eqnarray}\label{quade}
&& g_{\phi\phi}E^{2}+2 g_{t\phi}E L+g_{tt}L^{2}+\psi\left(g^{rr}p_{r}^{2}+g^{\theta\theta}p_{\theta}^{2}
+g_{\chi_{1}\chi_{1}}p_{\chi_{1}}^{2}\right.\nonumber\\
&& \left.+g^{\chi_{2}\chi_{2}}p\chi_{\chi_{2}}^{2}+ \cdots +g^{\chi_{N-1}\chi_{N-1}\chi}p_{\chi_{N-1}}^{2}+m^{2}\right)=0.
\end{eqnarray}
If the particle is constrained on the equatorial plane $\theta=\pi/2$, then $p_{\theta}=p_{\chi_{1}}=p_{\chi_{2}}=\cdots=0$. Equation (\ref{quade}) can be rewritten as
 \begin{equation}\label{quade1}
 \alpha E^{2}-2\beta E+\gamma +\delta p_{r}^{2} =0,
 \end{equation}
with
\begin{eqnarray}
\alpha &=& \frac{1}{\psi}\left(r^{2}+a^{2}+\frac{a^{2}h}{r^{2}}\left(r^{2}+\frac{\mu}{r^{N-2}}\right)+\frac{\mu a^{2}}{r^{N}}\right),\\
\beta &=& \frac{L}{\psi}\left(1+h\right)\frac{\mu a}{r^{N}},\\
\gamma &=& \frac{L^{2}}{\psi}\left(1+h\right)\left(1-\frac{\mu}{r^{N}}\right)-m^{2},\\
\delta &=& - \frac{r^{2}\left(1+h\right)}{\Delta +a^{2}h}p_{r}^{2}.
\end{eqnarray}
where $\psi = \left(1+h\right)\left(\Delta +a^{2}h\right)<0$ for $r>r_{+}$ (outer horizon). It is easy to check that $\psi=0\Leftrightarrow \Delta+ a^{2}h\sin^2\theta=0$.

\subsection{ Negative energy states in the Penrose  process}
In the Penrose process, we are interested in the region of
spacetime over which energy is negative and the orbit of particles
with negative energy in the ergosphere is very important to extract
energy from the black hole. As in the 4D Kerr, the negative energy states
occur due to counterrotating orbits. Unlike 4D Kerr or non-Kerr black holes,
where the black hole has two horizons, the HD non-Kerr black hole has just one
horizon where the 4-velocity of the counterrotating observer tends to
zero. In the Penrose process \cite{pen}, a particle falling onto a
black hole splits up into two particles at some $r>r_{+}$. The one particle
falling into a black hole has a negative energy (relative to $\infty$), and
 hence, the outgoing particle leaving the ergosphere has more energy than
the incident particle; thus energy is extracted. In fact, for the
$D$-momentum $p_{i}=m u_{i}$, the energy $E=-p^{i}\xi_{i}$ may
not be positive in the ergosphere; hence, one can extract energy from
the black hole by absorbing a particle with negative energy \cite{pen}. Now, we
shall focus on the effect of the deformation parameter and extra
dimension on the region of the negative energy state for the HD non-Kerr black hole.
In the HD non-Kerr black hole the orbit of the particle with the negative
energy obeys $\alpha>0$, $\beta<0$, $\gamma>\delta$.

\begin{figure*}
\begin{tabular}{c c}
\centering
\vspace{0cm}
\hspace{-1.1cm}
\includegraphics[scale=0.6]{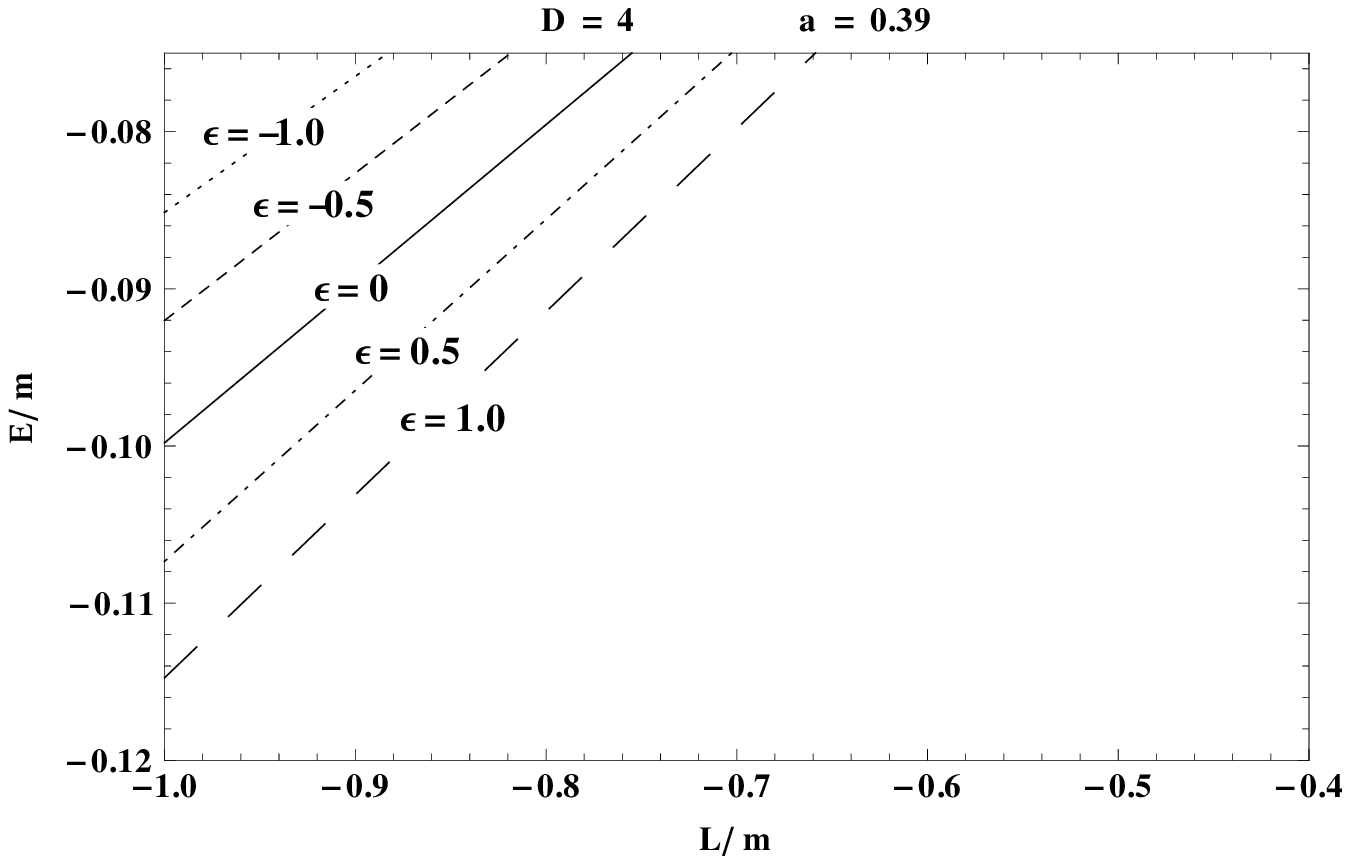}\hspace{-0.6cm}
&\includegraphics[scale=0.6]{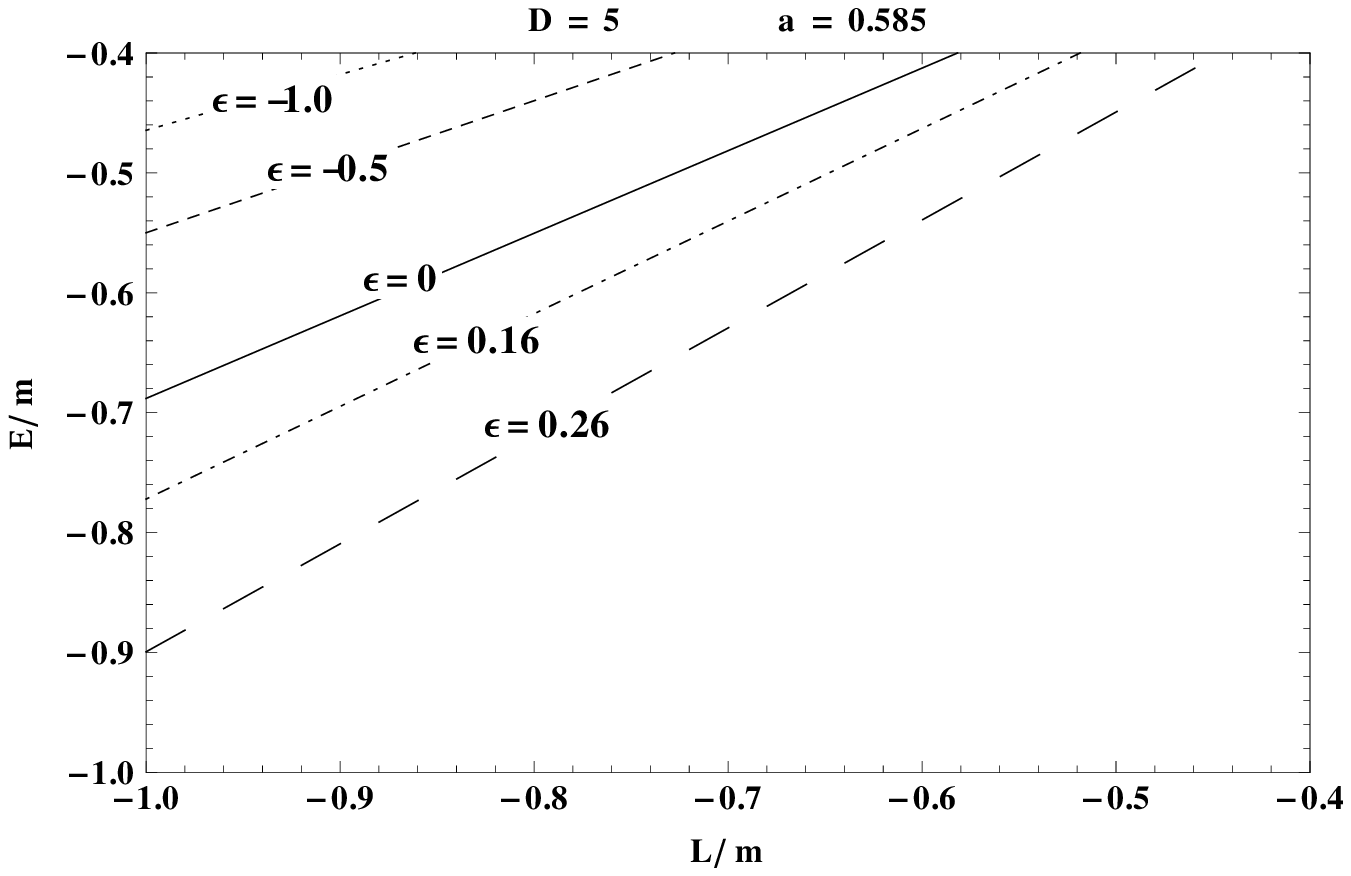}
\\
\hspace{-1cm}
\includegraphics[scale=0.6]{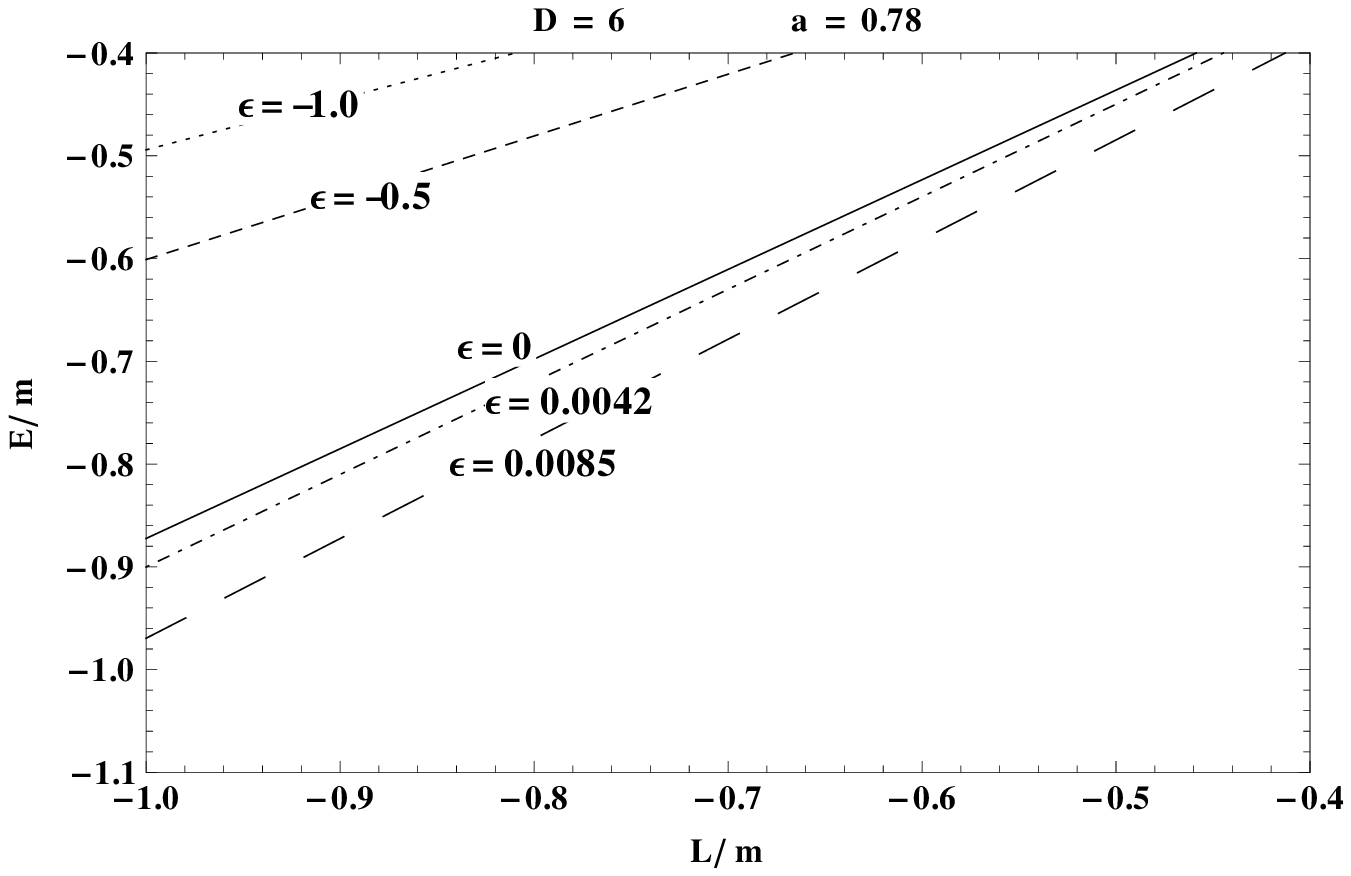}\hspace{-0.8cm}
&\includegraphics[scale=0.6]{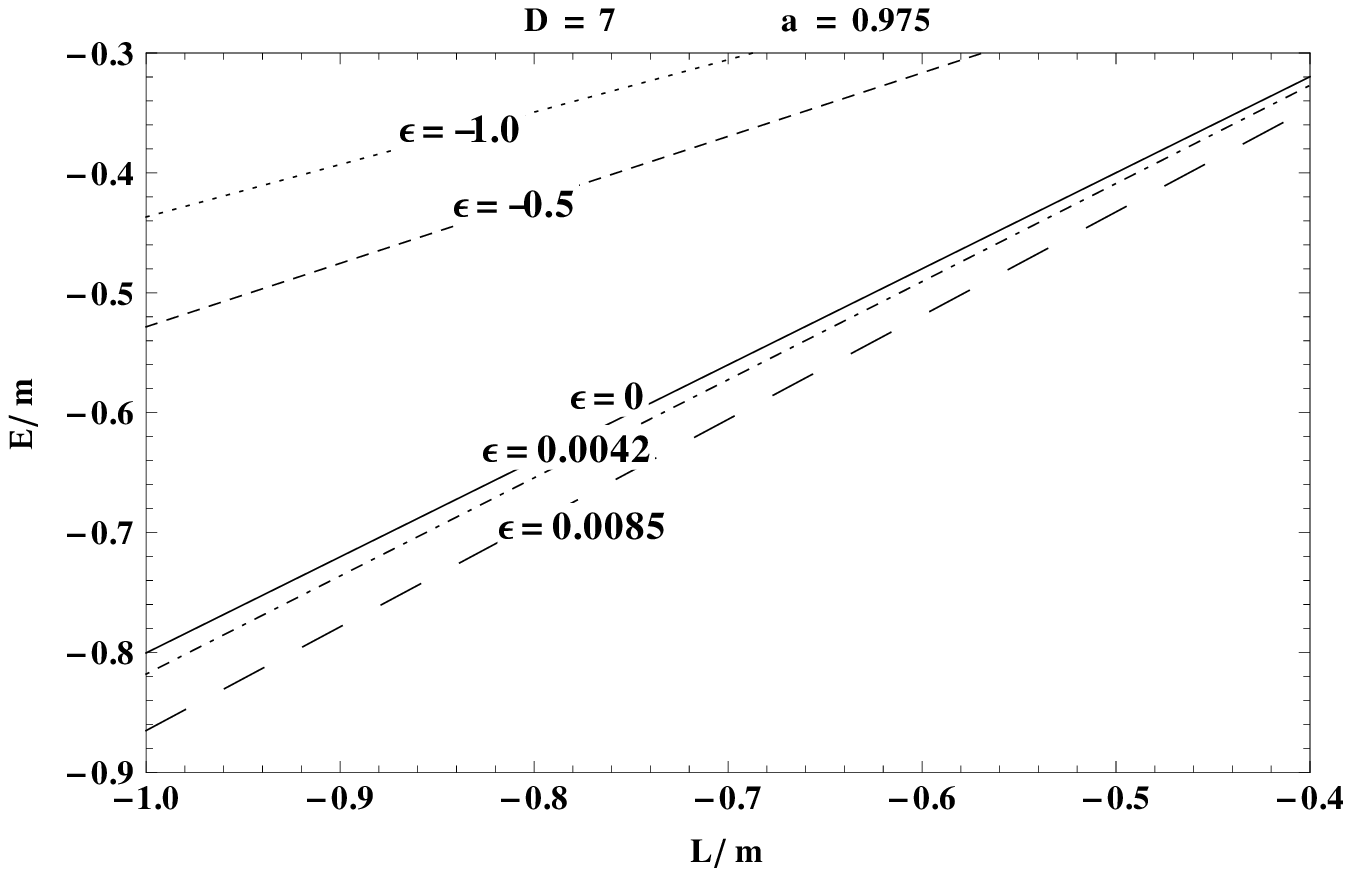}
\\
\hspace{-1cm}
\includegraphics[scale=0.6]{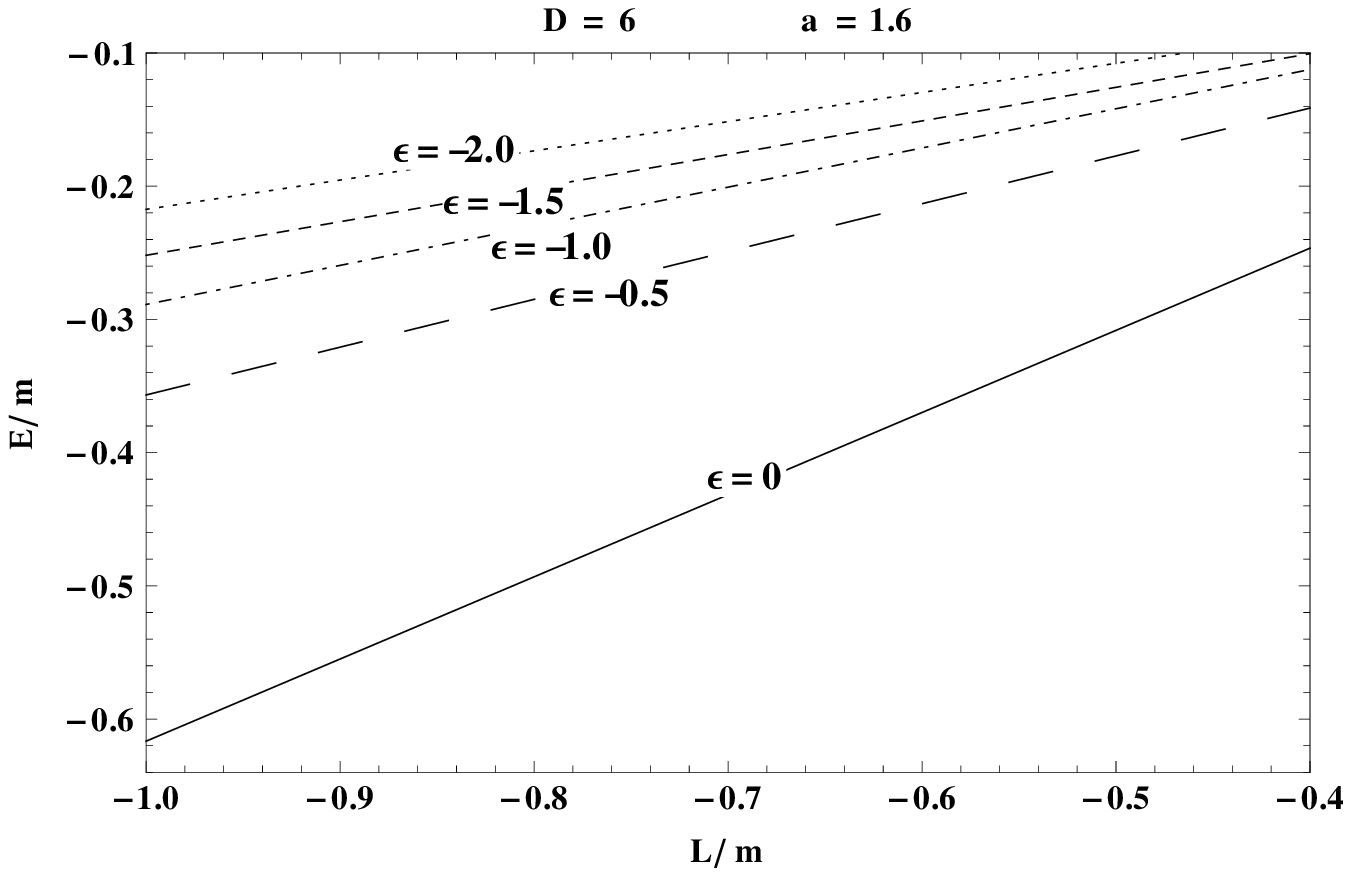}\hspace{-0.8cm}
&\includegraphics[scale=0.6]{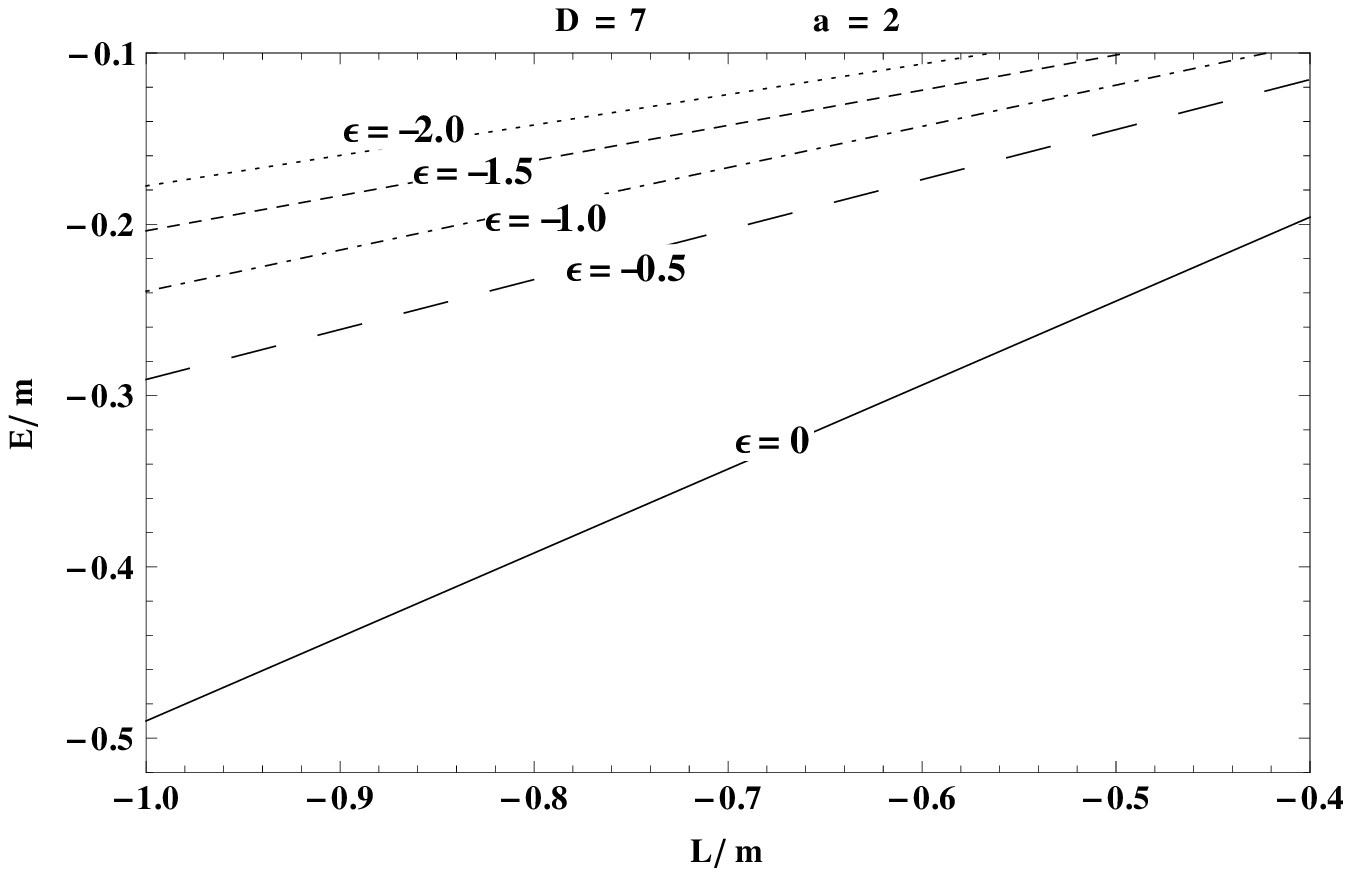}
\\
\hspace{-1cm}
\includegraphics[scale=0.6]{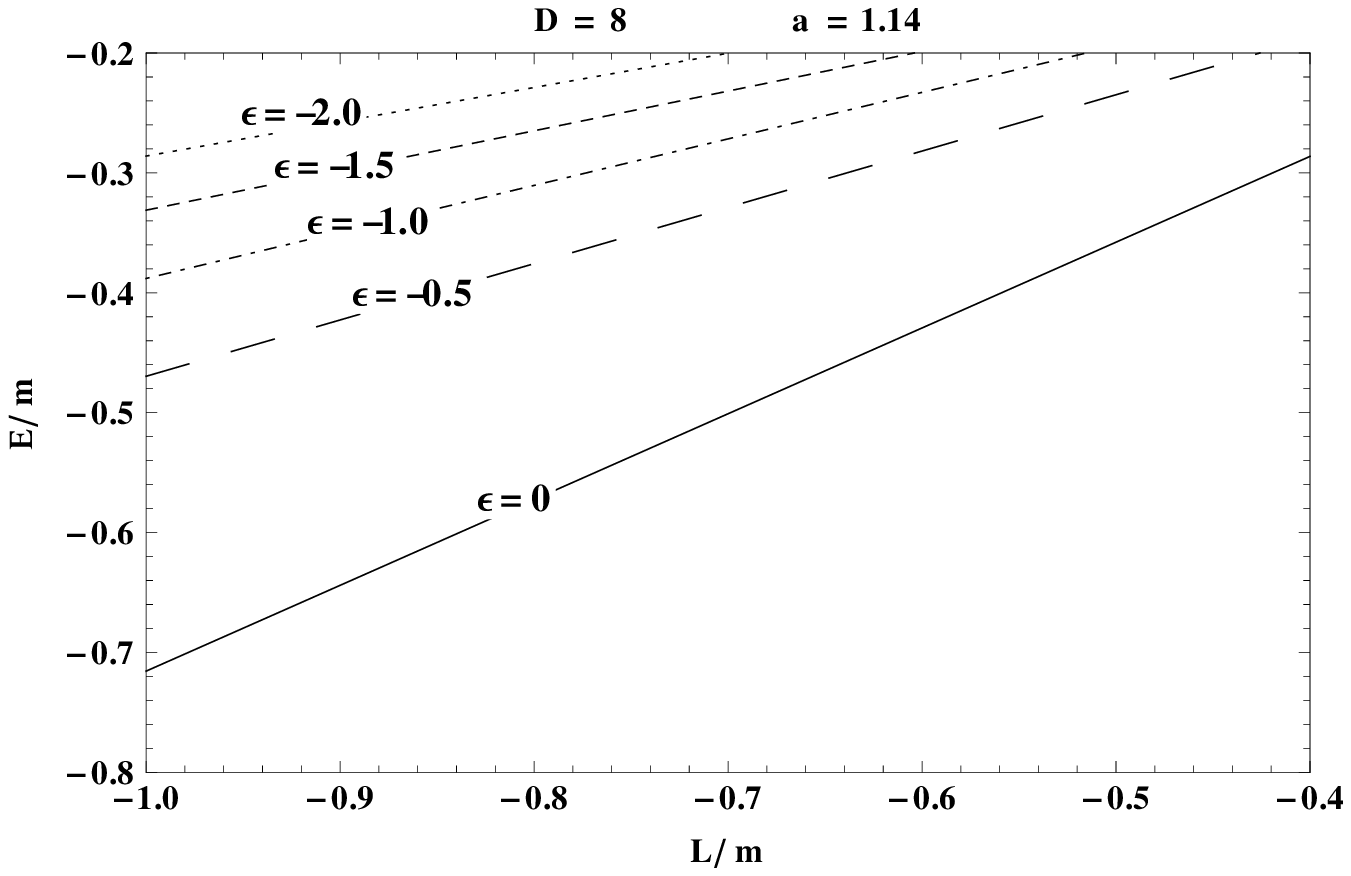}\hspace{-0.6cm}
&\includegraphics[scale=0.6]{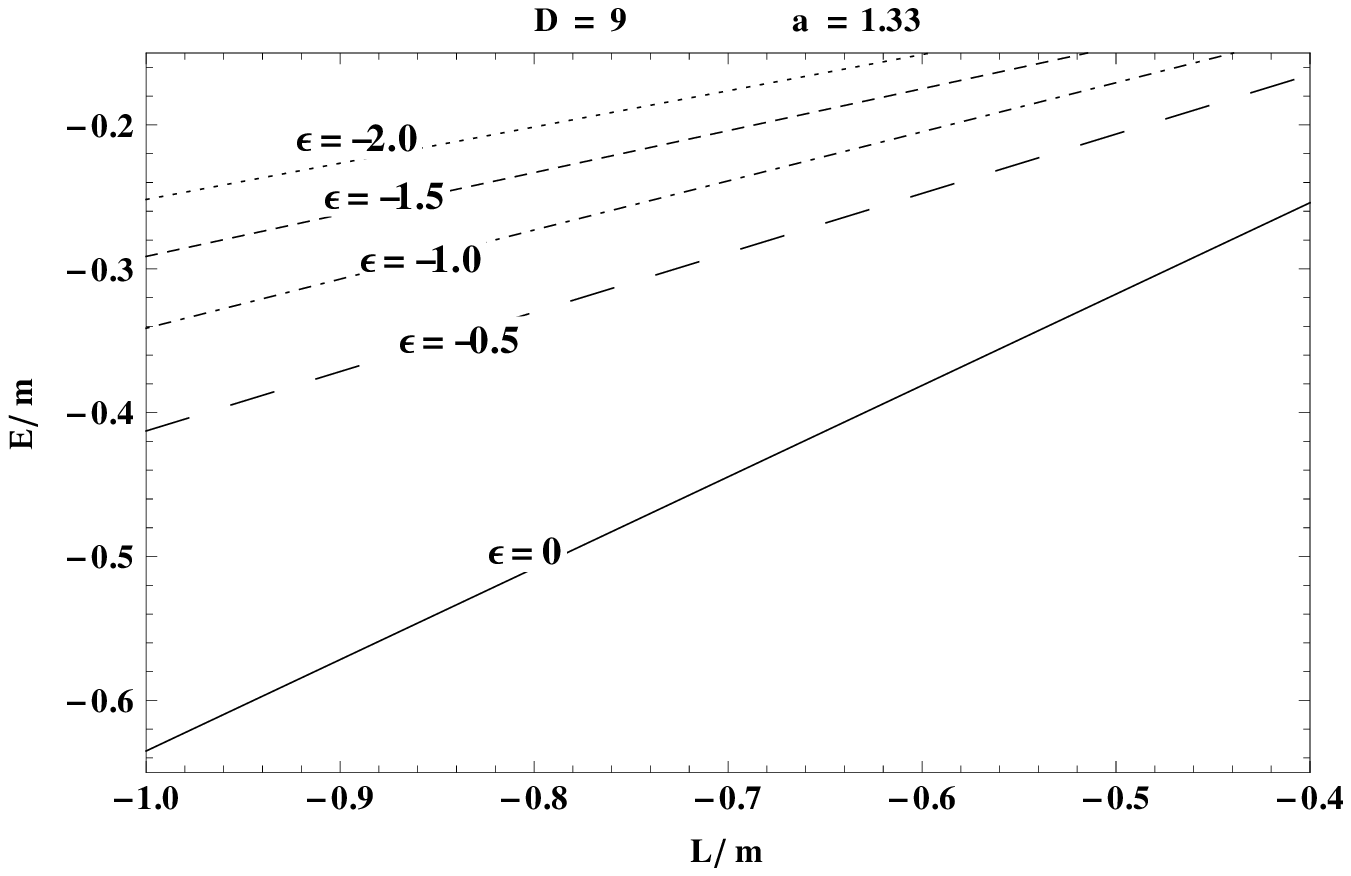}
\end{tabular}\vspace{-0.3cm}
\caption{The negative energy state $E$ allowed for the
angular momentum $L$ and rest mass $m$ for the particle in the
different dimensions ($D$ = 4,$\ldots$,9) with different values of deformation parameter
$\epsilon$ near the event horizon inside the
ergosphere.}\label{NESplot4a}
\end{figure*}

\begin{figure}
\includegraphics[scale=0.62]{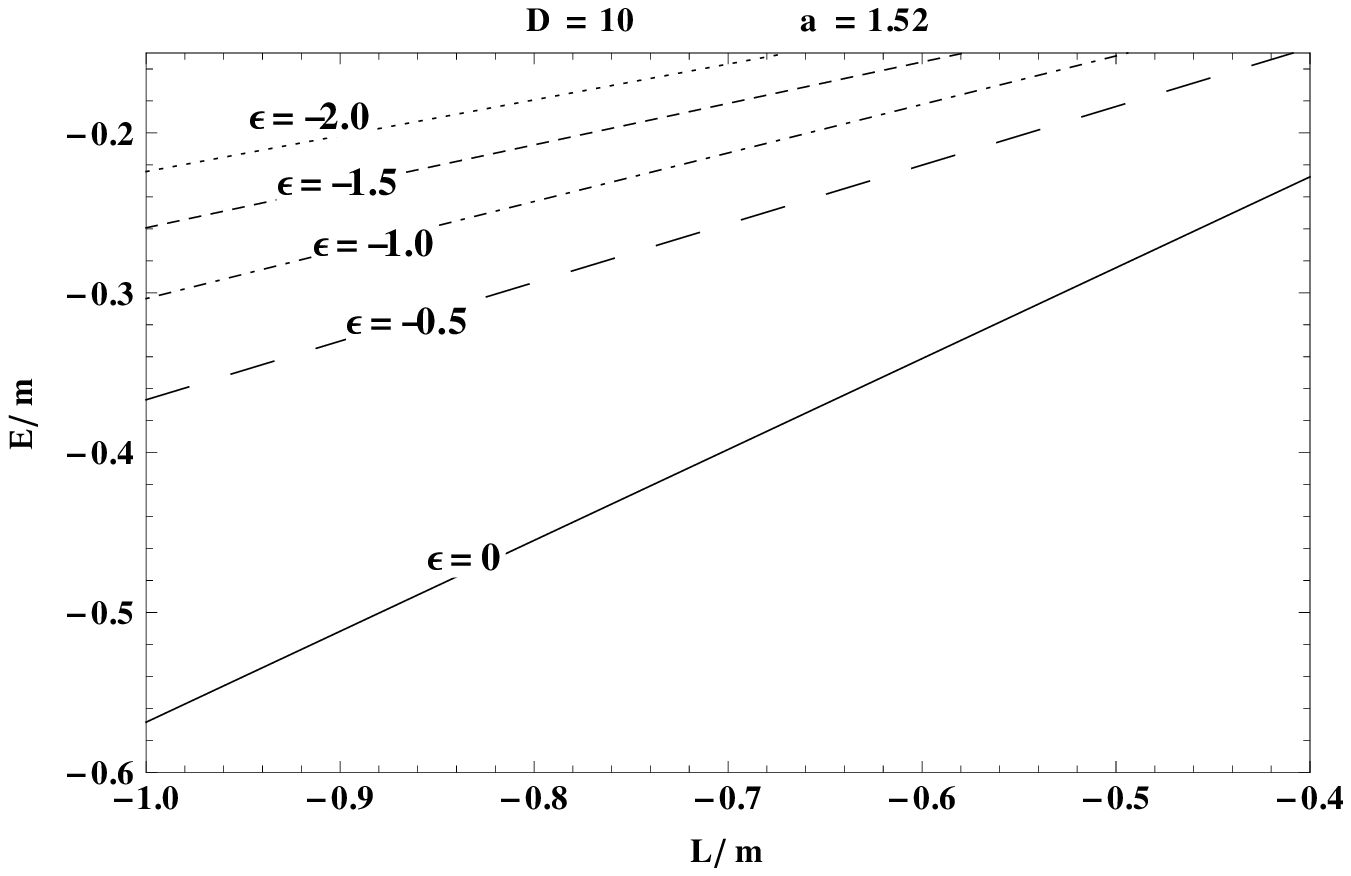}\hspace{-0.8cm}
\caption{The negative energy state $E$ allowed for the
angular momentum $L$ and rest mass $m$ for the particle in the
dimension ($D$ = 10) with different values of deformation parameter
$\epsilon$ near the event horizon inside the
ergosphere.}\label{NESplot4b}
\end{figure}

In Figs. \ref{NESplot4a} and \ref{NESplot4b}, we demonstrate that the negative energy states
near the horizon with different values of the deformation parameter in
different dimensions can be achieved if $La<0$. It is interesting to
note that the negative energy $E$ is sensitive to both deformation
parameter $\epsilon$ and extra dimension $D$. We see that the
extra dimension and deformation parameter $\epsilon$ favors negative
energy states; i.e., the negative energy $E$ increases with both an
increase in deformation parameter $\epsilon$ and extra dimension $D$.

\subsection{\label{efficiency}Efficiency of Penrose process}
One of the most interesting processes for extracting energy from a
rotating black hole is the Penrose process. As mentioned in the
Introduction, in the original Penrose paper \cite{pen}, we would
have to assume that the incoming particle in the ergosphere may be
decomposed into two subparticles and one of them with negative
energy will fall into the black hole, while the other is ejected to the exterior of the
ergosphere and will have more energy than original particle
\cite{pnfl,chr,bpt,bdd,pwdd}. Here we apply the recipe of energy
extraction proposed in \cite{pen} to the deformed HD Kerr black hole and use the
recipe provided by Bhat $et\; al.$ \cite{bdd}. In the energy extraction
process, we take the incident particle with the $D$-momentum
$p^{(in)}_{i}$, which breaks up into two particles $p^{(bh)}_{i}$ and
$p^{(out)}_{i}$ with the $D$-momentum in the ergosphere. We take the
total $D$-momentum as conserved at the point of break, which reads as
$p^{(in)}_{i} =p^{(bh)}_{i} + p^{(out)}_{i}$. Here the momentum of
the particles is non-spacelike  and  therefore lies inside the light
cone. To discuss the particles' energy, we consider the timelike
killing vector
\begin{eqnarray}
\xi^{(t)i}&=&\left(1,0,0,\ldots ,0 \right),\;\;
g_{ij}\xi^{(t)i}\xi^{(t)j}=g_{tt}<0,\nonumber \\
&&p^{(in)}_{i}\xi^{(t)i}=E^{(in)}=E^{(bh)}+E^{(out)}.\nonumber
\end{eqnarray}

Inside the ergosphere, the killing vector $\xi^{(t)i}$ becomes
spacelike vector and $g_{tt}>0$. Hence,
$E^{(bh)}=p^{(bh)}_{i}\xi^{(t)i}$ can possibly be negative. Thus
\begin{equation}
E^{(out)}=E^{(in)}-E^{(bh)}>E^{(in)}.
\end{equation}
and hence, we may say that the Penrose process extracts the rotational energy of the black hole.

\begin{figure*}
\begin{tabular}{c c}
\centering
\vspace{-0.3cm}
\hspace{-1cm}
\includegraphics[scale=0.6]{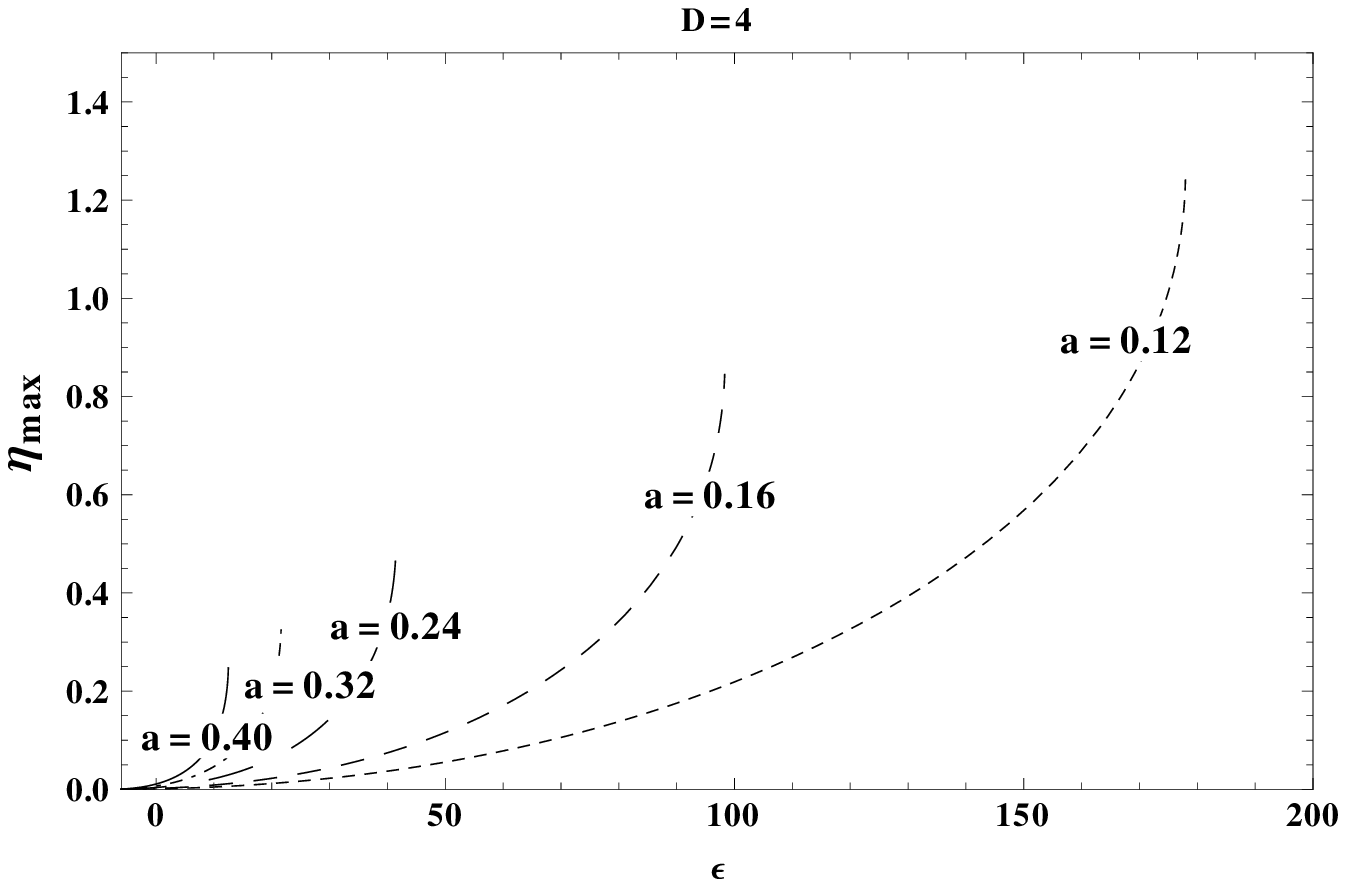}\hspace{-0.5cm}
&\includegraphics[scale=0.6]{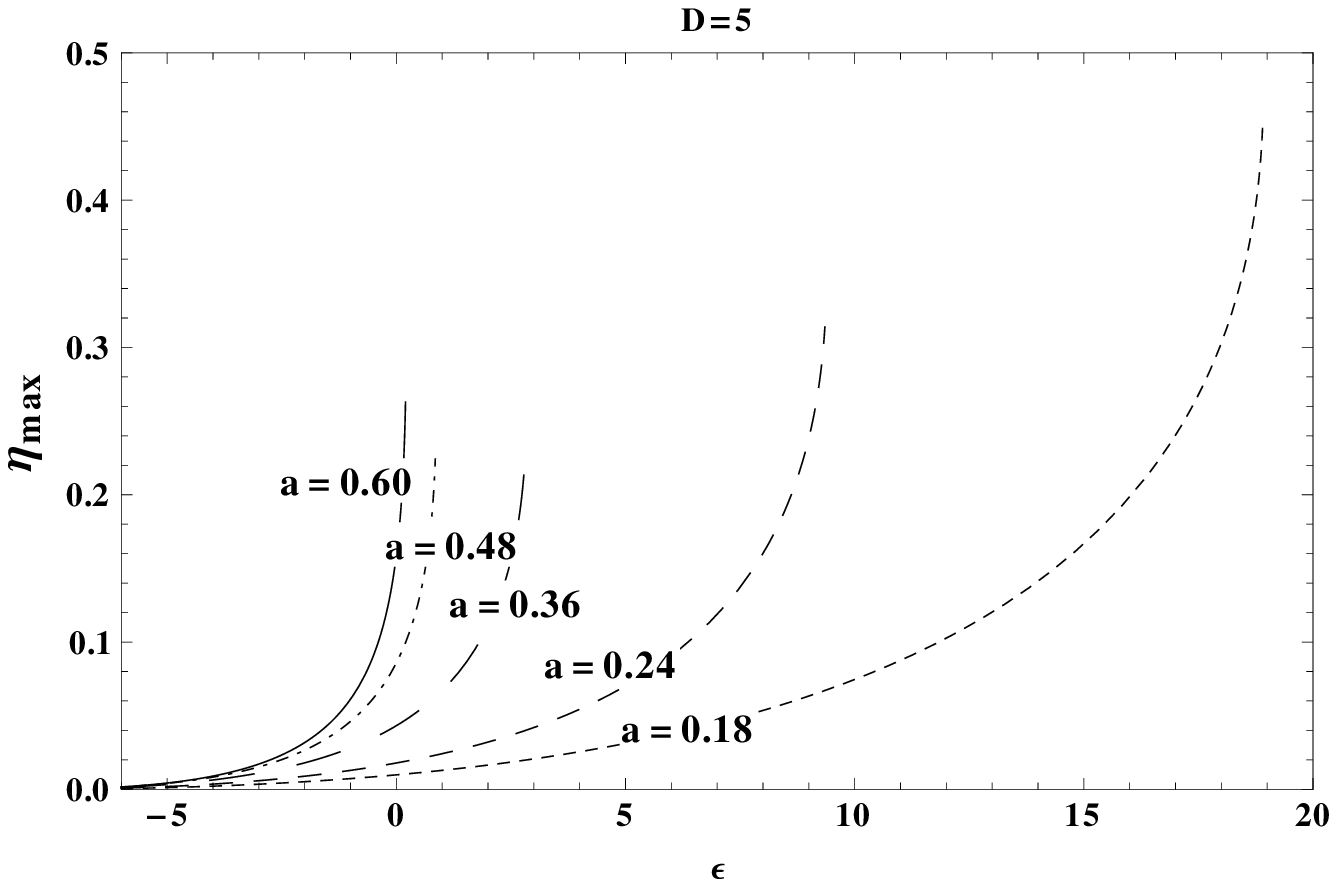}
\\
\hspace{-1cm}
\includegraphics[scale=0.6]{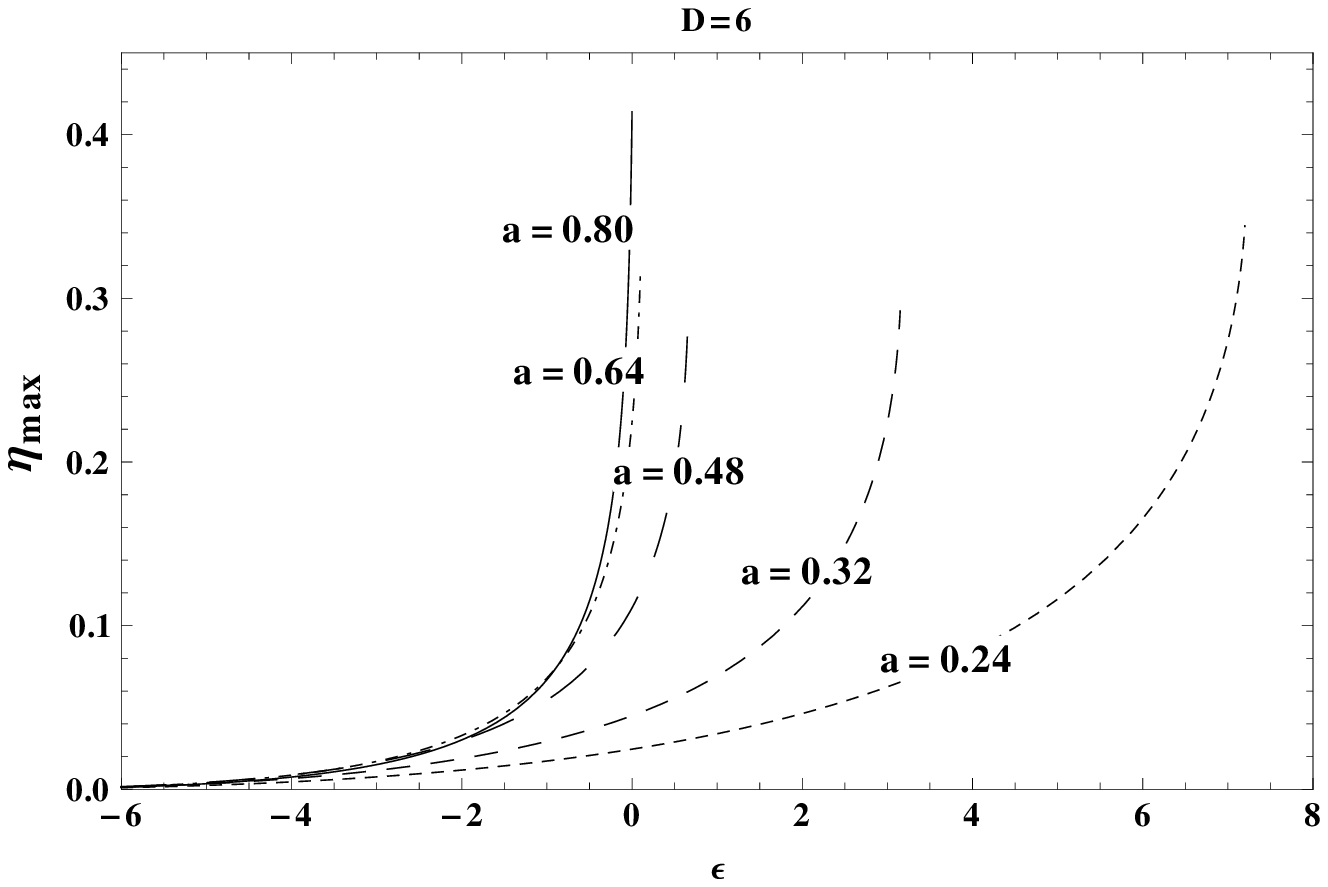}\hspace{-0.6cm}
&\includegraphics[scale=0.6]{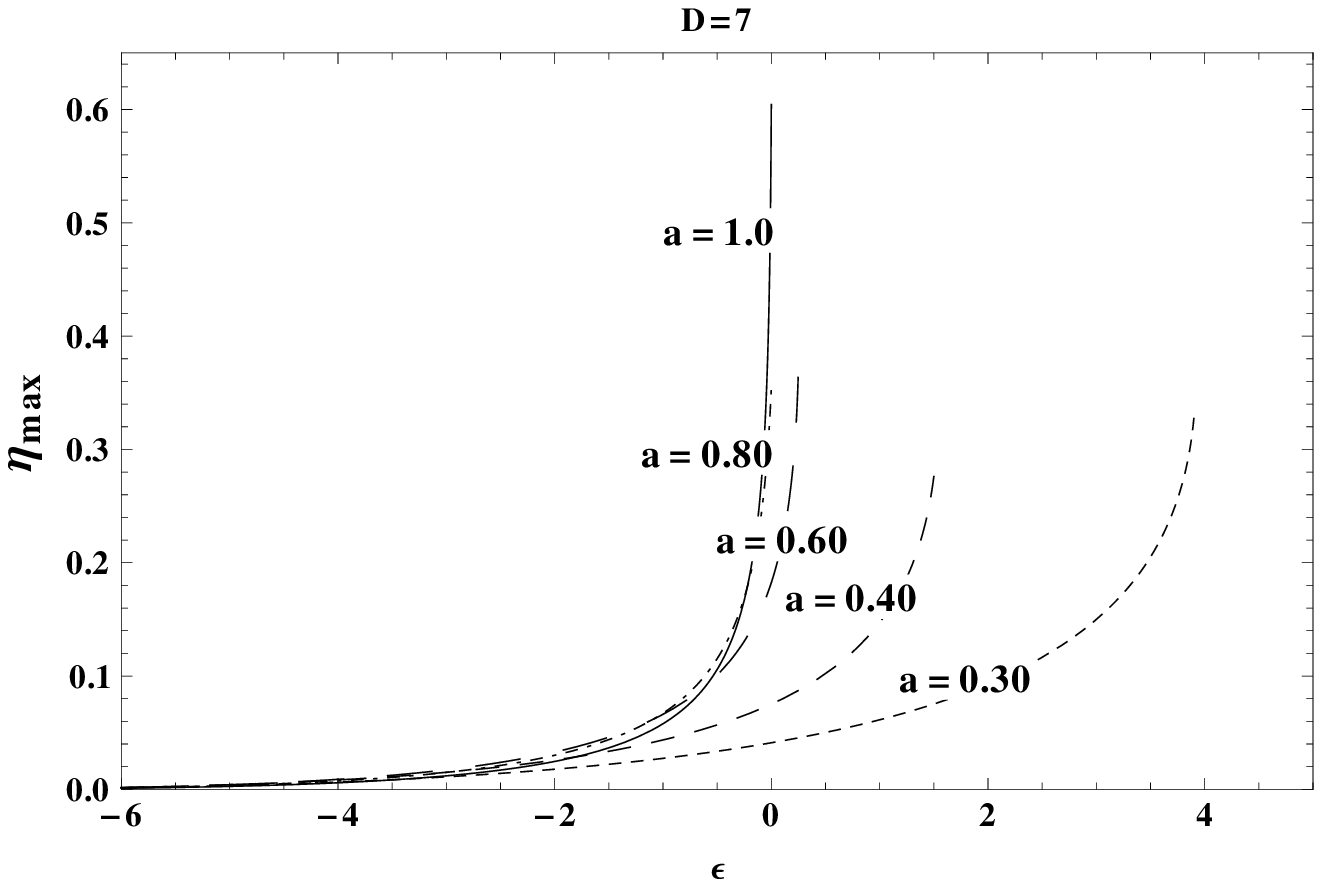}
\\
\hspace{-1cm}
\includegraphics[scale=0.6]{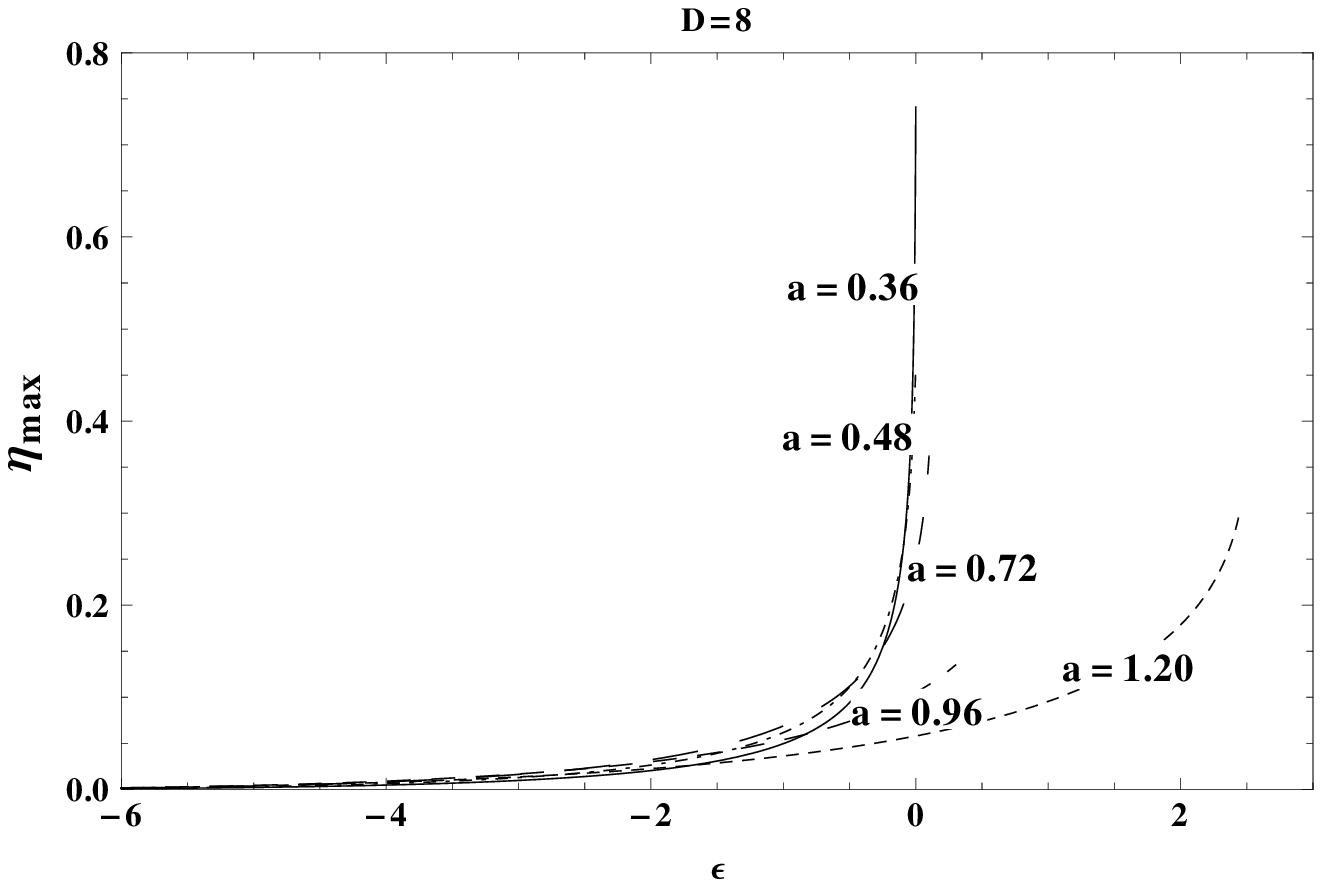}\hspace{-0.6cm}
&\includegraphics[scale=0.6]{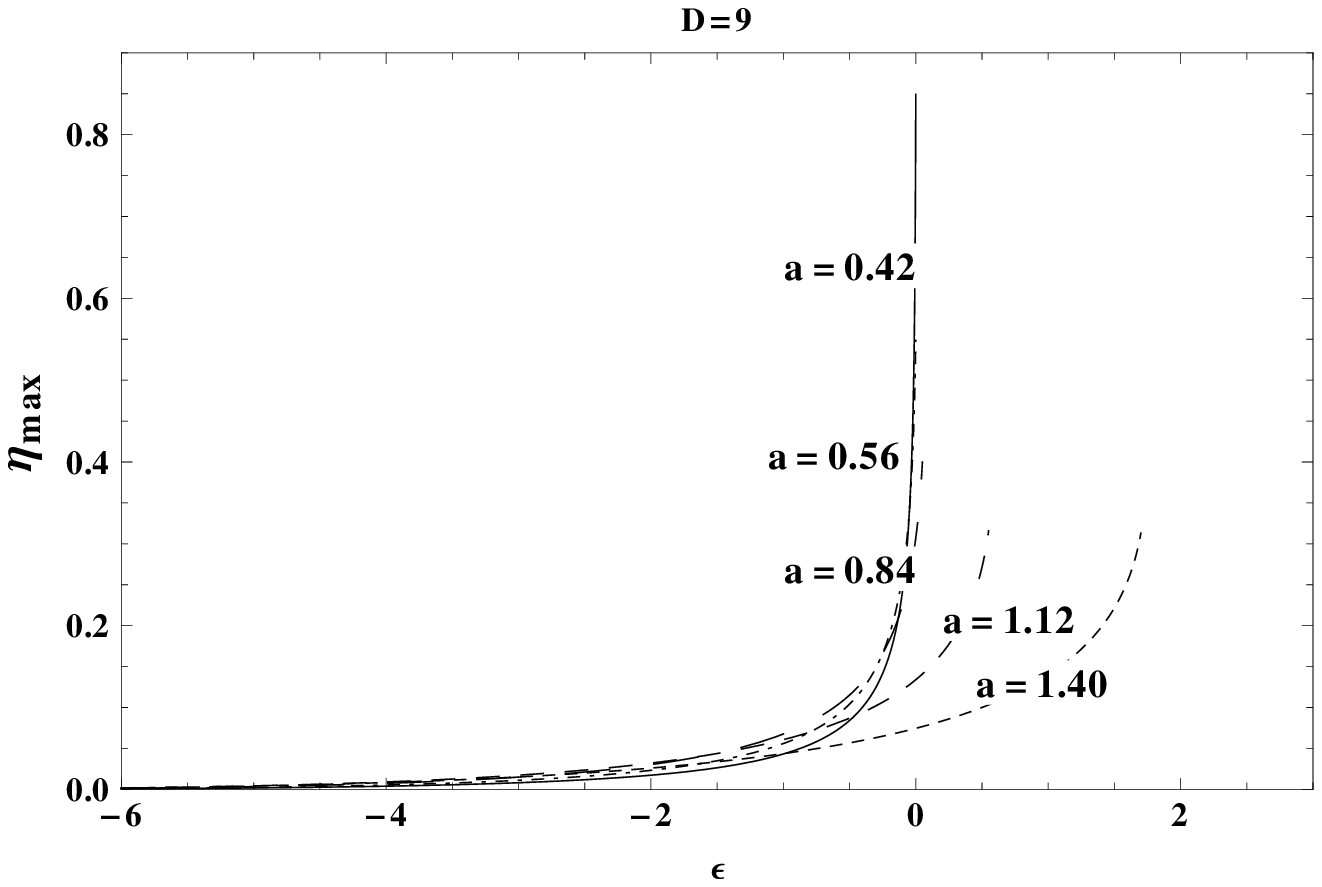}
\\
\hspace{-1cm}
\includegraphics[scale=0.6]{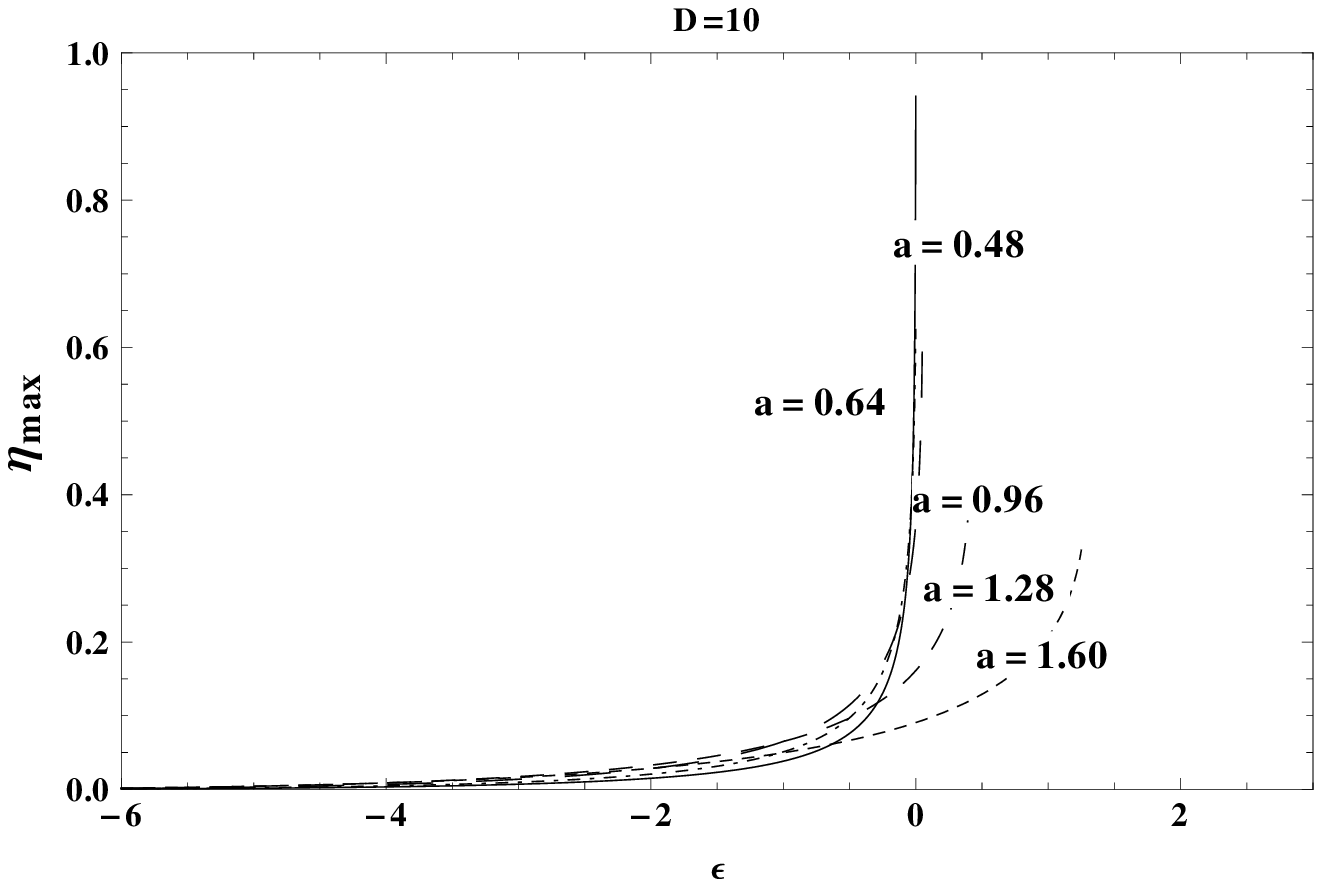}
\end{tabular}\vspace{-0.4cm}
\caption{The variation of the maximum efficiency of the energy
extraction process for different dimensions ($D$ = 4,$\ldots$,10) with the deformation
parameter $\epsilon$. }\label{efmax}
\end{figure*}

 We get the best result in the energy extraction process by choosing the rest mass $m^{(in)}$ 
 and energy $E^{(in)}$ of an incident particle equal to unity,
 which splits into two particles; the particle $(bh)$ absorbed by the HD non-Kerr
 black hole has $\Omega \rightarrow \Omega^{-}$ and the
particle $(out)$ escaping to infinity has $\Omega \rightarrow
\Omega^{+}$. From equation conservation of the energy and momentum
\begin{eqnarray}\label{cone}
E^{(in)} &=& E^{(out)}+E^{(bh)},\\
L^{(in)} &=& L^{(out)}+L^{(bh)}.
\end{eqnarray}
The particles falling into the black hole have energy $E^{(bh)}$ and angular momentum $L^{(bh)}$.
The particle falling into the black hole has negative energy, and hence the outgoing particle,
leaving the ergosphere, has more energy than the incident particle. The maximum efficiency is
obtained if we take the radial velocity component of velocities to be zero, at the point of the split.

Here, we are interested in the contribution of both the deformation parameter and the extra dimension on the
efficiency of the Penrose process for which we again rely on the prescription given in \cite{bdd}.
For this purpose, according to the conservation law of angular momentum, we have
\begin{equation}\label{consl}
U^{(in)} = m^{(bh)}U^{(bh)}+m^{(out)}U^{(out)},
\end{equation}
where $U^{(I)}_{i}(I=in,bh,out)$ denote the $D$-velocity of the particle at the point of the split, which can be expressed as
\begin{equation}
U^{(I)}_{i} = u_{t}\left(1,0,0,\Omega^{(I)},0,\ldots,0\right),
\end{equation}
where
\begin{equation}
u_{t} = \frac{E}{g_{tt}+g_{t\phi}\Omega^{I}},
\end{equation}
and
 \begin{widetext}
\begin{eqnarray}\label{ta}
\Omega^{(I)} =\frac{a\mu\left(1+h\right)\left[h\left(\mu-r^{N}\right)+\mu\right]\sqrt{-r^{2N}\left(1+h\right)\left[hr^{N}-\left(1+h\right)\mu\right]\left[a^{2}\left(1+h\right)r^{N}+r^{2}\left(r^{N}-\mu\right)\right]}}{a^{2}\left(1+h\right)\left[\left(1+h\right)\mu^{2}+r^{2N}+\mu r^{N}\right]+r^{N+2}}.
\end{eqnarray}
\end{widetext}
In the general relativity limit, $h\rightarrow0$, $N\rightarrow1$, Eq. (\ref{ta}) becomes
\begin{equation}
\Omega^{(I)}=\frac{\sqrt{2}r^{2}\sqrt{\frac{M}{r}\left(r^{2}-2Mr+a^{2}\right)}+4aM^{2}}{a^{2}\left(r^{2}+2Mr+4M^{2}\right)+r^{4}},
\end{equation}
Here $\Omega^{(I)}$ is the asymptotic angular velocity of the $i$th
particle.  The angular velocity of particles after the split
always lies within the future directed light cone and hence
is constrained in between  $\Omega^{-}<\Omega^{I}<\Omega^{+}$

,where
\begin{widetext}
\begin{eqnarray}\label{ct}
\Omega_{\pm}= \frac{-g_{t\phi}\pm \sqrt{-\psi}}
{g_{\phi\phi}}=\frac{a\mu\left(1+h\right)\pm
\sqrt{\left(1+h\right)\left[r^{2N}a^{2}\left(1+h\right)+r^{N+2}\left(r^{N}-\mu\right)\right]}}{a^{2}\mu
\left(1+h\right)+a^{2}r^{N}\left(1+h\right)+a^{2}r^{N+2}\left(1+h\right)}.
\end{eqnarray}
\end{widetext}
In the general relativity limit, $h\rightarrow0$, $N\rightarrow1$, Eq. (\ref{ct}) reads as
\begin{equation}
\Omega_{\pm} = \frac{2aM \pm \sqrt{\left(a r\right)^{2}+r^{3}\left(-2M+r\right)}}{r^{3} + a^{2} \left(2 M+r\right)}
\end{equation}
It turns out that the maximal output will be gained as $\Omega^{(bh)}\rightarrow \Omega^{-}$ and $\Omega^{(out)}\rightarrow \Omega^{+}$.

The important question in the black hole energy extraction process is the
efficiency of the process. It is supposed to be one of the many
important parameters in active galactic nuclei. Hence, it is very
relevant to examine the efficiency of the Penrose process. In 
Fig.~\ref{efmax}, we plot the maximum efficiency $\eta_{max}$ versus
the deformation parameter $\epsilon$. From the figure we conclude
that as the deformation parameter increases from $\epsilon<0$ to
$\epsilon>0$ the maximum efficiency $\eta _{max}$ of the energy
extraction process increases with the dimensions. We further see as
the dimension increases the maximum efficiency increases for the same
value of the spin parameter. In Tables~\ref{eff1}-\ref{eff4}, we show
the variation of maximum efficiency $\eta_{max}$ corresponding to
the different values of deformation parameter $\epsilon$ and the
spin parameter $a$. The value of the maximum efficiency increases in $4$D
for the increase in the value of deformation parameter $\epsilon$
(Table~\ref{eff1}). Similarly, the maximum efficiency also increases
with the increase in the spin parameter $a$ for the same value of
deformation parameter $\epsilon$. Also we see that the value of the
maximum efficiency first increases with the increase in value spin
parameter $a$ and then starts decreasing as the value of the spin
parameter $a>1$ (Tables~\ref{eff2}-\ref{eff4}). Further, we also
conclude that as the value of the spin parameter $a>0.75$ we get the
maximum efficiency $\eta_{max}$ only for the values of the
deformation parameter $\epsilon\leq0$.

Defining then the efficiency $\eta$ of the process as a gain in energy per input of energy, i.e.,
\begin{equation}
\eta = \frac{m^{(out)}E^{(out)}-E^{(in)}}{E^{(in)}}=m^{(out)}\frac{E^{(out)}}{E^{(in)}}-1.
\end{equation}

Now, using momentum conservation Eq. (\ref{consl}), we find that
\begin{equation}\label{eta1}
m^{(out)} \frac{E^{(out)}}{E^{(in)}}=\frac{\left(\Omega^{(in)}-\Omega^{-}\right)\bigg(1-\frac{\mu}{r^{N}}\left(1+\Omega^{+} a\right)\bigg)}{\left(\Omega^{+}-\Omega^{-}\right)\bigg(1-\frac{\mu}{r^{N}}\left(1+\Omega^{(in)} a\right)\bigg)}.
\end{equation}

Now in the limit when the split tends to $r_{+}$,
\begin{eqnarray}
\eta_{max} &=& \frac{\sqrt{1+g_{tt}}-1}{2}\bigg|_{r=r_{+}}\\
&=& \frac{1}{2}\left[\left(\frac{\mu}{r^{N}}\right)^{\frac{1}{2}}\sqrt{1-\frac{\epsilon \mu^{2}}{8 r^{2N}}\left(1+\frac{\mu}{r^{N}}\right)}-1\right]_{r=r_{+}}.
\end{eqnarray}
It is known that $\eta_{max}\sim 20.7\%$ for the extreme Kerr black hole
\citep{bdd}, which is amplified to $60\%$ for the deformed Kerr black hole.
However, there is no upper limit on $\eta$ in HD.

\begin{table*}[h]
\begin{center}
\caption{The maximum energy efficiency $\eta_{max}$ of
energy extraction in the HD non-Kerr black hole  for different values of deformation parameter  $\epsilon$ corresponding to dimension $D$ = 4. \label{eff1}}
\resizebox{\textwidth}{!}{%
\begin{tabular}{ c c c c c c c c c c c c c }
\hline\hline
$\epsilon$ & a = 0.1 & a = 0.2 & a = 0.3 & a = 0.4 & a = 0.5 & a = 0.6 & a = 0.7 & a = 0.8 & a = 0.9 & a = 1.0 & a = 1.1 \\
\hline               
-0.5 & 0.055\% & 0.222\% & 0.509\% & 0.926\% & 1.493\% & 2.237\% & 3.197\% & 4.424\% & 5.972\% & 7.838\% & 9.756\% \\
-0.4 & 0.056\% & 0.228\% & 0.524\% & 0.955\% & 1.545\% & 2.324\% & 3.344\% & 4.674\% & 6.416\% & 8.643\% & 11.099\% \\
-0.3 & 0.058\% & 0.235\% & 0.539\% & 0.985\% & 1.597\% & 2.414\% & 3.497\% & 4.944\% & 6.919\% & 9.650\% & 13.002\% \\
-0.2 & 0.059\% & 0.241\% & 0.554\% & 1.015\% & 1.651\% & 2.508\% & 3.659\% & 5.236\% & 7.498\% & 10.996\% & 16.212\% \\
-0.1 & 0.061\% & 0.247\% & 0.570\% & 1.046\% & 1.707\% & 2.604\% & 3.829\% & 5.554\% & 8.180\% & 13.061\% & 24.877\% \\
   0 & 0.062\% & 0.254\% & 0.585\% & 1.077\% & 1.764\% & 2.704\% & 4.008\% & 5.901\% & 9.009\% & 20.710\% &          \\
 0.1 & 0.064\% & 0.261\% & 0.602\% & 1.109\% & 1.822\% & 2.808\% & 4.197\% & 6.284\% & 10.072\% &          &          \\
 0.2 & 0.066\% & 0.267\% & 0.618\% & 1.141\% & 1.881\% & 2.915\% & 4.397\% & 6.709\% & 11.580\% &          &          \\
 0.3 & 0.067\% & 0.274\% & 0.635\% & 1.175\% & 1.943\% & 3.026\% & 4.609\% & 7.187\% & 14.593\% &          &          \\
 0.4 & 0.069\% & 0.281\% & 0.652\% & 1.209\% & 2.005\% & 3.141\% & 4.835\% & 7.732\% &         &          &          \\
 0.5 & 0.071\% & 0.288\% & 0.669\% & 1.243\% & 2.070\% & 3.261\% & 5.073\% & 8.366\% &         &          &          \\
\hline  
\end{tabular}}
\end{center}
\end{table*}

\begin{table*}[h]
\begin{center}
\caption{The maximum energy efficiency $\eta_{max}$ of
energy extraction in the HD non-Kerr black hole  for different values of deformation parameter  $\epsilon$ corresponding
to dimension $D$ = 5.\label{eff2}}
\resizebox{\textwidth}{!}{%
\begin{tabular}{ c c c c c c c c c c c c c c }
\hline\hline
 $\epsilon$ & a = 0.15 & a = 0.30 & a = 0.45 & a = 0.6 & a = 0.75 & a = 0.9 & a = 1.05 & a = 1.2 & a = 1.35 & a = 1.5 & a = 1.65 \\
\hline               
-0.5 & 0.587\% & 2.400\% & 5.457\% & 9.071\% & 11.289\% & 11.345\% & 10.295\% & 8.985\% & 7.747\% & 6.672\% & 5.769\%\% \\
-0.4 & 0.604\% & 2.491\% & 5.772\% & 9.937\% & 12.814\% & 13.107\% & 11.978\% & 10.489\% & 9.062\% & 7.816\% & 6.765\% \\
-0.3 & 0.622\% & 2.584\% & 6.111\% & 10.963\% & 14.819\% & 15.528\% & 14.323\% & 12.597\% & 10.912\% & 9.431\% & 8.175\% \\
-0.2 & 0.639\% & 2.680\% & 6.476\% & 12.209\% & 17.668\% & 19.214\% & 17.961\% & 15.893\% & 13.820\% & 11.978\% & 10.407\% \\
-0.1 & 0.658\% & 2.797\% & 6.870\% & 13.780\% & 22.390\% & 26.197\% & 25.079\% & 22.417\% & 19.618\% & 17.089\% & 14.910\% \\
   0 & 0.676\% & 2.882\% & 7.299\% & 15.887\% & 36.089\% & 183.785\% &         &         &      &          &          \\
 0.1 & 0.694\% & 2.987\% & 7.768\% & 19.073\% &          &           &         &         &     &          &          \\
 0.2 & 0.713\% & 3.096\% & 8.282\% & 26.289\% &          &           &         &         &     &          &          \\
 0.3 & 0.733\% & 3.208\% & 8.850\% &          &          &           &         &         &     &          &          \\
 0.4 & 0.752\% & 3.324\% & 9.484\% &          &          &           &         &         &     &          &          \\
 0.5 & 0.772\% & 3.444\% & 10.198\% &          &         &           &         &         &     &          &          \\
\hline  
\end{tabular}}
\end{center}
\end{table*}

\begin{table*}
\begin{center}
\caption{The maximum energy efficiency $\eta_{max}$ of energy
extraction in the HD non-Kerr black hole  for different values of
deformation parameter   $\epsilon$ corresponding to dimension $D$ = 6.
\label{eff3}}
\resizebox{\textwidth}{!}{%
\begin{tabular}{c c c c c c c c c c c c c}
\hline\hline
$\epsilon$ & a = 0.2 & a = 0.4 & a = 0.6 & a = 0.8 & a= 1.0 & a = 1.2 & a = 1.4 & a = 1.6 & a = 1.8 & a = 2 & a = 2.2 \\
\hline               
-0.5 & 1.437\% & 5.534\% & 10.131\% & 11.516\% & 10.373\% & 8.663\% & 7.111\% & 5.852\% & 4.859\% & 4.079\% & 3.461\% \\
-0.4 & 1.484\% & 5.844\% & 11.202\% & 13.245\% & 12.156\% & 10.247\% & 8.457\% & 6.985\% & 5.815\% & 4.891\% & 4.157\% \\
-0.3 & 1.532\% & 6.175\% & 12.489\% & 15.562\% & 14.652\% & 12.505\% & 10.398\% & 8.630\% & 7.210\% & 6.080\% & 5.179\% \\
-0.2 & 1.581\% & 6.529\% & 14.084\% & 18.925\% & 18.525\% & 16.106\% & 13.538\% & 11.319\% & 9.509\% & 8.053\% & 6.815\% \\
-0.1 & 1.631\% & 6.910\% & 16.150\% & 24.645\% & 26.004\% & 23.403\% & 20.063\% & 17.003\% & 14.428\% & 12.315\% & 10.591\% \\
   0 & 1.682\% & 7.320\% & 19.027\% & 41.401\% & 82.070\% & 149.458\% & 249.773\% & 387.476\% & 566.795\% & 792.207\% & 1068.410\% \\
 0.1 & 1.734\% & 7.763\% & 23.682\% &          &          &           &           &           &                                              &            &            \\
 0.2 & 1.788\% & 8.243\% &          &          &          &           &           &           &  &           &            \\
 0.3 & 1.842\% & 8.767\% &          &          &          &           &           &           &  &           &            \\
 0.4 & 1.898\% & 9.340\% &          &          &          &           &           &           &           &            \\
 0.5 & 1.954\% & 9.972\% &          &          &          &           &           &           &  &           &            \\
\hline  
\end{tabular}}
\end{center}
\end{table*}

\begin{table*}
\begin{center}
\caption{The maximum energy efficiency $\eta_{max}$ of energy
extraction in the HD non-Kerr black hole  for different values of
deformation parameter   $\epsilon$ corresponding to dimension $D$ = 7.
\label{eff4}}
\resizebox{\textwidth}{!}{%
\begin{tabular}{ c c c c c c c c c c c c c }
\hline\hline
 $\epsilon$ & a = 0.25 & a = 0.5 & 0.75 & a = 1.0 & a = 1.25 & a = 1.5 & a = 1.75 & a = 2.0 & a = 2.25 & a = 2.5 & a = 2.75 \\
\hline               
-0.5 & 2.362\% & 8.112\% & 11.476\% & 10.581\% & 8.573\% & 6.779\% & 5.392\% & 4.348\% & 3.562\% & 2.960\% & 2.494\% \\
-0.4 & 2.448\% & 8.738\% & 13.062\% & 12.415\% & 10.204\% & 8.133\% & 6.501\% & 5.261\% & 4.320\% & 3.597\% & 3.035\% \\
-0.3 & 2.537\% & 9.438\% & 15.111\% & 14.970\% & 12.544\% & 10.106\% & 8.134\% & 6.614\% & 5.450\% & 4.550\% & 3.847\% \\
-0.2 & 2.628\% & 10.227\% & 17.922\% & 18.891\% & 16.298\% & 13.344\% & 10.852\% & 8.889\% & 7.363\% & 6.173\% & 5.236\% \\
-0.1 & 2.722\% & 11.127\% & 22.193\% & 26.237\% & 23.925\% & 20.185\% & 16.736\% & 13.900\% &   11.635\% & 9.834\% & 8.394\% \\
   0 & 2.819\% & 12.166\% & 30.463\% & 60.445\% & 103.491\% & 159.506\% & 227.914\% & 308.220\% & 400\% & 503.320\% & 617.808\% \\
 0.1 & 2.918\% & 13.389\% &          &          &           &           &           &           &       &           &           \\
 0.2 & 3.021\% & 14.860\% &          &          &           &           &           &           &       &           &           \\
 0.3 & 3.126\% & 16.693\% &          &          &           &           &           &           &       &           &           \\
 0.4 & 3.234\% & 19.101\% &          &          &           &           &           &           &       &           &           \\
 0.5 & 3.345\% & 22.601\% &          &          &           &           &           &           &       &           &           \\
\hline  
\end{tabular}}
\end{center}
\end{table*}

\section{conclusions}
The non-Kerr black hole solution has an additional  deformation parameter
$\epsilon$ than the Kerr black hole, and it produces deviation from Kerr geometry
but with a richer configuration of the ergosphere. This motivates us to
reconsider the Penrose process in the non-Kerr black hole scenario as energy
is extracted from the ergosphere.

We have discussed the energy extraction via the Penrose  process from HD
non-Kerr black holes. However, there are problems with this method as well.
Penrose himself said that the method is inefficient \cite{pen},
although later \cite{cs} he showed that the theoretical efficiency
could reach $20\%$ extra energy up to $60\%$. We have studied in
detail the influence of the deformation parameter $\epsilon$ and
extra dimensions on the structure of horizons and the ergosphere of the Kerr
black hole. We can conclude that energy extraction via the Penrose process is
more realistic in the HD non-Kerr black hole as its efficiency is enhanced in the HD
non-Kerr case. The presence of the deformation parameter and extra
dimension influence the behavior of horizons, ergospheres, and
negative energy states. It is interesting to see that the HD non-Kerr black hole
($D\geq6$) can have two horizons for small values of deformation
parameters $\epsilon$, whereas the Myers-Perry black hole ($D\geq6$) has just
one horizon. The higher values of $\epsilon>0$ is not viable as it
leads to disconnected horizons (Figs. \ref{ergosphere1a}-
\ref{ergosphere2a}). We have also calculated the condition on $\epsilon$
for the proper horizons in the HD non-Kerr black hole. It has been demonstrated
that the Penrose process is more efficient than the Myers-Perry black hole or the HD
Kerr black hole ($\epsilon=0$) with no upper bound in the efficiency
(Tables~{\ref{eff2}, \ref{eff3}, and \ref{eff4}}) in contrast to the 4D
Kerr black hole where the maximum efficiency is just $20\%$. Further, it is
seen that the ergosphere is sensitive to both deformation parameter
$\epsilon$ and extra dimension; both of them individually lead
the enlargement of negative energy states and also facilitate the
Penrose process by enhancing the efficiency of the energy extraction
process. However, in the presence of both, i.e., in the HD non-Kerr black hole,
we expect only a tiny gain in efficiency of the Penrose process.

\acknowledgments
S. G. G. and P. S. thank IUCAA, Pune, for kind hospitality where a part
of this work was done, and one of the authors (S. G. G.) also thanks the
University Grant Commission (UGC) for the major research project grant F.
NO. 39-459/2010 (SR).

\end{document}